\documentclass[twocolumn,showpacs,superscriptaddress]{revtex4}

\usepackage{color}
\usepackage{graphicx}
\usepackage{lipsum}

\newcommand{\beq}{\begin{equation}}
\newcommand{\eeq}{\end{equation}}
\newcommand{\barr}{\begin{eqnarray}}
\newcommand{\earr}{\end{eqnarray}}

\newcommand{\bra}{\langle}
\newcommand{\ket}{\rangle}
\newcommand{\psis}{\psi_i^{\sigma}}

\begin{document}

\title{Hubbard-corrected DFT energy functionals: the LDA+U description
of correlated systems} 

\author{Burak Himmetoglu \footnote{present address: Materials Department, University of California,
Santa Barbara, CA 93106}}
\affiliation{Department of Chemical Engineering and Materials Science,
University of Minnesota, Minneapolis, MN 55455}

\author{Andrea Floris}
\affiliation{Department of Physics, King's College London,
Strand WC2R 2LS, London}
\author{Stefano de Gironcoli}
\affiliation{Scuola Internazionale Superiore di Studi Avanzati (SISSA),
via Bonomea 256,I-34136 Trieste, Italy}
\affiliation{CNR-IOM Democritos Simulation Center,
via Bonomea 256, I-34136 Trieste, Italy}
\author{Matteo Cococcioni}
\email{matteo@umn.edu}
\affiliation{Department of Chemical Engineering and Materials Science,
University of Minnesota, Minneapolis, MN 55455}

\date{\today}

\begin{abstract}
The aim of this review article is to assess the descriptive
capabilities of the Hubbard-rooted LDA+U method and to clarify 
the conditions under which it can be expected to be most predictive.
The paper illustrates the theoretical foundation of LDA+U
and prototypical applictions to the study of correlated materials,
discusses the most relevant approximations used in its formulation,
and makes a comparison with other approaches also developed for 
similar purposes.
Open ``issues" of the method are also discussed, including 
the calculation of the electronic couplings (the Hubbard $U$),
the precise expression of the corrective functional
and the possibility to use LDA+U for other classes of materials. 
The second part of the article presents 
recent extensions to the method and 
illustrates the significant improvements they have obtained in the description 
of several classes of different systems.
The conclusive section finally 
discusses possible future developments of LDA+U 
to further enlarge its predictive power and its range of applicability.
%
\end{abstract}

\maketitle


\section{Introduction}
After almost five decades from its 
formulation \cite{hohenberg64,kohn65}, Density Functional Theory (DFT)
still represents the main computational tool
to perform electronic structure calculations for systems
of realistic complexity. The possibility to express all the ground
state properties of a system as functionals of its electronic
charge density and the existence of a variational principle
for the total energy functional render DFT a practical computational
tool of 
remarkable simplicity and efficiency.
Unfortunately, the exact expression of the total energy functional
is unknown and approximations are needed
in order to use DFT in actual calculations.
Most commonly used approximate energy functionals for DFT calculations
are constructed as expansions around the homogeneous electron gas limit
and fail quite dramatically in capturing the properties of systems 
whose ground state is characterized by a more pronounced localization 
of electrons.
In fact, within these approximations the electron-electron interaction energy
is written as the sum of the classical Coulomb coupling between electronic
charge densities (Hartree term) 
and the so-called ``exchange-correlation" (xc) term that 
is supposed to contain all the corrections needed to recover 
the many-body terms of electronic interactions, missing from
the first.
Due to the approximations in the latter contribution and the intrinsic
difficulty in modeling its dependence on the electronic charge density,
approximate functionals generally provide a quite poor
representation of the many-body features of the N-electron 
ground state.
For these reasons, correlated systems (whose physical properties
are often controlled by many-body terms of the electronic interactions)
still represent a formidable
challenge for DFT and, despite the steady and notable progress in the 
definition of more accurate functionals and corrective approaches,
no single scheme has been defined that is able to capture entirely the 
complexity of the quantum many-body problem, while maintaining a 
sufficiently low computational cost to perform predictive
calculations on systems of realistic complexity.

While the quantitative entity of the inaccuracy of DFT functionals depends
on the details of their formulation, on the specific system being modeled,
and on the physical properties under investigation, 
on a more general and qualitative level, the failure in describing the 
physics of correlated systems can be ascribed to
the tendency of approximate xc functionals 
to over-delocalize valence electrons and to over-stabilize metallic 
ground states. Paradigmatic examples of 
problematic systems are Mott insulators \cite{mott70}
whose electronic localization on atomic-like states 
is missed by approximate DFT functionals which, instead,
predict them to be metallic.

To qualitatively understand the excessive delocalization of
electrons induced by approximate energy functionals it is convenient
to refer to the expression of the electron-electron interaction energy 
as the sum of Hartree and xc terms.
The over-delocalization of electrons 
can be attributed to the defective (approximate) account of exchange and 
correlation interactions
in the xc functional that fail to cancel out 
the eletronic self-interaction contained in the
classical Hartree term. In fact, 
the persistence of this (unphysical) self-interaction 
makes ``fragments" of the same electron
(i.e., portions of the charge density associated with it)
repel each other, thus inducing an excessive delocalization of the 
wave functions. 
In light of these facts, 
and based on the observation that HF is self-interaction free
many of the corrective functionals (e.g., hybrid),
formulated to improve the accuracy of DFT, aim to
eliminate the residual self-interaction of electrons through the 
explicit introduction of a (screened or approximate) Fock-exchange term.  
This correction often results in an insulating ground state associated
with a gapped Kohn-Sham (KS) spectrum.
However, 
two important aspects should be kept in mind. First, the KS 
single-particle energy spectrum is not bound to any physical quantity
(so that, for example, there is no guarantee that an insulator should
have a gapped KS band structure).
Second,
the above-mentioned difficulties arise
from both exchange and correlation terms 
of the energy and the lack of cancellation of the electronic 
self-Coulomb interaction is only 
the single-electron manifestation of their
approximate representation in current xc functionals.
%
A better treatment of correlation effects requires a more precise
description of the many-body terms of the electronic energy. 
Methods and corrective approaches able to handle these degrees
of freedom have been formulated and developed in the last decades.
DFT + Dynamical Mean Field Theory (DFT+DMFT)
\cite{metzner89,muller89,brandt89,janis91,georges92,georges96,pavarinibook11} 
and Reduced Density Matrix Functional Theory (RDMFT)
\cite{yang00,grossepl07,requist08,grossprb08,grosspra09} are two
notable examples in this class of computational methods.
Both these approaches improve quite significantly the
description of correlated systems compared to most DFT
functionals available. Unfortunately,
while still avoiding the prohibitive cost of wave function-based tractations
of the electronic problem (as, e.g., in quantum chemistry approaches), 
these methods are significantly more computationally intensive than 
DFT calculations performed with approximate energy functionals,
and are both outside the realm of DFT (or even generalized KS theory), 
thus requiring a significant effort to be implemented 
in (or to be interfaced with) existing DFT codes.

In recent years, the study of complex systems and phenomena has often been
based on computational methods complementing DFT with model Hamiltonians 
\cite{capelle13}.
LDA+U, based on a corrective functional inspired to the Hubbard model, 
is one of the simplest approaches that were formulated to improve
the description of the ground state
of correlated systems \cite{anisimov91,anisimov91_2,anisimov93_1,solovyev94,
anisimov97}. 
Due to the simplicity of its expression, 
and to its low computational cost, only marginally larger than that
of ``standard" DFT calculations, LDA+U 
(if not specified otherwise,
by this name we indicate a Hubbard, ``+U" correction to
approximate DFT functionals such as, e.g., LDA, LSDA or GGA) 
has rapidly become very popular in the ab-initio calculation
community.
Its use in high-throughput calculations
\cite{curta11,curta12,curta13} 
for materials screening and optimization is quite emblematic of both these
advantages the method offers. 
An additional and quite
distinctive advantage LDA+U offers
certainly consists in the 
easy implementation of energy derivatives as, for example, atomic
forces and stresses \cite{cococcioni10} (to be used in 
structural optimizations and molecular dynamics 
simulations \cite{sit06,sit07}), 
or second derivatives, as atomic force constants, 
(for the calculation of phonons\cite{floris11}) 
or elastic constants \cite{hsu11}.

While certainly important for its implementation,
the simplicity of the LDA+U functional requires 
a clear understanding of the approximations it is based on 
and a precise assessment of the 
the conditions under which it can be expected to provide quantitatively 
predictive results. This analysis is the main objective of this review article
together with the discussion of recent extensions to the corrective functional
and of their application to selected case studies.


The reminder of this review article is organized as follows.
In section \ref{basic} we will review the historical formulation
of LDA+U and the most widely used implementations, discussing
the theoretical background of the method in the framework of DFT.
In sections \ref{computeu}, \ref{bas_set}, and \ref{dc0} 
some open questions
of the LDA+U method, namely the calculation of the Hubbard $U$, the
choice of the localized basis set, and the formulation of the 
double counting term, will be discussed reviewing and comparing
a selection of different solutions proposed in literature to date.
In section \ref{extf} we will
present recent extensions to the LDA+U functional
that were designed to complete the Hubbard corrective Hamiltonian 
with inter-site and magnetic interactions.
Section \ref{derivae} will focus on
the calculation of first and second energy derivatives (forces, stress
and dynamical matrices)
of the LDA+U energy functional and will present, as an example, the 
calculation of the phonon spectrum of selected transition-metal oxides. 
Finally, in section
\ref{summary}, we will propose some conclusions and an outlook on the
possible future of this method.

\section{Theoretical framework, basic formulations and approximations}
\label{basic}

\subsection{General formulation}
\label{geninfo}

The idea LDA+U is based on is quite simple and consists in 
describing the ``strongly correlated" electronic
states of a system (typically, localized $d$ or $f$ orbitals)
using the Hubbard model~
\cite{hubbardI,hubbardII,hubbardIII,hubbardIV,hubbardV,hubbardVI}, 
while the rest of valence electrons are treated at the level 
of ``standard" approximate DFT functionals.
Within LDA+U the total energy of a system can be written as follows:
\barr
E_{LDA+U}[\rho({\bf \rm r})] &=&
E_{LDA}[\rho({\bf \rm r})] \nonumber \\
&+& E_{Hub}[\{n^{I\sigma}_{mm'}\}] -
E_{dc}[\{n^{I\sigma}\}].
\label{eu}
\earr
In this equation $E_{LDA}$ represents the approximate DFT total energy 
functional being corrected and E$_{Hub}$ is the term that
contains the Hubbard Hamiltonian to model correlated states.
Because of the additive nature of this correction, it is necessary to eliminate
from the (approximate)
DFT functional, $E_{LDA}$, the part of the interaction energy
to be modeled by $E_{Hub}$.
This task is accomplished through the subtraction of the so-called
``double-counting" (dc) term $E_{dc}$ that models the contribution
of correlated electrons to the DFT energy as a mean-field 
approximation of $E_{Hub}$. 
Unfortunately,
the dc functional is not uniquely defined (its definition is, 
indeed, an open issue
of LDA+U that will be discussed later in this review), 
and different possible formulations 
have been implemented and used in various circumstances.
The two most popular choices for the dc term have led to the 
so-called ``around mean-field" (AMF)\cite{anisimov91_2,czyzyk94,petukhov03,
anisimov07} and ``fully localized limit" (FLL)\cite{anisimov93_1,
liechtenstein95,anisimov97,dudarev98}
implementations of the LDA+U. As the names suggest, the first is more
suitable to treat fluctuations of the local density in systems characterized
by a quasi-homogeneous ditribution of electrons (as metals and weakly correlated
systems) while
the latter is more appropriate for materials whose electrons
are more localized on specific orbitals.
An exhaustive discussion on the characteristics of
both approaches and of their framing
within DFT has been presented in Ref \cite{anisimov07}.
We will briefly compare these two formulations in section \ref{doublec}.
Most of the rest of this review will focus, however, on the FLL 
LDA+U which, thanks to its better ability to capture Mott localizaton
and increase the width of band gaps in the KS spectrum,
has become far more popular and widely used than the AFM. 


The FLL formulation of LDA+U was introduced more than two 
decades ago in a series of seminal papers (see, for example, Refs.
\cite{anisimov93_1,solovyev94}) and consists of an 
energy functional that, consistently with Eq. (\ref{eu}) 
can be written as follows: 
\barr
\label{simpleu}
&&E_{LDA+U}[\rho({\rm \bf r})] = E_{LDA}[\rho({\rm \bf r})]  \nonumber \\
&&+\sum_{I}\left[
\frac{U^I}{2}\sum_{{m,\sigma}\neq{m',\sigma'}}
n_{m}^{I\sigma}n_{m'}^{I\sigma'}
- \frac{U^I}{2} n^{I}(n^{I}-1)
\right]
\earr
where $n_{m}^{I\sigma}$ 
are the occupation numbers of localized orbitals identified 
by the atomic site index $I$, state index $m$ (e.g., running over 
the eigenstates of $L_z$ for a certain angular quantum number $l$) 
and by the spin $\sigma$. 
Although the definition of these occupations depends on the specific 
implementation of LDA+U, in many DFT codes they are computed from the
projection of KS orbitals onto the states of a localized
basis set of choice (e.g., atomic states):
\beq
\label{occup}
n^{I\sigma}_{mm'}=\sum_{k,v}f_{kv}^{\sigma}\langle \psi_{kv}^{\sigma}|
\phi_{m'}^{I}\rangle\langle \phi_{m}^{I} | \psi_{kv}^{\sigma} \rangle
\eeq
where the coefficients $f_{kv}^{\sigma}$ represent the occupations of
KS states (labeled by k-point, band and spin indices),
determined by the Fermi-Dirac distribution of the corresponding
single-particle energy eigenvalues.
It is important to note that, upon rotation of atomic orbitals,
the quantity defined in Eq. (\ref{occup}) tranforms as a product of
atomic orbitals. Therefore, it can be treated as a tensor of rank two
(although this requires some care in case a non-orthogonal
basis set is used \cite{oregan11}). In this case a more
appropriate notation (e.g., with covariant and controvariant 
indexes) should be adopted as explained 
in Refs. \cite{yu06,oregan10,oregan11} 
where it resulted particularly useful for
defining the atomic occupations
based on non-orthogonal Wannier functions. 
However, in order to keep the notation simple and to avoid 
crowding superscripts, we will leave the indexing of
the occupation tensor as in Eq. (\ref{occup}) 
and, in consistency with abundant literature,
we will keep calling these quantities ``matrices". The same will be done also
for other quantities, as the response ``matrices" that will be introduced
in the linear response calculation of $U$.

In Eq. (\ref{simpleu}) the following definitions have been adopted:
$n^{I\sigma}_{m} = n^{I\sigma}_{mm}$
and $n^{I} = \sum_{m,\sigma} n^{I\sigma}_{m}$.
The expression of the corrective term in Eq. (\ref{simpleu}) as
a functional of the occupation numbers defined in Eq. (\ref{occup})
highlights how the Hubbard correction operates selectively on 
the localized orbitals of the system (typically the most correlated ones)
while all the other states continue to be treated at the level of
approximate DFT functionals.
It is important to stress that the one given in Eq. (\ref{simpleu})
is the simplest ``incarnation" of the Hubbard functional; in fact, it 
neglects all the interaction terms involving off-diagonal elements of 
the occupation matrix and all the exchange couplings.
The use of products of occupation numbers ($n_{ij}n_{kl} = 
\langle c^{\dagger}_i c_j\rangle \langle c^{\dagger}_k c_l \rangle$)  
instead of expectation values of 
quadruplets of creation and annihilation fermionic operators ($\langle
c^{\dagger}_i c^{\dagger}_k c_j c_l \rangle$) 
corresponds to a mean-field like approximation (Hartree-Fock factorization)
that is necessary to make the problem tractable within a computational
scheme based on single particle (Kohn Sham) wave functions, as DFT.
The second and the third terms of the right-hand side of Eq. \ref{simpleu} 
represent, respectively, the Hubbard and the 
double-counting terms specified in Eq. (\ref{eu}).

\begin{figure}[tbh!]
 \centering
 \vspace{0mm}
 \includegraphics[width=0.48\textwidth]{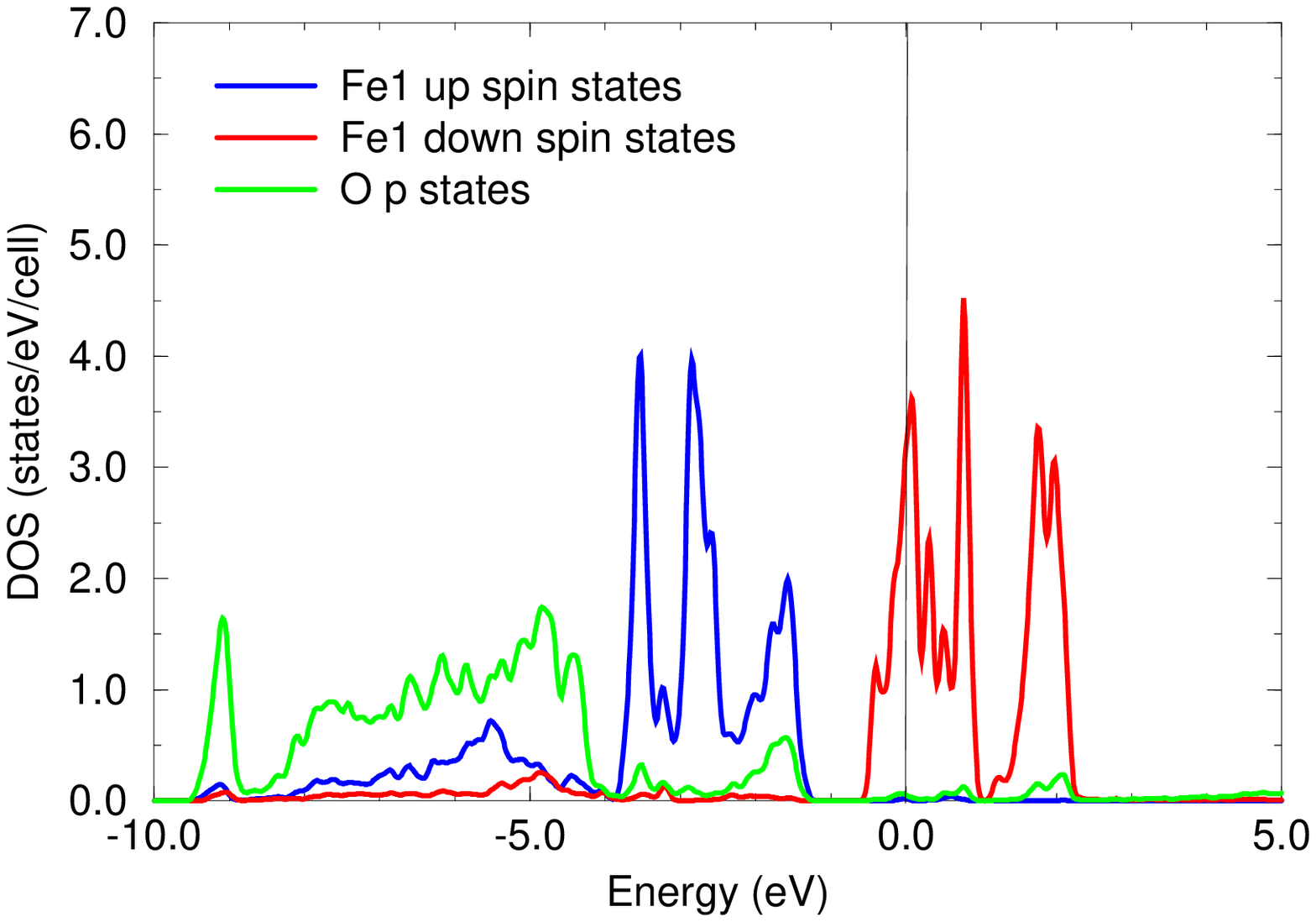}
 \includegraphics[width=0.45\textwidth]{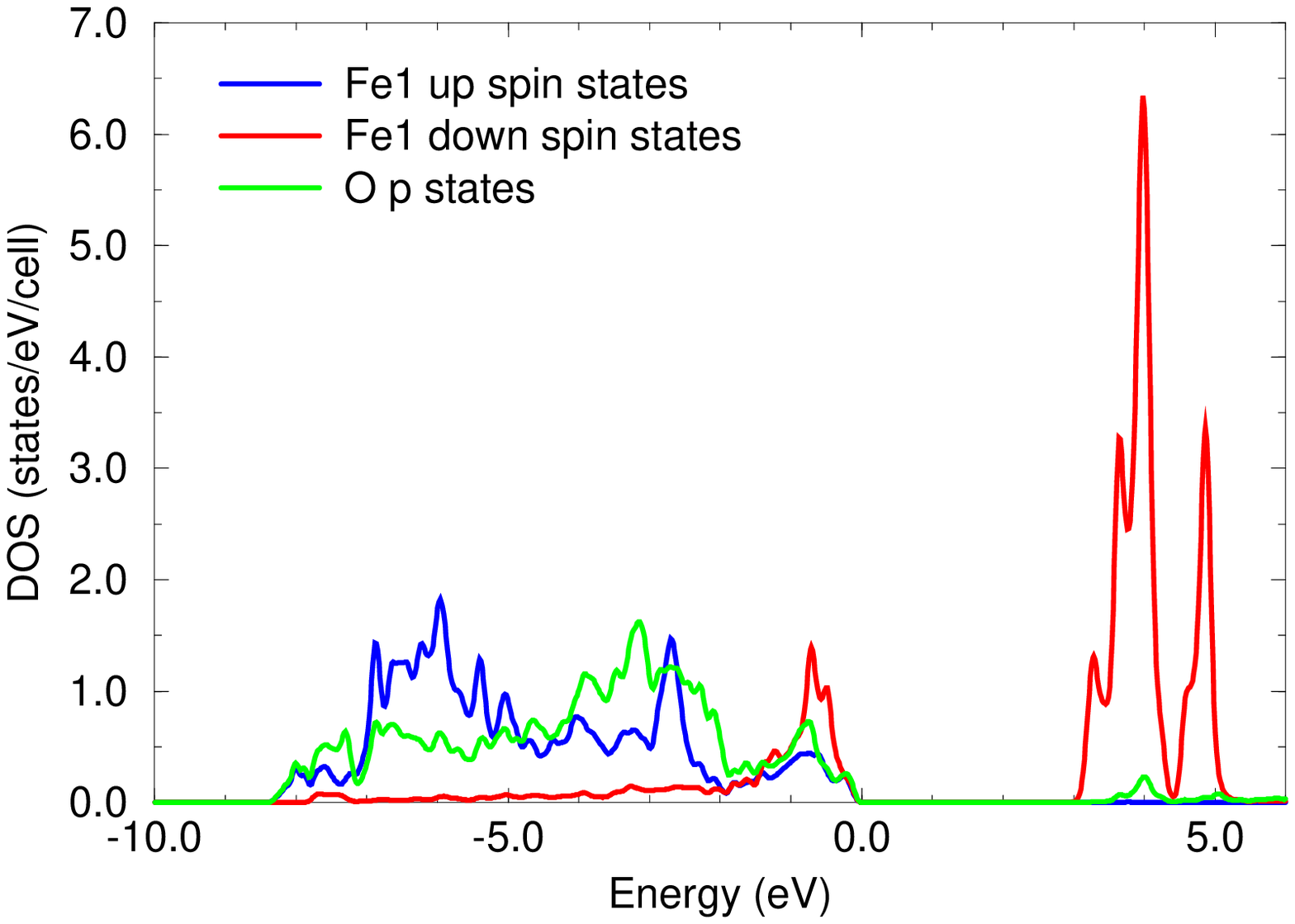}
\caption{(Color online; adapted from Ref. \cite{cococcioni05}) 
The density of states (DOS) 
of Fe$_2$SiO$_4$ fayalite obtained with GGA (top graph) and GGA+U
(bottom graph). For the sake of clarity, only the contributions from
the $d$ states of one of the Fe atoms and of the $p$ states of one of
the O atoms are shown. These data were obtained in a study published
in Ref. \cite{cococcioni05}.}
\label{fay}
\end{figure}

Using the definition of the atomic orbital occupations given 
in Eq. (\ref{occup}),
it is instructive to derive the Hubbard contribution to the KS potential. 
From the energy functional in Eq. (\ref{simpleu}) it is easy to obtain:
\begin{equation}
\label{simplepot}
V_{tot}^{\sigma} = V_{LDA}^{\sigma} +
\sum_{I,m}U^I\left(\frac{1}{2}-n_{m}^{I\sigma}\right)
\vert \phi^{I}_{m}\rangle \langle \phi^{I}_{m}\vert.
\end{equation}
Eq. (\ref{simplepot}) shows that the Hubbard potential is repulsive
for less than half-filled orbitals ($n_{m}^{I\sigma} < 1/2$), 
attractive in all the other cases. Therefore,
the Hubbard correction discourages fractional
occupations of localized orbitals (often indicating a substantial
hybridization with neighbor atoms) and favors the Mott 
localization of electrons on specific 
atomic states ($n_{m}^{I\sigma}\rightarrow 1$) while 
penalizing the occupation of others ($n_{m'}^{I\sigma}\rightarrow 0$). 
The difference between the potential
acting on occupied and unoccupied states, 
approximately equal to $U$,
corresponds to an effective discontinuity 
in correspondance of integer values of $n^{I\sigma}_m$.
This discontinuity in the potential, a feature of the exact 
DFT functional,
is responsible for the creation of an energy gap in the KS spectrum, 
equal to the fundamental gap of the system (i.e., the difference between
ionization potential and electron affinity in molecules, the HOMO-LUMO
gap in crystals) \cite{perdew82,perdew83,godby86}. 
A better representation of the potential discontinuity in DFT energy
functional was, in fact, one of the original purposes of 
LDA+U \cite{anisimov93_1}.
Fig. \ref{fay} compares the density of state
of Fe$_2$SiO$_4$ fayalite obtained with GGA and with GGA+U,
and illustrates how the Hubbard correction induces the opening
of a band gap in the KS spectrum.
Fayalite is the iron-rich end memebr of a family of iron-magnesium silicates
particularly abundant in the Earth upper mantle and, as many other transition
metal compounds, is a Mott insulator.
Approximate xc energy functionals result in a metallic single particle (KS)
spectrum and tend to over delocalize valence electrons (top panel
of Fig. \ref{fay}).
Through a more accurate description of on-site electron-electron interactions,
the Hubbard correction is able to re-establish an insulating ground state
with a band gap in the band structure of the material.
Occasionally, finite band gaps are obtained as a result of crystal field 
splittings or Hund's rule (as in NiO and MnO, respectively); 
however, even in these circumstances, they are 
underestimated by DFT, compared to experiments. In some cases (with degenerate
states at the top of the valence band) the opening
of a gap in the band structure through the Hubbard correction
requires lowering the electronic subsystem to have a lower symmetry
than the crystal, as discussed below.

\begin{figure}[tbh!]
 \centering
 \vspace{0mm}
 \includegraphics[width=0.45\textwidth]{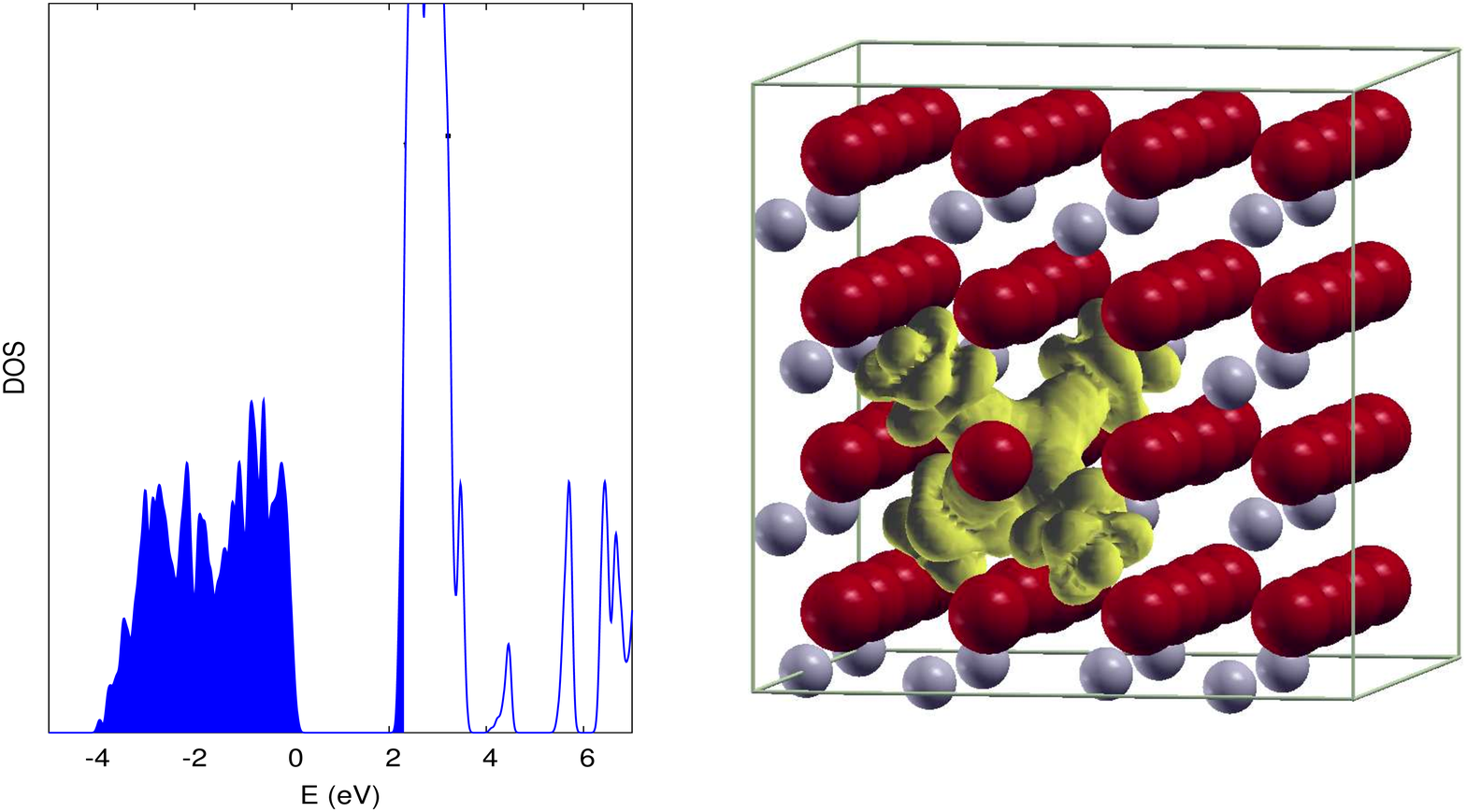}
 \vspace{0mm}
 \includegraphics[width=0.45\textwidth]{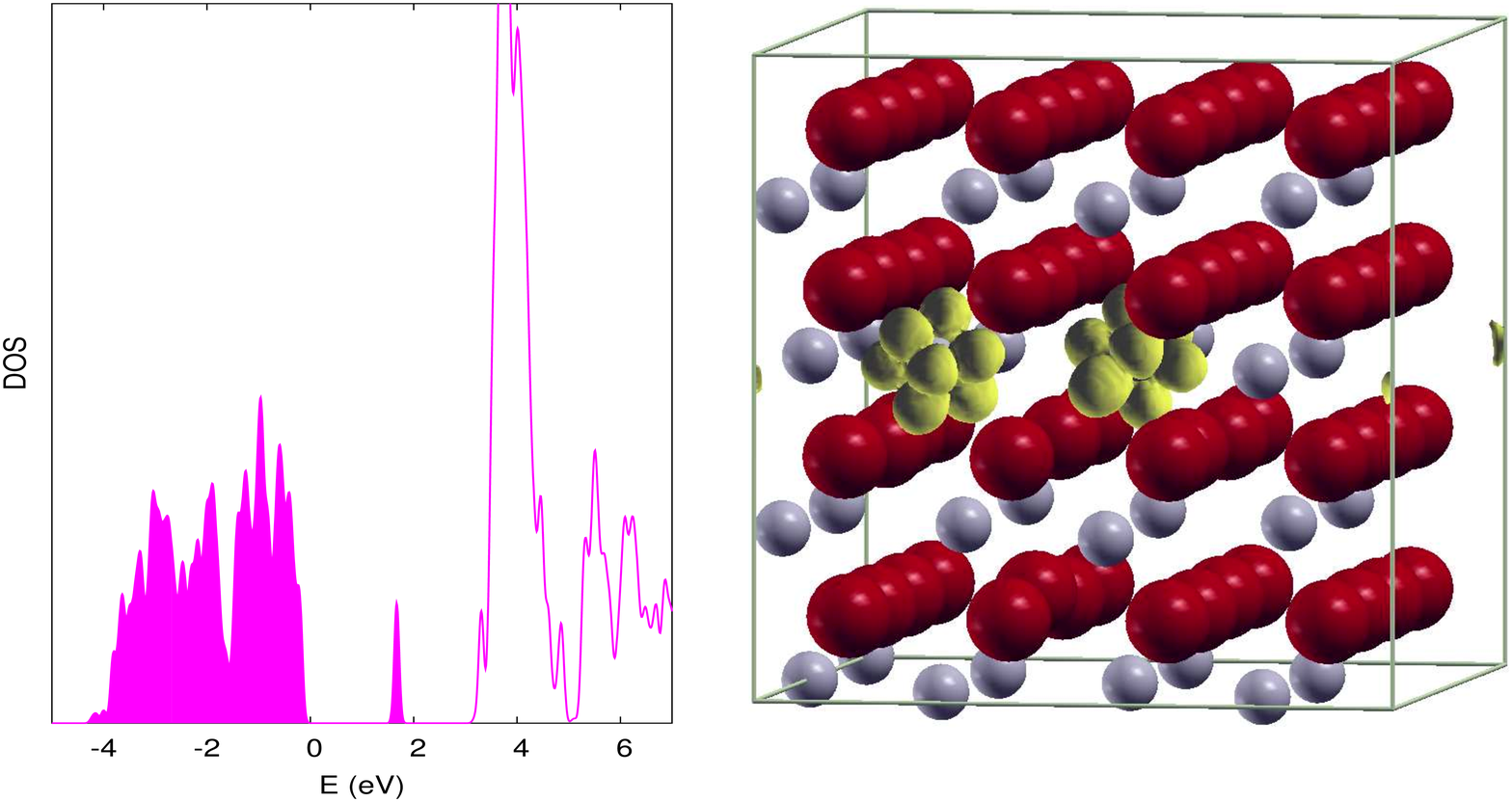}
\caption{(Color online) Density of electronic states (DOS) and charge density of
the gap state in a CeO$_2$ crystal with an oxygen vacancy 
(simulated by CeO$_{1.875}$). The shaded area in the DOS indicates 
occupied electronic states and the zero of the energy is fixed at the top of 
the valence band. The top and bottom panels show the results obtained 
with GGA and GGA+U, respectively. Results shown in this figure were obtained
in a study published in Refs. \cite{fabris05} and \cite{fabris05_1}.}
 \vspace{-0mm}
\label{ceo2}
\end{figure}

The opening of a gap in the band structure is only one particular
aspect of the effect the Hubbard correction has.
Consistently with the predictions of the Hubbard model, 
the explicit account of on-site electron-electron interactions 
also favors electronic localization and the onset of an insulating 
ground state (provided the on-site Coulomb repulsion prevails on the kinetic
term of the energy, minimized by electronic delocalization).
One example is shown in 
Fig. \ref{ceo2} that visualizes the density of state (DOS) and the 
charge density of the highest energy state in 
CeO$_2$ with an oxygen vacancy \cite{fabris05,fabris05_1}.
It is evident from the figure that, while GGA predicts the extra
charge induced by the oxygen vacancy to be spread among the four Ce atoms around
the vacancy and to be described by a delocalized state within
the conduction band (top panel), the Hubbard correction induces the localization
of the extra two electrons on the atomic $f$ orbitals of 
two of the Ce atoms around the defect that correspond to a state
well localized within the gap of the pristine material.
These results were obtained with a Wannier function-based implementation of the
LDA+U (to be discussed later in this article) 
that also predicted the crystal structure and the DOS of reduces
surfaces of CeO$_2$ and Ce$_2$O$_3$ in very good agreement with 
STM, AFM and photoemission experiments. If LDA or GGA were
used, instead, the extra charge in the system associated with 
the O defect would be
erroneously spread over the three outermost atomic layers, and the 
agreement with experimental results would significantly deteriorate
\cite{fabris05,fabris05_1}.
Similar calculations (albeit not based on Wannier functions) were also 
performed to study oxygen vacancies on reduced TiO$_2$ surfaces 
\cite{divalentin09,finazzi08,mattioli10}. These studies showed that the
Hubbard correction was necessary to capture the localization of the
extra charge on the $d$ states of the Ti atoms around the O vacancies and
the consequent deformation of the crystal in its neighborood 
(polaronic self-trapment), although the quality of the results depend
critically on the value of U and the way the Hubbard functional is
used (e.g., with $U$ only on Ti $d$ or on Ti $d$ and O $p$ states).

\subsection{Rotationally-invariant formulation}
\label{rotinva}
\index{rotational invariance}
While able to capture the main essence of the LDA+U approach,
the formulation presented in Eq. (\ref{simpleu}) 
is not invariant under rotation of the atomic
orbital basis set used to define the occupation numbers $n^{I}_{m\sigma}$.
Thus, calculations performed with this functional
are affected by an undesirable dependence on
the specific unitary transformation of the localized basis set chosen to define 
the atomic occupations, (Eq. (\ref{occup})).
A unitary-transformation-invariant formulation of LDA+U was introduced
in Ref \cite{liechtenstein95}. In that work 
$E_{Hub}$ and $E_{dc}$ were given a more general expression,
borrowed from the HF method:
\barr
\label{ub1}
&&E_{Hub}[\{n^{I}_{mm'}\}] = \nonumber \\
&&\frac{1}{2}\sum_{\{m\},\sigma,I}\{ \langle m,m''|V_{ee}|m',m'''\rangle
n^{I\sigma}_{mm'}n^{I-\sigma}_{m''m'''} + \nonumber  \\
&&(\langle m,m''|V_{ee}|m',m'''\rangle - \langle m,m''|V_{ee}|m''',m'\rangle)
\times \nonumber \\
&&n^{I\sigma}_{mm'}n^{I\sigma}_{m''m'''} \}
\earr 
\barr
\label{ub2}
&&E_{dc}[\{n^{I}_{mm'}\}] = 
\sum_{I}\{\frac{U^I}{2}n^{I}(n^{I}-1)- \nonumber \\
&&\frac{J^I}{2}
[n^{I\uparrow}(n^{I\uparrow}-1)+n^{I\downarrow}
(n^{I\downarrow}-1)]\}.
\earr
The invariance of the ``Hubbard" term (Eq. (\ref{ub1})) 
stems from the fact that 
the interaction parameters transform as quadruplets
of localized wavefunctions, thus compensating the variation of the
(product of) occupations they are associated with.
In Eq. (\ref{ub2}), instead, the invariance is due to the 
dependence of the functional on the trace of the occupation matrices.
In Eq. (\ref{ub1}) the $V_{ee}$ integrals represent
electron-electron interactions that are expressed as the
integrals of the Coulomb kernel on the wave 
functions of the localized basis set (e.g., $d$
atomic states), labeled by the index $m$:
\barr
\label{vee0}
&&\bra m,m''|V_{ee}|m',m'''\ket = \nonumber \\
&&\int d{\rm \bf r} \int d{\rm \bf r}' \psi^*_{lm}({\rm \bf r})
\psi_{lm'}({\rm \bf r})\frac{e^{2}}{|{\rm \bf r} - {\rm \bf r}'|}
\psi^*_{lm''}({\rm \bf r}')
\psi_{lm'''}({\rm \bf r}').
\earr
Assuming that atomic (e.g., $d$ or $f$) states are chosen as 
the localized basis, these integrals can be 
factorized in a radial and an angular contributions. 
This factorization stems from the expansion of the Coulomb kernel 
in spherical harmonics
(see Ref. \cite{liechtenstein95} and references quoted therein) and
yields:
\beq
\label{vee}
\bra m,m''|V_{ee}|m',m'''\ket  = \sum_{k}a_{k}(m,m',m'',m''')F^{k}
\eeq
where $0\le k \le 2l$ ($l$ being the angular quantum number of the
localized manifold with $-l\leq m\leq l$). The $a_k$ represent the
angular factors and correspond to
products of Clebsh-Gordan coefficients:
\barr
\label{vee1}
&&a_{k}(m,m',m'',m''') = \nonumber \\
&&\frac{4\pi}{2k+1}\sum_{q=-k}^{k}
\bra lm|Y_{kq}|lm'\ket \bra lm''|Y^{*}_{kq}|lm'''\ket.
\earr
The quantities $F^{k}$ 
are the (Slater) integrals involving the radial part of the atomic
wave functions $R_{nl}$ ($n$ indicating the atomic shell
they belong to).
They have the following expression:
\beq
\label{fk}
F^k = e^2 \int d{\rm \bf r} \int d{\rm \bf r}' r^2 r'^2 R_{nl}^2({\rm \bf r})
\frac{r^k_{<}}{r^{k+1}_>}
R_{nl}^2({\rm \bf r}')
\eeq
where $r_{<}$ and $r_{>}$ indicate, respectively, the shorter and the
larger radial distances between $r$ and $r'$.
For $d$ electrons only $F^{0}$, $F^{2}$, and $F^{4}$
are needed to compute the $V_{ee}$ matrix elements
(for higher $k$ values the corresponding $a_k$ vanish)
while $f$ electrons also require $F^{6}$. 
Consistently with the definition of the dc term (Eq. (\ref{ub2}))
as the mean-field approximation of the Hubbard correction (Eq. (\ref{ub1})),
the effective Coulomb and exchange interactions, $U$ and $J$, can be computed
as atomic averages of the corresponding Coulomb integrals over the localized 
states of the same manifold (in this example atomic orbitals of fixed $l$).
For $d$ orbitals it is easy to obtain:
\begin{equation}
\label{ueff}
U = \frac{1}{(2l+1)^{2}} \sum_{m,m'}
\bra m,m'|V_{ee}|m,m'\ket = F^{0}
\end{equation}
\begin{equation}
\label{jeff}
J = \frac{1}{2l(2l+1)} \sum_{m \ne m',m'}
\bra m,m'|V_{ee}|m',m\ket =
\frac{F^{2}+F^{4}}{14}.
\end{equation}
\index{Hubbard $U$} 
Although strictly valid for atomic states and {\it unscreened}
Coulomb kernels,
these equations have often been adopted to evaluate {\it screened} 
Slater integrals: once $U$ and $J$ are computed from the ground state
of the system of interest, the $F^k$ parameters (and the 
$V_{ee}$ integrals) are extracted using
Eqs. (\ref{ueff}) and (\ref{jeff}) 
based on the assumption that the ratio between them has the same
value as for atomic states (e.g., $F_2/F_4 = 0.625$).
The limits of this assumption were thoroughly discussed in Ref \cite{vaugier12}.

\subsection{A simpler formulation}
\label{simpler}
The one presented in section \ref{rotinva} is the most complete formulation
of the LDA+U, with fully orbital-dependent electronic
interactions. 
However, in many occasions, a simpler expression of the Hubbard correction
($E_{Hub}$), introduced in Ref. \cite{dudarev98}, 
is actually adopted and implemented.
This simplified functional can be obtained from the full formulation
discussed in section \ref{rotinva} by retaining only the lowest order
Slater integrals $F^{0}$ and neglecting all the higher order ones:
$F^{2} = F^{4} = J = 0$. 
This simplification corresponds 
to assuming that $a_{0}(m,m',m'',m''') =
\delta_{m,m'}\delta_{m'',m'''}$. 
Using these conditions in Eqs.  (\ref{ub1}) and (\ref{ub2}), 
one easily obtains:
\begin{eqnarray}
\label{our1}
E_{U}[\{n^{I\sigma}_{mm'}\}] &=& E_{Hub}[\{n^{I}_{mm'}\}] -
E_{dc}[\{n^{I}\}] = \nonumber \\
&=&\sum_{I} \frac{U^I}{2} \left[\left(n^I\right)^2 - \sum_{\sigma}
{\rm Tr}~[({\rm \bf n}^{I\sigma})^2]\right] \nonumber \\
&-& \sum_{I}\frac{U^I}{2}n^{I}(n^{I}-1) 
\nonumber \\
&=&\sum_{I,\sigma}\frac{U^I}{2}
{\rm Tr}~[{\rm \bf n}^{I\sigma}(1-{\rm \bf n}^{I\sigma})].
\end{eqnarray}
It is important to stress that the simplified functional in
Eq. (\ref{our1}) still preserves the rotational invariance of the one in 
Eqs. (\ref{ub1}) and (\ref{ub2}), through its dependence
on the trace of occupation matrices and of their products.
On the other hand, the formal resemblance to the HF energy
functional is lost in this formulation 
and only one interaction parameter ($U^I$) is needed to 
specify the corrective functional.
It is also worth remarking that, when a non-orthogonal basis set is used to
define atomic occupations, the rotational (tensorial) invariance of the
Hubbard energy requires the use of a covariant-controvariant formulation
(which won't be detailed in this article), as explained in Ref. 
\cite{oregan11}.

The simplified version of the Hubbard correction, Eq. (\ref{our1}), 
has been succesfully used in several studies and for most materials it yields 
similar results as the fully rotationally invariant one (Eq. (\ref{ub1}) 
and (\ref{ub2})).
Some recent literature has shown, however, that the explicit inclusion of
the Hund's rule coupling $J$ is crucial to describe the
ground state of systems characterized by non-collinear magnetism 
\cite{spaldin10,pickett06}, to capture correlation effects in
multiband metals \cite{demedici11,georges11}, or to study 
heavy-fermion systems, typically characterized by $f$ valence electrons
and subject to strong spin-orbit couplings \cite{spaldin10,pickett06,
bultmark09}.
A recent study \cite{nakamura09} also showed that in some Fe-based 
superconductors a sizeable
$J$ (possibly exceeding the value of $U$) 
is needed to reproduce (within LDA+U) the 
experimentally measured magnetic moment of Fe atoms.
Several different flavors of corrective functionals with exchange 
interactions were also discussed in Ref. \cite{ylvisaker09}.

Due to the spin-diagonal form of the simplified LDA+U approach in Eq. 
(\ref{our1}), it has become customary to attribute the
Coulomb interaction $U$ an effective value that incorporates the exchange
correction: $U_{eff} = U - J$. This practice has been introduced in the 
original formulation of the simplified functional, in Ref. \cite{dudarev98}. 
As discussed in section \ref{dftpj},
this assumption is actually not completely justifyable as the resulting 
functional neglects other interaction terms (proportional to $J$)
that are of the same order as the ones responsible for
the negative correction to the on-site Hubbard $U$ for parallel-spin electrons.

\subsection{Theoretical background and practical remarks}
\label{concept}
The previous parts introduced the general formulation of the 
LDA+U functional and reviewed
the most widely used implementations. 
This section is devoted 
to clarifying in a more detailed way its theoretical foundation 
(possibly in comparison 
with other corrective methods), to discussing the possibility to use this
tool for the study of various classes of systems and to
assessing the conditions under which
it can be expected to be most predictive. 
While useful for a more precise theoretical framing of the method,
this part is not essential to understand how LDA+U is implemented
in DFT codes and how it works in actual calculations.

\subsubsection{LDA+U vs Hartree-Fock and Exact Exchange}
The expression of the full rotationally invariant Hubbard
functional (Eqs. (\ref{ub1})) shows a quite clear resemblance
with the Hartree-Fock (HF) energy. Therefore, what the LDA+U correction
does could be understood as a substitution of mean-field-like density-density
electronic interactions, contained in the approximate exchange-correlation (xc)
functional, with a HF-like Hamiltonian.
This is much in the same spirit of hybrid functionals in which the
exchange part of the xc functional is shaped as a Fock operator (multiplied
by a screening factor) constructed on KS states. 
Some notable differences from HF are, however, to be stressed: 
$i)$ the effective interactions in the LDA+U functional are {\it screened}, 
rather than based on the {\it bare} Coulomb kernel (as in HF); $ii)$ 
the LDA+U functional only acts on a subset of states (e.g., localized 
atomic orbitals of $d$ or $f$ kind), rather 
than on all the states in the system;
$iii)$ due to the marked localization of the orbitals the Hubbard functional 
acts on, the effective interactions are often assumed to be orbital-independent
so that, in the simpler formulation of 
Eq. (\ref{our1}), they are substituted by
(or computed from) their atomic averages, Eqs. (\ref{ueff}) and 
(\ref{jeff}). This assumption, justified by the fact that more localized
states retain their atomic character (and spherical symmetry)
to a higher extent, (partially) looses its
validity in presence of crystal field or spin-orbit interactions
that can lift the degeneracy (and equivalence) of localized
orbitals. 
Although the use of Fock integrals make hybrid functionals appear
a more systematic and accurate method to correct some of the
above-mentioned flaws of approximate DFT,
their calculation incurr in significantly higher computational costs.
Furthermore, hybrid functionals also depend on 
a parameter (as the Hubbard $U$ is often seen for LDA+U), 
namely the amount of Fock-exchange (mixing coefficient)
to be included in the xc functional.
The quality of the results can depend sensibly on this parameter
that needs to be chosen for each system.
This quantity is generally determined 
semi-empirically (e.g., through fitting of the properties of a large
variety of different systems)\cite{feng04}, or through a 
material-dependent optimization,
(e.g., by an iterative procedures, as proposed in Ref. \cite{marom12}).
Although this mixing coefficient results usually in the 0.2 - 0.3 range,
there is no universal value that can be used with all the systems, nor
a precise physical meaning attached to it except, 
perhaps, the not so precisely quantified attenuation of the exchange 
interaction due to the correlation part of the functional.

The formal similarity with a HF functional may arise some suspect
about the possibility to use LDA+U (and hybrid functional) to 
improve the description of correlated systems. In fact, by definition, 
HF does not account for electronic correlation and it is quite
unrealistic that the complexity of the many-body problem 
can be captured by the screening of the effective
electronic interactions.  
However, it should be noted that the LDA+U functional, besides still
containing a correlation term in the LDA part, operates the Hubbard
correction on KS wave functions. These are not associated to any 
physical meaning other than being constrained to reproduce the exact
charge density of the system. On the other hand, the 
single particle wave functions that are optimized during the self-consistent
solution of the Hartree-Fock equations are 
the ones that minimize the energy of the system in the hypothesis 
that the ground state many-body wave function is the single Slater determinant
that can be constructed out of them.
While this is an important difference, the question of whether a HF-like
corrective functional acting on the KS orbitals can effectively improve 
(with respect to approximate exchange correlation functionals)
the description of the ground state of correlated systems remains open. 
Aiming more to a qualitative argument than a conclusive answer, 
we can observe that if a gap is present in the single-particle 
(KS) energy spectrum of a system,
the occupations of the corresponding energy levels are all
1 or 0, depending on whether the state is in the valence or in the
conduction manifold, respectively, and 
the ground state charge density
can be obtained from a single Slater determinant constructed with the
fully occupied orbitals.
In these circumstances, it is reasonable to expect that a correction
formally shaped as a HF energy functional could be effective in improving
the representation of the correlated ground state by
tuning the width of the energy gap in the single-electron energy spectrum 
(possibly incorporating the xc potential discontinuity) 
and favoring integer occupations of the states 
at the edge of valence and conduction bands.
This action can be expected to affect also other physical
properties (as, for example, the equilibrium crystal structure and
the vibrational spectrum) of the material
through the modifications it brings to its electronic structure 
(charge density).
As documented in abundant literature
(see, for example, Refs. \cite{anisimov91_2,anisimov93_1,dudarev98,
bengone00,martin02,feng04,towler94}),
LDA+U and HF (or hybrid functionals) obtain, in fact,
a quite good representation of the ground state properties of 
correlated systems (e.g., transition metal oxides), provided
a gap is present in the KS spectrum (e.g., because of
crystal field), as in NiO and MnO. 
When this is not the case 
and the degeneracy of frontier valence states (closest to the
Fermi level) results in fractional
occupations and absence of band gaps, 
a preliminary symmetry breaking is usually required to 
create the optimal conditions under which these corrections
are most effective. However, this preliminary ``preparation" of the system
has some theoretical and practical implications that will be discussed
for the case of FeO in one of the following sections.

\subsubsection{Potential discontinuity, band gap and energy linearization}
Improving the estimate of the band gap 
is one of the original objectives of the LDA+U \cite{anisimov93_1,solovyev94} 
and can be shown to also address (albeit in an approximate way) well-known
flaws of approximate energy functionals, such as 
the lack of a discontinuity in the
xc potential (as discussed after Eq. (\ref{simplepot})). 
To see this, let us consider a $N$-electron isolated system. 
The fundamental gap is defined 
as the difference between the ionization potential $I$ and the electron
affinity $A$:
\barr
\label{fgap}
E_g &=& I - A \nonumber \\
&=& [E(N-1) - E(N)] - [E(N) - E(N+1)] \nonumber \\
&=& E(N+1) + E(N-1) - 2 E(N)
\earr
where $E(N)$, $E(N+1)$ and $E(N-1)$ are the total energy of the system
in its neutral state and with one electron added to or removed from
its orbitals, respectively.
It is important to note that the one in the last line of Eq. (\ref{fgap})
is a finite-difference approximation of the second derivative of the total
energy with respect to the number of electrons.
This observation will be useful to understand some approaches to compute
the effective interaction $U$ of the Hubbard functional that will be discussed
in section \ref{computeu}.
Based on the expression of the DFT total energy
it can be shown (see, e.g., Ref \cite{grosslibro}) that: 
\barr
\label{fgap1}
E_g = \Delta_{KS} + \Delta_{xc}
\earr
where the first term corresponds to the energy gap between the
HOMO and the LUMO states from the KS
energy spectrum,
\beq
\label{dks}
\Delta_{KS} = \epsilon^{LUMO}_{N} - \epsilon^{HOMO}_{N},
\eeq
while the second represents the discontinuity in the exchange-correlation
potential computed for the neutral system \cite{grosslibro}:
\beq
\label{dxc}
\Delta_{xc} = \frac{\delta E_{xc}[n]}{\delta
n({\rm \bf r})}\vert_{N+\delta} - \frac{\delta E_{xc}[n]}{\delta
n({\rm \bf r})}\vert_{N-\delta},
\eeq
where the derivatives are evaluated for densities that integrate 
to $N+\delta$ and $N-\delta$ electrons, respectively, and the limit 
$\delta \rightarrow 0+$ is implied.
The discontinuity of the xc potential is a property
of the exact DFT functional 
which is of fundamental importance to describe, for example, molecular
dissociations and electron-transfer processes \cite{perdew82,perdew83}.
In extended systems a discontinuity in the xc potential is 
also expected for insulating ground states. The fundamental gap can be 
defined in a similar way as for isolated systems, as the difference 
between the total energies obtained 
from calculations with a fraction of electron per unit cell 
in eccess or in defect with
respect to the neutral crystal and compensated by a jellium charge 
\cite{grosspra09,chan10}.
Most of approximate exchange-correlation functionals, however, 
miss the discontinuity of the xc potential
$\Delta_{xc}$ and yield an analytical dependence of the total energy
on $N$.

As illustrated in the discussion after 
Eq. (\ref{simplepot}), 
the Hubbard correction introduces a discontinuity in the potential acting
on the orbitals of the localized basis set, whose amplitude is approximately
$U$. Therefore, if these localized states are the ones at the borders
of valence and conduction manifolds (usually the case for systems this
correction is applied to), and the value of $U$ is appropriately
chosen, the Hubbard energy functional can be used to reintroduce
the discontinuity in the exchange-correlation potential. In particular,
since the correction modifies the KS potential,
the discontinuity is introduced in the single particle
spectrum as well and the KS band gap obtained with
the corrected functional can be expected to 
match the fundamental gap: 
$\epsilon_{LUMO}^{LDA+U} - \epsilon_{HOMO}^{LDA+U} \approx 
\Delta_{KS}+\Delta_{xc}$.
It is important to remark that this is not a feature of the exact
KS theory. Because of this difference, LDA+U could be classified as as
a ``generalized Kohn-Sham" theory.

%
The introduction of the exchange-correlation potential discontinuity
is also related to (and is the necessary condition for)
the linearization of the total energy 
profile as a function of the number of electrons.
As explained in abundant literature (see, for example, Refs. 
\cite{grosslibro,perdew82,levy82,yang98,yang00}) 
a piece-wise linear profile of the energy is characteristic of
systems able to exchange electrons with a reservoir of charge.
In this context, a fractional number of electrons 
on the orbitals of these systems is to be interpreted 
as resulting from a mixture of states with different integer occupations.
With the exact xc functional the resulting ground state
energy would be the linear combination of those corresponding
to nearest integer number of electrons.
 
The linearizing action of the Hubbard functional on the approximate
DFT energy is more evident in its simpler formulation, Eq. (\ref{our1}),
that consists of subtracting a quadratic term and adding a linear one.
It is important to stress that
the ``+U" correction linearizes the energy with respect to on-site occupations,
rather than the total number of electrons. However, localized orbitals
(e.g., $d$ or $f$) can be thought of as belonging to isolated atoms 
immersed in a ``bath" of delocalized states. 
In addition, they typically belong to open 
shells and thus represent the frontier states whose occupation changes
when the total number of electrons is varied. Therefore, 
the linearization of the energy with respect to the atomic occupations 
is a legitimate operation.

The elimination of the (spurious) curvature of the energy profile
also makes the Hubbard functional look similar to a self-interaction 
correction (SIC) \cite{perdew81}. 
In fact, a SIC functional could be easily obtained from
the diagonal term of the exact exchange contained in hybrid functionals,
whose analogy with LDA+U has already been highlighted. 
This similarity has also been amply discussed in the 
context of Koopmans-corrected DFT functionals \cite{dabo09,dabo10}
and won't be further expanded here. 
It is worth to stress, however, that LDA+U only corrects localized
states, for which self-interaction is generally expected to be stronger.
The formal similarity with SIC and hybrid functionals suggests that LDA+U should
be also effective in correcting the underestimated band gap of covalent
insulators (e.g., Si, Ge, or GaAs), for which the former have often been 
successfully used.
Indeed, while the ``standard" 
``+U" functional (Eq. \ref{our1}) is not effective on these systems, 
a generalized formulation of the Hubbard correction
with inter-site couplings proves able to achieve this result, as 
will be discussed in section \ref{upv}.

\subsubsection{Degenerate ground states: the case of FeO}
\label{feo}
The orbital independence of the effective
electronic interaction, allows to regard the positive-definite 
``+U" correction in
Eq. (\ref{our1}) as a penalty functional that 
forces the on-site occupation matrix (Eq. (\ref{occup})) 
to be idempotent. This action 
corresponds to favoring a ground state described by a set of KS states 
with integer occupations (either 0 or 1). 
and to imposing a finite
gap in the single-particle (Kohn-Sham) energy spectrum. 
While this is another way to see how
the ``+U" correction helps improving the description of insulators,
it should be kept in mind that the 
linearization of the energy as a function
of orbital occupations is a more general and important effect to
be obtained. In fact, in case of degenerate ground states, 
fractional occupations (resulting, effectively, in a metallic KS 
system) can, in principle, 
represent linear combinations of insulating ground states,
each having different sets of equivalent single-particle
states occupied (and a lower symmetry than their sum).
In these cases the total energy should be equal to the corresponding 
linear combination
of the energies of the single insulating ground states.
In these situations, an insulating KS system should not 
be expected/pursued unless
the symmetry of the electronic state is decreased and the system
``prepared" in one of the equivalent insulating ground states of
lower symmetry.
%
%
\begin{figure}[tbh!]
 \centering
 \vspace{0mm}
 \includegraphics[width=0.4\textwidth]{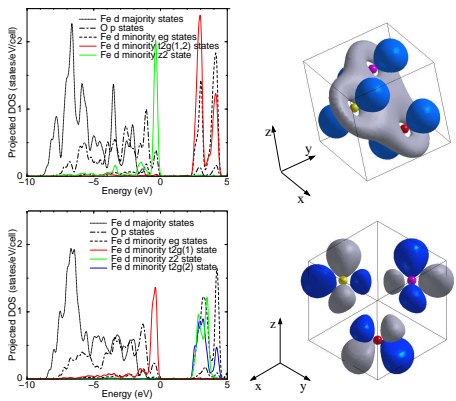}
\caption{(From Ref. \cite{cococcioni05}; color online).
Projected density of states and highest energy occupied orbital
of FeO (top) in
the unbroken symmetry (upper panels) and broken symmetry states (lower panels).}
\label{feo}
\end{figure}
Transition metal oxides with nearly half-filled or full $d$ shells, 
as FeO and CuO, are well known examples of materials with degenerate 
insulating ground states. 
The case of CuO will be discussed in section \ref{dftpj}.
In FeO all the Fe ions are in their high magnetization configuration
with (nominally) five $d$ electrons in the majority-spin states and one
in their minority spin counterparts. Because of the rhombohedral symmetry of 
the crystal and the consequent crystal field splitting of the $d$
energy levels, the minority spin electron occupies two almost
degenerate groups of states composed, respectively, by a $z^2$ state
($z$ being the rhombohedral axis) and a doublet of states (mostly of
$x^2-y^2$ and $xy$ symmetry with $x$ and $y$ on the (111) planes of the
lattice). This degeneracy leads to a ground state
associated with a metallic KS system.
If LDA+U is used, the total energy is minimized
when the doublet degeneracy is lifted (through lowering the rhombohedral 
symmetry by a tripartition of the metal sublattice) 
and the minority spin electron is hosted on a combination
of $d$ states extending on the (111) planes \cite{cococcioni05}, 
as illustrated in Fig. \ref{feo} (lower panel).
This solution is actually not unique: in fact, at least three distinct 
linear combinations (orbital orders)
of $d$ states exist, all hosting the minority spin electrons
of Fe on (111) planes, that are equivalent and degenerate.
Each of them is predicted to be insulating thanks to the lifting of the
degeneracy between the states in the doublet group and to the 
discontinuity in the potential introduced by the Hubbard correction. 
However, the ground state of the system should be
regarded as a linear combination (with equal weights) of these 
solutions which, in fact, recovers the full rhombohedral symmetry of the
crystal.
An insulating state, with the minority spin $d$ electrons of Fe
hosted on the $z^2$ state (along [111]), preserves the rhombohedral
symmetry, but has higher energy.
In Ref. \cite{cococcioni05} it was shown that each of the equivalent
ground states with broken symmetry
can predict the rhombohedral distortion of the crystal under
pressure in better agreement with experiments \cite{willis53,yagi85}
than DFT or GGA+U 
ground states with full rhombohedral symmetry. This result is shown in 
Figure \ref{feo1} that reports the variation of the rhombohedral angle of
FeO under pressure and compares the results obtained with GGA (red line),
GGA+U with rhombohedral symmetry (green line) and GGA+U in the broken symmetry
phase (blue line). 
This conclusion is in agreement with the results 
of Ref. \cite{gramsh03}, although
the ground state stabilized by LDA+U seems to have a different
symmetry (and different orbital order)
than the one predicted in Ref. \cite{cococcioni05}.  
\begin{figure}[tbh!]
 \centering
 \includegraphics[width=0.45\textwidth]{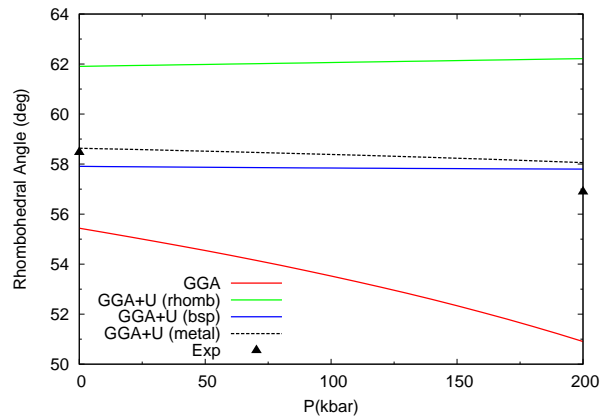}
 \vspace{-0mm}
\caption{(Adapted from Ref. \cite{cococcioni05}; color online).
The rhombohedral angle of FeO as a function
of pressure. The solid blue line describes GGA+U results in the broken-symmetry
phase (bsp), the red one GGA, the green GGA+U in the rhombohedral symmetry.
The dotted line is from GGA+U calculations with a metallic
KS spectrum (see text). Diamonds represent
the (extrapolations from) experimental data \cite{willis53,yagi85}.}
\label{feo1}
\end{figure}
%

The necessity to break the symmetry of the electronic system to reach
its insulating ground state has also been stressed in Ref. \cite{dorado09} 
where LDA+U is used to study the electronic and structural properties
of UO$_2$. In this work it is also shown that, favoring the anisotropy of
localized states, LDA+U often induces the formation of metastable phases
in which the system can get trapped. To avoid this inconvenience, a 
preconditioning of the occupation matrices (and of the Hubbard potential)
is sometimes needed.

If the rhombohedral symmetry is not broken to allow for the localization
of the minority spin electron of Fe on one of the (111)-planar $d$ states, this
electron is equally shared by them resulting in a metallic KS spectrum.
If the GGA+U is used on this ground state the dotted line of Fig. \ref{feo1}
is obtained. The good agreement with experimental data and with other
GGA+U results in the broken symmetry phase is a confirmation of the 
idea that the linearization of the energy with
respect to the occupation of degenerate states
is effective even in cases where no gap appears in the
KS single-particle spectrum.
A more accurate theoretical analysis shows that this less
``orthodox" use of LDA+U (on the fully symmetric and metallic ground
state) is less accurate and should be trusted only in cases where
the degeneracy that is responsible for the metallic character is not
lifted by the deformation.

In some cases, where the degenerate states are
quantistically entangled, the breaking of symmetry 
could have negative consequences on the description of some physical
properties and should be imposed with care (if at all). 
In these cases the use of
the Hubbard correction on a degenerate (metallic) state could
actually be a better option. A typical example of this type of situations
is represented by open-shell singlet molecules, typically affected by the
problem of spin contamination. Section
\ref{upv} reports the case of the Ir(ppy)$_3$ dye
(discussed more extensively in Ref. \cite{himmetoglu12_2})
whose open shell excited singlet state is best captured 
(in consistency with the Slater half-occupation theorem \cite{slater72})
by a configuration having half electron of each spin promoted to the 
LUMO of the molecule which, in a KS (or band structure) picture,
corresponds to a metallic state.

In the case of FeO, discussed above, after the spin symmetry is broken
and an antiferromagnetic ground state is obtained, 
a finite gap in the KS spectrum is obtained after lowering the rhombohedral
symmetry of the crystal and breaking the equivalence of 
$d$ orbitals on the same (111) plane. 
In CuO, described in section \ref{dftpj} of this review, 
the breaking of the symetry is somewhat harder to obtain as
spin and orbital degeneracies reinforce each other.
Other transition metal mono-oxides, as NiO,
only require spin symmetry breaking (AF ground state) as
a (small) gap appear in their KS spectrum due to crystal field.
Spin degeneracies also need to be lifted in paramagnetic insulators
if a gap in the KS band structure is to be obtained with LDA+U.

The necessity to lower or break the symmetry of the electronic
system to obtain an insulating single particle spectrum descends from
the degeneracy of the ground state of many correlated materials
(e.g., transition metal oxides)
and in the multi-reference character of their wave function \cite{dimarco12}, 
whose implications cannot be capture by the straight
use of LDA+U on the fully symmetric ground state.
More sophisticated methods and corrective approaches, as
RDMFT and DFT+DMFT are able to describe degenerate insulators 
without explicitly lowering the symmetry of the system
and can be used to capture metal-to-insulator phase transitions
(see, for example, Refs. \cite{held01,shorikov10,sharma13}). 
It is significant that these corrective approaches can be shown to also 
introduce a finite discontinuity in the chemical 
potential of the system (the corresponding quantity
of the potential in a KS framework) \cite{grossepl07,grosspra09}.

%
\subsubsection{Other flaws and merits of LDA+U}
At this point, it is important to highlight also other approximations 
inherent to the LDA+U description of correlated ground states. 
Atomic states are treated as effectively localized
and their dispersion is neglected, as is the k-point dependence of 
the effective interaction (the Hubbard $U$). 
This limit can be alleviated, in part, by taking into account
inter-site electronic interactions as explained in Ref. \cite{campo10}
and in section \ref{upv} of this review.
Another aspect to remark is the fact that LDA+U corresponds 
to a static correction. In fact, 
it neglects the frequency dependence of the
effective electronic interaction (i.e., of its screening).
The numerical difference between statically and dynamically screened
effective interactions was already pointed out in Ref. \cite{springer98},
that focused on bulk Ni as a case study.
In Ref. \cite{sasioglu11} the constrained RPA approach \cite{arya04}
is used to evaluate the Hubbard $U$ in transition
metals and to stress the importance of its dependence on frequency for these
materials. 
The variation of $U$ with $\omega$ suggests
that the static LDA+U is probably not very accurate for
many systems of this type. 
A possible correction to static models that allows to (partially)
account for the frequency dependence of $U$ at low energies
($\omega < \Lambda$ where $\Lambda$ represents the bandwidth of the system)
was already proposed in Ref. \cite{arya04}. 
Ref. \cite{casula12} has recently discussed the meaning and the importance
of using a frequency-dependent Hubbard $U$.
In order to account for dynamical effects on correlation (e.g., the frequency 
dependence 
of the screening of effective
interaction by delocalized electrons) more sophisticated
approaches are needed such as, for example, DFT+DMFT
\cite{metzner89,muller89,brandt89,janis91,georges92,georges96},
that has been shown to be able to describe 
metals and Mott insulators and to capture 
correlation-driven phenomena as metal-to-insulator transitions (see, e.g.,
Refs. \cite{held01,shorikov10}).
LDA+U, that can be considered the static (Hartree-Fock-like)
limit of DFT+DMFT, can not capture dynamical fluctuations
and can lead to qualitatively wrong results in systems, as many rare-earth
compounds, where these play an important role in determining both
ground and excited state properties as, for example,
the strength of hybridization between orbitals or the quasi-particle
excitation energies \cite{pourovskii05,suzuki09}.
Unfortunately, DFT+DMFT is significantly more computationally demanding than
LDA+U and, while inherently superior in describing multi-reference
ground states, it is hardly usable or quite impractical
for large systems, for molecular dynamics or for
large-scale calculations as, for example, those screening and comparing the
total energy of large numbers of different materials and
structural phases.
Furthermore, DFT+DMFT accounts for electronic correlation
using a Hubbard model (solved within the DMFT approximation)
wherein each correlated atom is treated 
as an Anderson impurity in contact with the
``bath" represented by the rest of the crystal.
Therefore, it shares with LDA+U the dependence of its results
on the choice of the interaction parameter $U$ and on the
specific double-counting term used to compensate for the 
correlation energy already contained in the LDA Hamiltonian.
Other problematic aspects of DFT+DMFT are instead inherent to the 
approximations made in solving the Anderson impurity model 
such as, the finite sampling of the Greens functions on the frequency
axis, the lack of self-consistency over the charge density, or
the overlook of spatial fluctuations, sometimes cured through the
cluster (or cellular) DMFT approach (cDMFT) \cite{lic00,kotliar01,senechal12}.
The effects of these approximations will not be further discussed here and
the reader is encouraged to review publications dedicated to the DFT+DMFT method
(e.g., Ref. \cite{pavarinibook11} and references quoted therein).

The small computational cost of LDA+U and 
the significant improvement it brings to the KS eigenvalues towards
their interpretation as single-particle excitation energies 
have promoted its use in conjunction with methods
to compute excitation energies: time-dependent DFT (TDDFT)
\cite{runge84,burke05} and GW \cite{hedin65}.
The use of TDDFT on extended (crystalline) systems can be quite challenging
due to the inability of the (approximate) interaction kernels
to capture important long-range interactions \cite{sottile05,botti07}.
Starting TDDFT calculations from a LDA+U functional has proven 
effective to circumvent this problem and to compute the bound
$d-d$ Frenkel excitons in NiO (using a Wannier function basis set) \cite{ku10}
in quite good agreement with experimental results \cite{larson07,muller08}.
The theoretical relationship between LDA+U and GW methods
has been discussed in Ref. \cite{anisimov97}.
The incorporation of the potential discontinuity in the KS gap
has opened the possibility to interpret LDA+U wave functions and KS energies 
as zeroth-order estimates of their quasi-particle counterparts.
Therefore, when applied to a LDA+U reference Hamiltonian, 
the GW correction, needed to recover the physical value of these quantities,
is smaller than with approximate DFT functionals
and the simplest approximations 
(most commonly, G$_0$W$_0$) become inherently more accurate. 
In fact, LDA+U/G$_0$W$_0$ has been succesfully used 
to calculate the quasi particle spectrum of several systems
\cite{jiang09,scheffler10,toroker11,kanan12,patrick12,liao11,isseroff12,
kioupakis08}, often improving the results of LDA/G$_0$W$_0$.

The negligible computational overload associated with LDA+U 
also makes it a precious (often the only affordable) 
method for ab initio calculations aimed 
at screening large sets of correlated materials to 
either scout new compounds and phases or 
to optimize the properties of existing ones for target applications. 
A typical approach to computational materials design, the
high-throughput (HT) technique is a clear example of this type of
application of LDA+U. HT is based on the efficient construction of
a database of known/computed materials and on a smart data mining
technique to select or design optimal candidate systems for target properties
\cite{curta03,fischer06}. LDA+U can be
easily implemented and used in HT searches based on DFT calculations.
A recent implementation of LDA+U in HT \cite{curta11,curta12,curta13} 
has demonstrated that a better description of electronic correlation
is very useful to make reliable predictions on the properties of
correlated materials. 
Although a qualitative improvement of results over approximate DFT
functional is often obtained for correlated materials, the 
quantitative outcome of LDA+U
calculations depends on the value of the Hubbard $U$. For a full 
exploitation of the potential of HT calculations, an automatic
(and run-time) evaluation of this interaction parameter
would be highly desirable.
Some approaches to obtain the value of $U$ from ab initio calculations
are discussed and compared in the next section.


\section{Computing the Hubbard $U$}
\label{computeu}
\subsection{The necessity to compute $U$}
From the expression of the Hubbard functionals discussed
in previous sections, it is natural to expect the results of the
LDA+U method to sensitevely depend on the numerical value of
the effective on-site electronic interaction, the Hubbard $U$.
A tendency wide-spread in literature is to use this approach 
for a rough assessment of the role of electronic correlation; 
therefore, it has become common practice to tune
the value of $U$ in a semiempirical way, through seeking agreement with
available experimental measurement of certain properties and using
the so determined value to make predictions on other aspects of the 
behavior of systems of interest.
Besides being not satisfactory from a conceptual point of view, this
practice does not allow to appreciate the variations of the on-site 
electronic interaction $U$ during chemical reactions, 
structural/magnetic transitions or, in general, under changing 
physical conditions.
As demonstrated in literature \cite{kulik06,hsu11}, instead, to capture the
variation of the electronic interactions is crucial for modeling
in a quantitatively predictive way the above mentioned situations. 
Therefore, in order to exploit all the potential of this approach it is
very important to define a procedure to compute the Hubbard $U$ in 
a consistent and reliable way. The interaction parameters should be
calculated for every atom the Hubbard correction is to be
used on, for the considered 
crystal structure and the specific magnetic ordering of interest.
The obtained value depends not only on the atom, its crystallographic position
in the lattice, the structural and magnetic properties of the crystal, 
but also on the localized basis set used to define the on-site
occupation in the ``+U" functional. Therefore, contrary
to another quite common practice, the effective
interactions have limited portability and their values should not be 
extended from one crystal to another, or from one implementation 
of LDA+U to another but, rather, recomputed each time.

\subsection{A brief literature survey}
In several works on LDA+U (see, e.g., Ref. \cite{anisimov91}), 
based on the use of localized basis
sets and on the Atomic Sphere Approximation (ASA),
the Hubbard $U$ is computed from the variation of the total energy
upon changing by one electron the population of the localized
(e.g., 3$d$) states of a single atom:
\barr
\label{ufd}
U&\approx& E\left(\frac{n}{2}+1;\frac{n}{2}\right) - E\left(\frac{n}{2};\frac{n}{2}\right)
\nonumber \\
&-&E\left(\frac{n}{2}+1;\frac{n}{2}-1\right) + E\left(\frac{n}{2};\frac{n}{2}-1\right).
\earr
In this equation 
the two numbers in between parenthesis represent the population of
the two spin manifolds and the original configuration is spin unpolarized
with $n$ electrons on the $d$ shell of each atom.
In practice, this quantity is evaluated (thanks to the Janak
theorem \cite{janak78}) from the difference between $3d$ energy levels:
\beq
\label{ufd1} U\approx \tilde \epsilon_{3d}\left(\frac{n}{2}+\frac{1}{2};\frac{n}{2}\right) - \tilde \epsilon_{3d}\left(\frac{n}{2}+\frac{1}{2};\frac{n}{2}-1\right)
\eeq
where $\tilde \epsilon(x,y) = \epsilon(x,y) - \epsilon_F(x,y)$
($\epsilon_F$ representing the Fermi level). In the expression of Eq. 
(\ref{ufd1}) the screening from 
the other (e.g., $s$ and $p$) states is automatically included
by letting their population reorganize when changing the number of electrons
on $d$ states.
From a comparison between Eqs. (\ref{ufd}) - (\ref{ufd1}) and
Eq. (\ref{fgap}) it is easy to realize that the $U$ is computed as the 
effective second derivative of the energy with respect to the occupation of
the $d$ orbitals. To ensure that the computed $U$ does not contain contributions
from the hopping terms (explicitly accounted for at run-time) 
the hopping between the $d$ states of the perturbed atom and other
states in the crystal is explicitly eliminated. This procedure
ensures that the computed $U$ corresponds to the amplitude of the potential 
discontinuity, $\Delta_{xc}$, and that the gap in the LDA+U KS spectrum has 
a width equal to the fundamental gap of the system.
The possibility to change the occupation of $d$ states and to cut hopping terms
with other states are quite specific to implementations that use
localized basis sets (e.g., LMTO); other implementations (e.g., using
plane waves) require different procedures to compute the effective
interaction parameters \cite{pickett98} that will be discussed below.

Another method to compute the Coulomb and exchange parameters for DFT+U
calculations has been recently proposed in Ref. \cite{mosey07}.
In this work $U$ and $J$ are evaluated by projecting unrestricted HF
molecular orbitals 
onto atomic orbitals
and retaining only on-site (intra-atomic) terms from the Hartree Fock
interactions, averaged over the states (of specific angular
momentum) of the same atom. 
While consistent with the HF-like expression of the DFT+U
corrective functional, this method yields values for $U$ and $J$ that are 
somewhat higher than those obtained from other methods, probably due
to the use of unscreened Coulomb (and exchange)
integrals from UHF. Screening is instead accounted for in other approaches
described below.

One of the latest methods to compute the effective (screened) Hubbard $U$
is based on constrained RPA (cRPA) 
calculations \cite{springer98,arya04,arya06,vaugier12,sasioglu12} and has
become particularly popular within the DFT+DMFT community.
This approach yields a fully frequency-dependent interaction 
parameter that is efficiently screened by ``non Hubbard" orbitals. 
If the polarization of the
system is written as the sum of a term from localized (e.g., $d$) states,
and one from delocalized ones: $P({\rm \bf r},{\rm \bf r}') = 
P_d({\rm \bf r},{\rm \bf r}') + P_r({\rm \bf r},{\rm \bf r}')$,
the inverse dielectric function can be factorized as follows:
$\epsilon^{-1} = \epsilon_r^{-1}\epsilon_d^{-1}$.
The effective interaction acting on the $d$ (localized) manifold can then
be computed from the screening of the electronic interaction kernel due to
the reorganization of electrons on
extended states. The dielectric function, responsible 
for this screening, can be defined as follows:
\beq
\label{er}
\epsilon_r({\rm \bf r},{\rm \bf r}') = \delta({\rm \bf r},{\rm \bf r}') - \int P_r({\rm \bf r},{\rm \bf r}'')
f_{Hxc}({\rm \bf r}'',{\rm \bf r}')d{\rm \bf r}''.
\eeq 
In this expression $f_{Hxc}({\rm \bf r}'',{\rm \bf r}')$ is the kernel of the
Hartree and xc interactions: 
$f_{Hxc}({\rm \bf r},{\rm \bf r}') = \frac{1}{|{\rm \bf r}-{\rm \bf r}'|}+
\frac{\delta v_{xc}({\rm \bf r})}{\delta \rho({\rm \bf r}')}$ \cite{vaugier12}.
Based on this definition, the effective interaction $W_r$ acting on the 
$d$ (localized) manifold can be computed as:
\beq
\label{crpa}
W_r = \epsilon_r^{-1} f_{Hxc} = 
\frac{f_{Hxc}}{1-P_r f_{Hxc}}.
\eeq
The Hubbard $U$ is obtained as the expectation value of 
$W_r$ on the wave functions
of the localized basis set \cite{arya04,arya06}.
In actual calculations, based on the explicit evaluation of the
polarization $P$ \cite{springer98,arya04}, only the Coulomb kernel 
is used (hence the name
``constrained RPA"). This approximation is based on the assumption
that the xc kernel, whose inclusion would make
the procedure much more involved and demanding, is numerically less important
than the Hartree one and can be safely neglected.
From the procedure outlined above
$U$ results the effective interaction partially screened by the
degrees of freedom not explicitly included in the model
Hamiltonian it is used in. 
In fact, the polarization $P_r$, necessary to compute 
$W_r$ (and $U$), is obtained subtracting from the total polarization $P$ 
the term $P_d$ due to $d-d$ transitions (the transitions between correlated
$d$ states and non correlated ones are still included).
The screening of the interaction due to $P_d$ is performed
at run-time when solving the DFT+DMFT equations.
From the definition of the dielectric
function it is easy to show that, when the
screening from $P_d$ is applied to $W_r$, the fully screened interaction is 
obtained: $W = \epsilon^{-1} f_{Hxc} = \epsilon_d^{-1} W_r$.

\subsection{Computing $U$ from linear-response}
\label{linearr}
\index{calculation of U from linear response}
\subsubsection{Technical aspects and computational procedure}
\begin{figure}[h]
 \vspace{-1mm}
 \centering
 \includegraphics[width=0.5\textwidth]{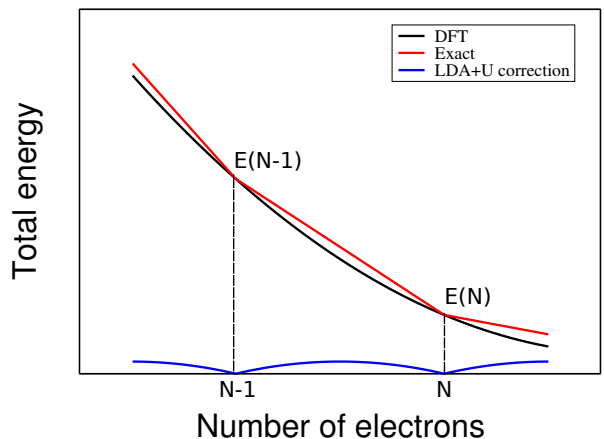}
 \vspace{-4mm}
 \caption{(
Color online) 
Sketch of the total energy profile as a
function of the number of electrons in a generic atomic system in contact
with a reservoir. The black line represents the DFT energy, the red the exact
limit, the blue the difference between the two. 
The discontinuity in the slope of the red line for integer occupations, 
corresponds to the difference between ionization potential and 
electron affinity and thus measures the fundamental gap of the system.}
\label{parab}
 \vspace{-2mm}
\end{figure}

In this section we describe the linear response approach
to the calculation of the effective Hubbard $U$ that was introduced in 
Ref. \cite{cococcioni05}. 
This method (inspired to the one proposed in Ref \cite{pickett98}) 
has been implemented
in the plane-wave pseudopotential total-energy
code of the Quantum-ESPRESSO package \cite{giannozzi09}.
As in the first method outlined above, and consistently
with the definition and the intent of the Hubbard corrective functional,
the $U$ is calculated from the spurious curvature of the (approximate DFT)
total energy of the system
as a function of the number of electrons on its localized (atomic) orbitals.
In fact, as was briefly discussed in section \ref{concept},
when these localized states 
exchange electrons with the rest of the crystal (acting like a charge
reservoir), the total energy obtained from approximate DFT xc functionals
varies in an analytic way
and its derivative (the effective potential acting on them)
misses or significantly underestimates 
the discontinuity at integer occupations that corresponds
to the fundamental gap of the system (Eq. (\ref{fgap1})).
As demonstrated by quite abundant literature 
\cite{perdew-par82,levy82,perdewlevy83},
the energy profile should consist, instead,
of a series of straigth segments joining the energies corresponding
to integer occupations.
A visual comparison between the exact (piece-wise linear) and the 
approximate energy (as functions of the localized states
occupations) is made in Fig. \ref{parab} where the latter
is modeled by a parabola.
If the curvature of the approximate energy profile is
assumed to be constant (actually a very good
approximation within single intervals between integer
occupations \cite{dabo10}), the expression of
the additive correction needed to recover the exact behavior
(bottom line in the cartoon of Fig. \ref{parab})
can be easily worked out 
as the difference between a parabola and a straight
line and results to have 
the same expression of the Hubbard correction in Eq. (\ref{our1}),
provided the $U$ is equal to the (spurious) curvature of 
the DFT energy profile.
Therefore, the main objective of this calculation is to evaluate
the second derivative of the total energy of the system with
respect to the occupation of the localized states, as defined in
Eq. \ref{occup}.

In codes that are not based on localized basis sets, (and use, 
e.g., plane waves and pseudopotentials) 
the on-site occupations 
are obtained as an outcome from the calculation after projecting
Kohn-Sham states on the wave functions of the localized basis set 
(Eq. (\ref{occup})). Therefore, to compute the second derivative of the
total energy with respect to the occupations, a different 
approach was adopted that is based on Legendre transforms \cite{cococcioni05}.
The first step consists in applying a shift
to the external potential that only acts on the localized
orbitals of a Hubbard atom $I$ through a projection operator:
\beq
\label{dv}
v^p_{ext} = v_{ext}
+\alpha^I \sum_m \vert \phi_m^I\rangle \langle \phi_m^I \vert
\eeq
(the superscript ``{\it p}" standing for ``perturbed").
In this equation 
$\alpha^I$ represents the amplitude of the perturbation (usually
chosen small enough to maintain a linear response regime).
The potential in Eq. (\ref{dv}) is the one used to solve the KS equations.
They yield a $\alpha$-dependent ground state charge density
and total energy:
\barr
\label{ealfa}
E(\alpha^I) &=& \sum_{i,\sigma} \epsilon_i^{\sigma}(\alpha^I) 
- \frac{1}{2} \int v_H[\rho_{\alpha^I}({\rm \bf r})]
\rho_{\alpha^I}({\rm \bf r}) d{\rm \bf r} \nonumber \\
&+& E_{xc}[\rho_{\alpha^I}] - \int v_{xc}[\rho_{\alpha^I}({\rm \bf r})]
\rho_{\alpha^I}({\rm \bf r}) d{\rm \bf r}
\earr
where $\epsilon_i^{\sigma}(\alpha^I)$ are the single-particle 
energies obtained from the solution of the KS problem.
An occupation-dependent total energy functional can be recovered from 
the expression in Eq. (\ref{ealfa}) using a Legendre transform:
$E[\{n^I\}] = E(\alpha^I) - \alpha^I n^I$ (where
$n^I$ indicates the value of the on-site occupation corresponding
to the perturbed ground state).
Based on this definition, the first and
second derivatives of the energy are, respectively,
\beq
\label{d1e}
\frac{d E}{d(n^I)}=-\alpha^I 
\eeq
and
\beq
\label{d2ee}
\frac{d^2 E}{d(n^I)^2}=-\frac{d\alpha^I}{d(n^I)}. 
\eeq
In actual calculations the latter quantity is obtained by solving the 
Kohn-Sham equations for a range of values of 
the parameter $\alpha^I$ (on every Hubbard atom) centered
around 0 and collecting the response of
the system in terms of the variation of the total occupations $n^J$
of all the atoms.
The operation is repeated perturbing each Hubbard atom separately.
The quantity that can be directly measured from this series of calculations 
is the response matrix 
\beq
\label{rm}
\chi_{IJ} = \frac{d(n^I)}{d\alpha^J}, 
\eeq
where $I$ and $J$
are site indices that label the Hubbard atoms.
The curvature of the total energy (Eq. \ref{d2ee}) 
with respect to the occupations
$n^I$ is thus obtained as the inverse of the response matrix:
$d^2 E/d(n^I)^2 = -(\chi^{-1})_{II}$.
This quantity is not the effective $U$. 
In fact, applying a perturbation as the one in Eq. (\ref{dv}) to a non
interacting electron system, also results in a response of 
the occupations (due to the rehybridization of the electronic wave functions)
that contributes a finite term to the
second derivative of the total energy. Based on Eq. (\ref{d2ee})
and on the definition of the response matrices,
this ``non-interacting" contribution 
can be expressed as $d^2 E/d(n^I)_0^2 = -(\chi_0^{-1})_{II}$,
$d(n^I)_0$ being the variation of occupation due to the above mentioned
rehybridization.
Being not related to electron-electron
interactions, this term 
should be subtracted from the second derivative of the
total energy. The Hubbard $U$ is thus obtained as:
\beq
\label{ucalc}
U^I = (\chi_0^{-1} - \chi^{-1})_{II}.
\eeq
As the response of a non interacting electron gas (with the same density of 
the interacting one), $\chi_0^{-1}$ is sometimes interpreted as the
kinetic (or single-body) 
contribution to the second derivative of the energy \cite{pickett98}. 
In order to understand how $\chi_0$ is actually computed, 
it is useful to realize that
it measures the response of the system to a variation of 
the {\it total potential} (while $\chi$ is the response to the variation
to the {\it external potential}). In other words:
\beq
\label{lrt}
\Delta n^I = (\chi_0)_{IJ} \Delta V^{tot}_J = \chi_{IJ} \Delta V^{ext}_J = 
\chi_{IJ} \Delta \alpha^J.
\eeq
The calculation of $\chi_0$ 
requires special care in the iterative solution of the Kohn-Sham equations
at finite $\alpha^I$.
In fact, $\chi_0$ is calculated from the variation of 
the atomic occupations immediately
after the {\it first diagonalization} of the Hamiltonian resulting from
the sum of the self-consistent (unperturbed)
KS Hamiltonian and the perturbative potential
in Eq. (\ref{dv}). At this initial step of the perturbed run,
the variation of the {\it external potential} has not yet been screened by the
response of the electronic charge density (through the Hartree and xc 
potentials) and is thus coincident with the variation of the {\it total
potential}: $\Delta V^{tot}_J = \Delta V^{ext}_J = \Delta \alpha_J$.
Thus, from Eq. (\ref{lrt}), one easily obtains: 
$(\chi_0)_{IJ} = (\Delta n^I / \Delta \alpha^J)_{first}$.  
Therefore, the first diagonalization of the electronic Hamiltonian
in the perturbed run must be very precise in order to obtain
an accurate non-interacting response matrix $\chi_0$.
$\chi$ is instead computed after the perturbed calculation has 
reached self-consistency: $\chi_{IJ} = (\Delta n^I / \Delta \alpha^J)_{last}$.

It is important to stress that the linear-response calculations to 
compute the response matrices $\chi$ and $\chi_0$
are performed in a supercell of the crystal (whose size is determined
from the convergence of the obtained Hubbard $U$ \cite{cococcioni05})
where only one atom is perturbed each time. In fact, consistently with the
treatment of localized states as isolated atomic orbitals in contact
with a bath (represented by the rest of the crystal), 
and with the definition of the Hubbard $U$
as the energy cost associated with the double occupancy of the orbitals of
{\it a single atom}, it is necessary to isolate
the atom with perturbed states and to avoid the interaction with its
periodic images. If the separation between perturbed atoms 
(i.e., the size of the supercell) is not
large enough, the resulting $U$ is screened, to some extent, by the
residual coupling between the perturbations on equivalent atoms
and, when used in LDA+U calculations (in the actual unit cell of the 
crystal), it incurs in some double screening. 
The charge redistribution induced by the perturbation in the external 
potential, Eq. (\ref{dv}), usually involves multiple Hubbard atoms. 
This is the reason for which $U$ is obtained from inverting 
the entire response matrices rather than their single (diagonal) elements.
In Ref. \cite{cococcioni05} the matrices $\chi$ and $\chi_0$
are constrained to represent the response of a system to a neutral perturbation.
This amounts to impose the sum of the matrix elements on the same row and
the same column to be zero by adding a neutralizing
``background" (described by an extra row and an extra column in 
each of these matrices). However this condition would be legitimate
to impose only if there was a perfect overlap between the Hilbert spaces
spanned by atomic and KS states (never exactly the case, in practice).
Recently, it was realized that a better way to account for the charge reservoir
Hubbard atom exchange electrons with, while screening the external
perturbation, is to explicitly
include in $\chi$ and $\chi_0$ the collective response of ``non Hubbard"
atoms and states (also collected in an extra row and an extra column)
and to also consider the response of the system to their collective 
perturbation (obtained from imposing the same $\alpha$ to all these 
states at the same time).
This refinement was found to have beneficial effects on the convergence 
of the calculation with the size of the chosen supercell.

Following the same procedure illustrated above for the Hubbard $U$,
the intra-atomic exchange interaction 
$J$ could, in principle, be 
obtained in a similar way, adding a perturbation that couples with the on-site 
magnetization $m^I = n^{I\uparrow} - n^{I\downarrow}$:
\beq
\label{dvm}
v^p_{ext} = v_{ext}
+\beta^I \sum_m 
\left[ \vert \phi_m^{I\uparrow}\rangle \langle \phi_m^{I\uparrow} \vert - 
\vert \phi_m^{I\downarrow}\rangle \langle \phi_m^{I\downarrow} \vert \right]. 
\eeq
However, since the total energy is not variational with respect
to the magnetization, and
the magnetization of a system often reaches its
saturation value (compatibly with the number of electronic states),
a perturbative approach is generally not viable.
Also, in these circumstances, $n^I$ and $m^I$ are not independent variables
(in fact, only one spin population can be perturbed, the other spin states
being fully occupied and typically removed from the Fermi level)
and only linear combinations of $U$ and $J$ can be
obtained from LR, not their separate values.
A possible way around this problem consists in perturbing 
a ground state whose absolute magnetization has been constrained to be 
lower than its saturation value so that 
$n^I$ and $m^I$ can be varied independently.
However, calculations of this kind require effective constraints on the
atomic magnetization of atoms and turn out to be technically
difficult to perform and very delicate to bring to convergence. 

Similar problems arise when computing $U$ for fully occupied or empty 
states. In fact, the linear-response approach discussed above is suitable
to calculate the effective electronic coupling of manifolds of states that are 
either in the vicinity of the Fermi level (and thus partially full), or result
from the hibridization of atomic orbitals of different atoms
(as, e.g., the valence states in bulk elemental semiconductors).
If the manifold is completely full (e.g., the O $p$ states in some
transition metal oxides) and distant in energy from the Fermi level or the
top of the valence band, their response is very small and 
may easily fall within the
numerical noise of the calculation. In these cases the reliability 
of the obtained $U$ is questionable (values of 30 eV or higher are not
uncommon). Whether or not a preliminary shift of the manifold closer
to the Fermi level could be a solution, depends on the specific material
and on the entity of the collateral effects this shift has on its 
electronic structure and its physical properties.


\subsubsection{The analytic expression of $U$ and the problem of screening}
It is useful, at this point, to study the analytic expression
of the Hubbard $U$, obtained, as detailed in appendix \ref{app1}, from the
(linear) response of atomic occupations to a perturbation
in the potential acting on localized orbitals that is a generalization
of the one given in Eq. (\ref{dv}).
If the definition of the occupation matrix is extended to contain
off-diagonal terms with atomic orbitals belonging to different 
sites $I$ and $J$, 
$n^{IJ}_{i,j}=\sum_{n}f_{n}\langle \psi_{n} |
\phi_{j}^{J} \rangle\langle \phi_{i}^{I} | \psi_{n}\rangle$ 
(this extension will be also needed for the LDA+U+V functional, 
discussed in section \ref{upv}),
and the perturbation to the external potential is generalized accordingly to
$\Delta V_{ext}^{LM} = \alpha^{LM}_{lm} |\phi^L_l\rangle\langle\phi^M_m|$  
a four-index response matrix can be defined as follows:
\beq
\label{chigen}
\tilde\chi^{JKML}_{jkml} = \frac{dn^{JK}_{jk}}{d\alpha^{LM}_{lm}}
\eeq
where upper case letters indicate atomic sites, lower case letters label
atomic states. 

The matrix $\tilde {\rm \bf U}$, that is obtained from the inversion
of this matrix (and its non-interacting analog $(\tilde\chi_0)^{JKML}_{jkml}$), 
as indicated in Eq. (\ref{ucalc}),
consists of the expectation values of the Hartree and 
exchange-correlation interaction kernels over the states of the atomic
basis set:
\barr
\label{umat0}
&&\tilde U^{OPSR}_{opsr} = (\tilde\chi_0^{-1} - \tilde\chi^{-1})^{OPSR}_{opsr} = \nonumber \\
&&\int \int (\phi^O_o({\rm \bf r}))^*\phi_p^P({\rm \bf r})
f_{Hxc}({\rm \bf r},{\rm \bf r}')
\phi^S_s({\rm \bf r}')^* \phi^R_r({\rm \bf r}')~d{\rm \bf r}~d{\rm \bf r}'.
\earr
This expression
might surprise for the lack of screening. A similar result
was obtained in Ref. \cite{anisimov07} where the definition of an
orbital dependent functional, able to eliminate the spurious
curvature of the DFT energy and to re-establish the finite
discontinuity of the potential, was based on the same {\it unscreened} 
interaction kernel as the one in Eq. (\ref{umat0}). The specialization of this
correction to a fixed basis set of Wannier functions also resulted
in a final expression resembling closely the LDA+U one with 
effective interactions computed as in Eq. (\ref{umat0}).
It is important to remark that the Hubbard $U$ used in actual
LDA+U calculations is not the one given in Eq. (\ref{umat0}) but,
rather, the one calculated as in Eq. (\ref{ucalc}),
which is based on the response matrices measuring the variation of 
the {\it total} on-site occupations $n^I$ (Eq. (\ref{rm})) in
response to (diagonal) perturbations acting on all the states of
each atom.
While linear response equations do not have a closed form
for the two-atomic-indexes response matrices, the following
formal relationship can be derived (appendix \ref{app1})
between the effective Hubbard $U$ and the one in Eq. (\ref{umat0}):
\beq
\label{umat1I}
{\rm \bf U} = ({\rm \bf \chi_0})^{-1}{\rm \bf A}~{\rm \bf \chi}^{-1}.
\eeq
where the response matrix, defined in Eq. (\ref{rm}), 
can be obtained from the contraction of its four indices analog 
in Eq. (\ref{chigen}): 
$\chi^{IR} = \sum_{ir} \tilde \chi^{IIRR}_{iirr}$ and
the matrix {\rm \bf A} is defined as follows:
\beq
\label{ars}
A^{RS} = \sum_{rs} \sum_{KQTZ}\sum_{kqtz}\left(\tilde\chi_0\right)^{RRQK}_{rrqk}
\tilde U^{KQTZ}_{kqtz}\tilde\chi^{ZTSS}_{ztss}
\eeq
(refer to Eq. (\ref{cu1}) for the expression of {\rm \bf U} in terms of 
$\tilde {\rm \bf U}$ with explicit sums over state and site indexes). 
It is instructive, at this point, to compare the effective $U$
obtained from the linear-response (LR) method outlined above, 
Eq. (\ref{umat1I}), with the one computed from cRPA\cite{arya04} 
(neglecting the frequency dependence of the dielectric constant),
Eq. (\ref{crpa}).
The difference between the two results is in the way the screening
is performed. If all the electronic states were treated explicitly, a bare
(i.e., unscreened) interaction (Eq. (\ref{umat0})) is obtained 
with both methods. This case has been discussed in appendix \ref{app1} for LR,
and would correspond to putting $\epsilon = \epsilon_d$ (i.e., 
$\epsilon_r = 1$) in the cRPA method.
As described earlier, within cRPA the (kernel of the) effective interaction is 
computed as $\epsilon_r^{-1} f_{Hxc}$, through the screening operated by
all the electronic degrees of freedom not treated explicitly in the 
model Hamiltonian (e.g., by $d-s$ or $s-s$ transitions, $s$ indicating
non Hubbard states). An analogous approach in LR would require
writing $\epsilon^{-1} = \chi_0^{-1}\chi$ as the product of
two contributions, from localized ($d$) and delocalized ($s$)
states or, equivalently, to write $\chi_0$ as the sum
of $d$ and $s$ terms, $\chi_0 = \chi_0^d + \chi_0^s$.
However, this is not possible, due to the ``coarse-grained" 
nature of the response matrices employed.
In LR an effective screening of the electronic 
interaction is operated by the matrix multiplications in Eq. (\ref{ars})
that contain summations over transitions
between $d$ states of distinct atomic sites, between $d$ and $s$, and 
between $s$ and $s$ states.
These summations lead to a significant contraction of the 
computed interactions whose value decreases 
from 15-30 eV, typical of the unscreened quantitity, to the
2-6 eV range of the effective one. A qualitative argument to understand 
this result is as follows: when an electron (or a fraction of it) is 
moved on to a specific atomic site and increases its occupation, it is
drawn from other states and orbitals, resulting in negative 
state- and site-off-diagonal elements of the response matrices 
in the multiplication of Eq. (\ref{ars}). In other words, the effective energy
cost of double occupancy of the considered site is reduced by the 
decreased weight of other terms of the electron-electron interaction, 
mostly involving off-diagonal terms of the occupation
matrices.
From these observations, further detailed at the end of appendix \ref{app1},
we can conclude that the effective $U$ obtained from LR can be best 
understood as resulting from 
the downfolding of the electron-electron interaction to the
$d$ (localized) states, after the elimination of higher order
off-diagonal $d-d$, $d-s$ and $s-s$ transitions.

\subsubsection{Ab-initio LDA+U: examples}
The calculation of the effective Hubbard $U$, described above, 
renders the LDA+U an ab initio method, 
eliminating any need of semi-empirical evaluations of the interaction
parameters in the corrective functional.
It also introduces the possibility to compute the values of these
interactions in consistency with the choice of the localized
basis set, the crystal structure, the magnetic
phase, the crystallographic position of atoms, etc. This ability 
proved critical to improve the predictive capability of LDA+U and
the agreement of its results with available experimental data
for a broad range of different materials and conditions.
The capability to compute the interaction parameters 
significantly improves the description of the structural, electronic and
magnetic properties of a variety of transition-metal-containing crystals 
and was particularly useful in presence of structural transformations
\cite{cococcioni05,hsu09}, 
magnetic transitions \cite{hsu11} and chemical reactions 
\cite{zhou04-1,zhou04-2}.
In Ref. \cite{hsu11} the use of a Hubbard $U$ 
recomputed for different spin configurations
allowed to predict a ground state for the (Mg,Fe)(Si,Fe)O$_3$ perovskite with
high-spin Fe atoms on both A and B sites, and a pressure-induced
spin-state crossover of B-site Fe atoms that couples with a
significant volume contraction, an increase in the quadrupole splitting
(consistent with recent X-ray diffraction and M\"ossbauer spectroscopy
measurements) and a marked anomaly in the bulk modulus of the material
under pressure.
The calculation of the Hubbard $U$ also
improved the energetics of chemical reactions \cite{kulik06,scherlis07},
and electron-transfer processes \cite{sit06}.
A recent extension to the linear response approach has further increased
its reliability through the self-consistent calculation of the $U$ from an
LDA+U ground state \cite{campo10,kulik06}. This method 
is mostly useful for systems where the LDA and LDA+U ground states are
qualitatively different. It is based on a similar calculation to the one 
described above except that the perturbative LDA+U calculation is performed
with the Hubbard corrective potential {\it frozen} to its self-consistent
unperturbed value. This strategy guarantees that the ``+U" part does 
not contribute to the response of the system
and, consistently to its definition, 
the Hubbard $U$ is measured as the curvature of the LDA
energy in correspondance of the LDA+U ground state.
Using the Hubbard $U$ computed at the previous step to induce the LDA+U
ground state for the next, the calculation is repeated cyclically
until when the input and otput values are numerically consistent.
The procedure usually reaches convergence in few cycles (less than five
in most cases).
Recently a similar self-consistent calculation of $U$ has been also 
implemented for the cRPA approach \cite{karlsson10}.

\section{Choosing the localized basis set}
\label{bas_set}
The choice of the localized basis set to define the occupation matrix
and the possible dependence of the results on this choice remain open issues
of the LDA+U method. Very often this choice is dictated by the specific
implementation of DFT being used (e.g., based on gaussian functions, 
muffin-tin orbitals and augmented plane waves, etc).
In principle, if the localized basis set were $i)$ orthonormal and $ii)$
complete, 
$iii)$ the effective interactions had full orbital dependence, and $iv)$
their numerical value was chosen/computed consistently with the basis set,
the results obtained from LDA+U calculations would not depend
on the choice of the basis set. Since, in practice, the first three 
conditions (and often the forth too) are never verified, some care must 
be used in the selection of the localized orbitals.
In fact, when basis sets are finite, switching from one to another
only generates an equivalent description if the two span the
same Hilbert space.
Consistently with our plane wave, pseudopotential implementation of 
LDA+U \cite{giannozzi09}, in this section we will only discuss basis sets
consisting of atomic orbitals (e.g., from the pseudo potentials) or
Wannier functions.

The choice of atomic orbitals (e.g., solutions of the radial Schr\"odinger
equation for isolated atoms, multiplied by spherical harmonics) is 
somewhat ``natural" to LDA+U since
it is based on the Hubbard model that was designed to capture the Mott
localization of electrons on atoms. In addition, in its simplest version, 
it contains only ``on-site" interaction parameters accounting for the
Coulomb repulsion betwen electrons on the same atom.
However, as discussed in section \ref{concept}, the Hubbard functional 
can be associated to a broader scope and it can be regarded
as a simple correction designed to impose to the exchange correlation 
potential the discontinuity it is supposed to have (the Hubbard $U$ is
actually the amplitude of the discontinuity) and to obtain a Kohn-Sham
HOMO-LUMO gap equal to the fundamental gap of the considered system.
In its original formulation, it is most effective when the gap is to 
result from lifting the (nearly exact) degeneracy of 
{\it localized} atomic states (typically $d$ or
$f$) of open-shell systems. The localized character of these states
is indeed what justifies 
orbital-, k-point-independent and (usually) atomically averaged
effective interactions. 
This approximation remains indeed well justified
even for systems where electronic localization occur on more general
orbitals (centered, for example, on bonds). In these cases, however,
the correction loses its atomic character and the effective interactions
should be recomputed accordingly.

\subsection{Atomic orbitals}
Atomic orbitals are usually obtained from the solution of the
radial Schr\"odinger equation for isolated atoms (the angular part is 
added when performing calculations for the systems of interest).
These orbitals are represented by wave functions centered on the 
single atoms and decaying with the distance from its nucleus:
$\phi_m^{IR} = \phi_m({\rm \bf r} - {\rm \bf \tau_I} - {\rm \bf R})$. In this
expression ${\rm \bf \tau_I}$ is the position of the $I^{th}$
atom in each unit cell of the crystal 
(or in the molecule), ${\rm \bf R}$ designates the unit cell.
The atomic orbital occupations, Eq. \ref{occup}, should be defined, in 
principle, by specifying the unit cell the atomic wave 
function is localized in.
However, the periodicity of the crystal allows us to drop this 
index from the expression of the occupation matrices and to use
the definition in Eq. (\ref{occup}).
Consequently, the Hubbard energy (per unit cell) does not depend on
${\rm \bf R}$ and can be computed from a single unit cell.
The expression in Eq. (\ref{our1}) can be understood as obtained from
the unit cell at ${\rm \bf R} = {\bf 0}$: 
$\frac{1}{N}\sum_{{\rm \bf R}I}\frac{U^I}{2} Tr\left[
{\bf \rm n}^{{\rm \bf R}I\sigma}\left(1-{\bf \rm n}^{{\rm \bf R}I\sigma}\right)\right] =
\sum_{I}\frac{U^I}{2} Tr\left[{\bf \rm n}^{{\rm \bf 0}I\sigma}
\left(1-{\bf \rm n}^{{\rm \bf 0}I\sigma}\right)\right]=
\sum_{I}\frac{U^I}{2} Tr\left[{\bf \rm n}^{I\sigma}
\left(1-{\bf \rm n}^{I\sigma}\right)\right]$.
It is important to stress that the projection of the atomic wave function
on a Kohn-Sham state at a given k-vector ${\rm \bf k}$ selects the 
Fourier component of the localized atomic orbital at the same k-vector
(and at all the ${\rm \bf k+G}$ points, ${\rm \bf G}$ being a fundamental
vector of the reciprocal lattice). Therefore, the calculation of atomic
occupations in Eq. \ref{occup} would give exactly the same result
if, instead of localized atomic orbitals, their Bloch sums were used.
This observation will be important when computing the derivatives 
of the occupation matrices to obtain, for example, forces and stresses
(see section \ref{derivae}).

A problem that arises with atomic wave functions is the finite overlap between
orbitals belonging to neighbor atoms. This fact compromises the summation
rules of atomic occupations in Eq. \ref{occup} (some portion of electrons
are counted more than once) and make the Hubbard energy and potential
less well defined (some occupations may exceed 1).
The problem can be solved performing a preliminary orthogonalization of the
atomic basis set using, for example, L\"owdin decomposition. Within
this scheme an orthonormal basis set can be obtained as
\beq
\label{ortho}
\phi_i^{orth} = \sum_j O^{-1/2}_{ij}\phi_j
\eeq
where $O_{ij} = \langle \phi_i \vert \phi_j \rangle$ is the overlap matrix
between orbitals of the original basis set (the low case indexes $i$ and $j$
are comprehensive of site and state labels).
The mixing of orbitals from different sites through the overlap matrix
in Eq. \ref{ortho} leads to a loss of the atomic character of the 
wave functions; however, the contribution from neighbor sites is usually 
small and their use in a Hubbard-modeled correction with atomic interactions
is still largely legitimate. 

It is important to stress that the use of an orthogonalized basis
set makes the calculation of energy derivatives significantly more 
challenging. In fact, the overlap matrix in Eq. (\ref{ortho})
does not usually commute with its derivative (nor do its powers, obviously)
so that the derivative of ${\rm \bf O}^{-1/2}$ can not be easily obtained 
from that of ${\rm \bf O}$ (a numerical solution to this problem could
be obtained exploiting the
fact that ${\rm \bf O}$ should contain a relatively small deviation from
the unit matrix ${\rm \bf O} = ({\rm \bf 1} + {\rm \bf t})$
and a series expansion on ${\rm \bf t}$ should converge rapidly). 
As a consequence, when derivatives of the energy are needed,
a non-orthogonal basis set is generally used.

It is also useful to keep in mind that
the effective interaction parameter to be used in the 
Hubbard functional is sensitive to the specific localized basis set used
to define the atomic occupations and the difference between orthogonalized
and non-orthogonalized wave functions is sufficient to cause an 
appreciable variation in its value.
Therefore, the Hubbard $U$ should be recomputed consistently 
e.g., using the linear response technique discussed in section \ref{computeu}, 
with the same basis set employed in the costruction of the functional.

Another possible way to eliminate or significantly alleviate the 
orthogonalization problem consists in truncating the atomic wave function
at the core radius of the pseudopotential of the atoms they belong to.
In this way, the integration of on site occupations is restricted
within the regions around the atomic cores and the Hubbard potential
amounts to a renormalization of the coefficients $D^I_{ij}$ of the non-local
pseudopotential: 
\barr
\label{nlpu}
V_{tot} &=& V_{loc} + V_{NL} + V_{Hub} \nonumber \\
&=& V_{loc} + \sum_I\sum_{ij}\left(D^I_{ij}+\Delta^I_{ij}\right)
\vert\beta_i^I\rangle\langle\beta^I_j\vert.
\earr
In this expression $\Delta^I_{ij}$ contains the expectation value 
of the Hubbard potential
on the all-electron partial waves of the pseudopotential, corresponding
to the projector waves $\beta^I_i$. While Eq. (\ref{nlpu}) uses a formalism
that resembles that of ultrasoft pseudopotentials \cite{vanderbilt90}, 
it can be used with general non-local pseudopotentials. 
This implementation of the Hubbard functional was first 
introduced in a projector-augmented wave framework
in Ref. \cite{bengone00} where it was also shown that the charge excluded
from the atomic cutoff spheres is usually small and contributes negligible
corrections to the Hubbard functional. In Ref. \cite{sclauzero13} this 
pseudopotential implementation of LDA+U was adapted to general non-local 
pseudopotentials and used to study the ballistic electron transport
in Au monoatomic chains.
A similar method was also used to construct an atomic self-interaction 
correction and to effectively embed it in the pseudopotential 
\cite{filippetti03,filippetti03_1,sanvito07,filippetti09}.

\subsection{Wannier functions}
If electrons localize on states that are not centered on atoms, a Hubbard
corrective functional based on atomic states and on-site only interactions
is not likely to improve the description of the corresponding ground state.
A possible solution to this
problem consists in generalizing the expression of the Hubbard corrective
functional to include interaction terms (e.g., between electrons
on different atoms) that are usually neglected in the on-site
formulations and to partially 
recover the invariant second-quantization expression of
the Hubbard functional \cite{hubbardI,hubbardII,hubbardIII,hubbardIV,hubbardV,
hubbardVI} with orbital- and/or site-dependent effective electronic 
interactions.
This approach is the one followed in the formulation of the LDA+U+V
correction that, through including inter-site interactions, proves
able to capture the localization of electrons on the $sp^3$ bonds of 
covalent semiconductors (e.g., Si). 
This generalization is presented in section \ref{upv}
and won't be further discussed here.

An alternative approach to this problem 
consists in adopting a basis set of 
orbitals particularly suitable to capture the localization
of electrons in the considered system.
In the case of elemental band semiconductor (e.g., Si) this would imply 
to use, for example, Wannier functions centered around the Si-Si bonds.
While this choice of basis functions guarantees the possibility to still
use a ``localization-center-diagonal" interaction term, it requires a 
preliminary knowledge about the localization centers of the electrons
or an additional mathematical criterion (e.g., maximal localization
\cite{marzari97}) to precisely define the basis set.
Wannier functions have indeed become a quite popular choice in recent years
to define corrective functionals and computational schemes to improve
the description of electronic localization in strongly correated systems.
In Refs \cite{lechermann06,trimarchi08}, for example, maximally-localized
Wannier functions (MLWF) \cite{marzari97} were used to  
facilitate the identification of correlated orbitals
and to construct a more flexible and general interface between 
DFT and DMFT that allows to construct a DFT+DMFT scheme from all 
possible implementations of DFT as, for example, 
those based on plane waves and psudopotentials.
In Ref. \cite{anisimov07} Wannier functions were instead employed to construct
a LDA+U - like correction to the DFT total energy aimed at restoring
the discontinuity in the exchange-correlation potential, generally
missing in approximate functionals. 
While MLWF are a popular choice for the definition of 
many of these functionals based on
Wannier functions, other schemes have also been employed in literature. 

In Refs. \cite{oregan10,oregan11,oregan12} non-orthogonal generalized
Wannier functions (NGWF) were used to define a LDA+U scheme compatible
with linear-order-scaling ($O(N)$) DFT. Within this implementation, 
based on a generalized covariant-controvariant
definition of the occupation matrix, the total energy
is minimized with respect to both the kernel of the density matrix and
with respect to the coefficients of the expansion of the NGWF
on the Kohn-Sham states. This ``internal" minimization, while
adding a negligible overload to the calculation, leads to a variational 
optimization of the localized basis set used in the definition of the 
Hubbard correction. The WF basis obtained in this way
thus results optimally adapted to 
capture electronic localization and to produce a density matrix as
close as possible to be idempotent.

An alternative WF-based LDA+U approach was 
proposed in Refs \cite{fabris05} and \cite{fabris05_1} where
the Wannier functions were defined (from the Kohn-Sham states 
of the system) by maximizing their overlap with the atomic
wave functions of ``Hubbard" atoms. This LDA+U scheme was
used to study the electronic structure around an
oxygen vacancy on the surface of CeO$_2$, a material often 
employed in the catalytic purification of exhaust gases resulting from various 
processes. It was found that LDA+U (with 
the Hubbard interaction computed from linear response\cite{cococcioni05})
favors the reduction of two of the Ce atoms 
around the oxygen vacancy (first neighbors) from 4+ to 3+, inducing
the localization of the two excess electrons on their $f$ states.
This redistribution of charge is indeed in better agreement with 
experiments and chemical intuition than 
a metallic state with excess elecrons
spread among the $f$ orbitals of all the Ce atoms surrounding
the vacancy, as predicted by non corrected DFT functionals. 
More importantly, it was found that the Wannier function-based LDA+U scheme
works better, in this case, than one using occupations defined on atomic
orbitals. In fact, because of the optimal overlap with 
the atomic states of ``Hubbard" atoms, the use of Wannier functions allowed to
effectively separate the fully occupied (valence) manifold from the empty 
(conduction) one. With all the occupations equal
to either 0 or 1, the total energy of the system does
not depend on the value of the Hubbard $U$, as it can be easily
understood from Eq. (\ref{our1}), and the agreement of the 
computed reduction energy with availale experiments was significantly
improved. It is important to notice that the effective 
interaction $U$ still controls the position of the Hubbard bands (the Ce $f$
states in this case) with respect to the conduction and valence manifolds.
In fact, the Hubbard potential (Eq. (\ref{simplepot})) does not vanish
in this case
and rigidly shifts the energy of the unoccupied states with respect to
that of unoccupied ones.
Therefore, even in cases where the energy does not depend
on $U$, its calculation is still important to accurately
describe the electronic properties of the system and its chemical 
reactivity. 

\section{The double-counting ``issue" and the LDA+U for metals}
\label{dc0}
\subsection{Comparison between the FLL and the AMF approaches}
\label{doublec} 
The choice of the double counting term certainly represents
an open issue of the LDA+U method. 
The lack of an explicit expression of the xc energy
makes it difficult to model how electronic correlation is accounted for 
in approximate DFT energy functionals. 
As a result, simple dc functionals, like the ones in Eqs. (\ref{ub2}) and
(\ref{our1}), are not general and flexible enough to work equally well for all 
kinds of systems. 

In section \ref{basic} and \ref{computeu} it was shown
how the ``fully-localized" (FLL) formulation of the LDA+U is 
constructed to impose a finite discontinuity to the xc
potential. 
Since this discontinuity also represents an important
term of the fundamental gap of a system, it is natural to expect
that this specific formulation is particularly effective to improve
the description of semiconductors and insulators, 
but not well suited to treat metals or ``weakly correlated" materials in
general.
In fact, the eccessive stabilization of occupied
states due to the ``+U" corrective
potential (see Eq. \ref{simplepot}) can lead to a description of the
ground state inconsistent with experimental data and
to quite unphysical results
(such as, e.g., the enhancement of the Stoner factor \cite{chioncel03}) or 
serious discrepancies with available experimental evidence (e.g., in the
equilibrium lattice parameter or in the bulk modulus \cite{cococcioni05})
that seriously question its applicability in these cases.

The ``around mean-field" (AMF) formulation of LDA+U was introduced
to alleviate these difficulties and to improve the description of
correlation in systems where electronic localization is less pronounced
or for which a metallic behavior is expected.
The AMF LDA+U was actually the first one to be introduced
\cite{anisimov91_2} and in its simplest formulation the 
energy functional can be expressed as follows:
\barr
\label{eamf}
&&E_{LDA+AMF} = E_{LDA} \nonumber \\
&&+ \frac{1}{2}\sum_{m m' \sigma} U 
(n_{m\sigma} - \langle n \rangle)(n_{m'-\sigma} - \langle n \rangle)
\nonumber \\
&&+ \frac{1}{2}\sum_{m m' \sigma}^* (U-J)
(n_{m\sigma} - \langle n \rangle)(n_{m'\sigma} - \langle n \rangle)
\earr
where the asterisk on the second summation indicates it runs over 
all the $m$ and $m'$ such that $m \neq m'$ and 
$\langle n \rangle = \frac{1}{2(2l+1)}\sum_{m\sigma} n_{m\sigma}$.
The idea that inspired this formulation is quite different from 
the one behind the FLL atomic limit \cite{czyzyk94}. While the latter
can be viewed as introducing a finite energy cost for occupations
of localized orbitals deviating from integer values,
the AMF corrective functional can be regarded as a penalty 
against deviations (fluctuations) of the occupations from their 
mean value. This latter approach corresponds to considering 
the approximate DFT total energy as containing
a mean field approximation of the electron-electron interaction.
This is easily seen from the identity \cite{czyzyk94}: 
$n_{m\uparrow}n_{m'\downarrow} = n_{m\uparrow}\langle n_{\downarrow}\rangle + 
\langle n_{\uparrow}\rangle n_{m'\downarrow} - \langle n_{\uparrow}\rangle
\langle n_{\downarrow}\rangle +
(n_{m\uparrow} - \langle n_{m\uparrow} \rangle)
(n_{m\downarrow} - \langle n_{m\downarrow} \rangle)$. It is evident 
that, if the LDA (or any approximate) xc functional contains 
the first three terms on the rhs of this expression (the mean-field
approximation of the quantity on the lhs), the AMF correction, proportional
to the last term on the rhs (Eq. (\ref{eamf})) is exactly what
is needed to recover the product of occupation matrices from its 
mean field value.

To better understand the differences between the FLL and the AMF 
formulations of LDA+U it is convenient
to follow Ref. \cite{czyzyk94} and to rewrite both 
functionals in the form:
\beq
\label{gencorr}
E_{LDA+U} = E_{LDA} + H_{Hub} - \langle H_{Hub} \rangle.
\eeq
In both flavors, $H_{Hub}$ contains electron-electron interactions as 
modeled in the Hubbard Hamiltonian:
\barr
\label{hhub}
H_{Hub} &=& \frac{1}{2} \sum_{mm'\sigma} U n_{m\sigma}n_{m'-\sigma}\nonumber \\
&+& \frac{1}{2} \sum_{mm'\sigma}^* (U-J) n_{m\sigma}n_{m'\sigma}
\earr
(rotational invariance is neglected here) where the asterisk has the 
same meaning as in Eq. (\ref{eamf}).
The difference between the FLL and the AMF formulations thus amounts to
a different dc term $\langle H_{Hub} \rangle$. In fact, one can recover
the two corrective functionals using, in Eq. (\ref{gencorr}), the following
two expressions for $\langle H_{Hub} \rangle$ \cite{czyzyk94}:
\barr
\label{dc2}
\langle H_{Hub} \rangle_{AMF}= UN_{\uparrow}N_{\downarrow}
+\frac{1}{2}(U-J)\frac{2l}{2l+1}(N_{\uparrow}^2 + N_{\downarrow}^2) \nonumber \\
\langle H_{Hub} \rangle_{FLL} = \frac{U}{2}N(N-1) - \frac{J}{2}\sum_{\sigma}
N_{\sigma}(N_{\sigma} - 1)
\earr
where $N_{\sigma} = \sum_m n_{m\sigma}$ and $N = N_{\uparrow}+N_{\downarrow}$.
A comparative analysis of these dc terms (in particular, their like-spin 
parts) highlights the different ``philosophy"
behind them: while in the AMF every electron interacts with all the 
electrons in the system (suggesting a spread charge density) and the 
self-interaction is eliminated through the rescaling (by a factor
$\frac{2l}{2l+1}$) of the effective interaction parameter, in the FLL
limit, due to a more pronounced localization, each electron interacts with
the other N-1.
An exhaustive discussion about the main ideas at the basis
of both the FLL and the AMF formulations, 
their theoretical framework within
density functional theory, and their specific characteristics,
has been also presented in Ref. \cite{anisimov07}.
A comparison between these two flavors was also offered in
Ref. \cite{eschrig2003}.

An attempt to obtain a general correction that bridges the AMF and
the FLL formulations and is able to treat a broad range of systems 
with intermediate degrees of electronic localization has been made 
in Ref. \cite{chioncel03}. This work proposes 
a linear combination of the AMF and the FLL flavors of LDA+U, 
based on a mixing parameter that has to be determined for each material 
and is a function of various quantities related to its electronic
structure. This approach has been used to study intermetallic \cite{chioncel03}
and selected rare earth compounds \cite{bultmark09} showing promising
results and a significant improvement with respect to either 
functional.
In spite of the desirable capability to improve the prediction
of properties related to electronic localization (such as, e.g., magnetization)
without compromising the description of delocalized electrons,
this approach, as well as the AMF itself, has had limited popularity
due, perhaps, to the diffusion of DFT+DMFT. In fact, this 
scheme offers a more rigorous treatment of dynamical effects
(particularly important for metallic system) and is able to capture, 
within the same theoretical framework, the physics
of systems and phases characterized by widely different levels of
electronic localization.
At the same time
the LDA+U method has been identified almost exclusively with its FLL limit
(because of a closer adherence to the
formulation of the Hubbard model)
and has been mostly used for systems with strongly localized
electrons where the main consequence of (static) electronic correlation
usually consists of the onset of an insulating ground state.

Is the FLL formulation only suitable for re-establishing the discontinuity
of the xc potential and inserting it in the Kohn-Sham
spectrum of a system? 
In view of its ability to increase the separation in energy between 
full and empty states, the FLL corrective functional could still be 
useful to selectively correct the energetics of localized states
(i.e., the position of the corresponding bands) 
while leaving more itinerant ones, or those in the vicinity of the
Fermi level uncorrected.
In bulk transition metals (generally cubic and often magnetic), 
for example, it was recognized that
the $e_g$ subgroup of the $d$ orbitals form bands
significantly more localized than the t$_{2g}$ ones~\cite{goodenough60}.
Consistently with this observation, 
the use of the FLL ``+U" functional to correct only the $e_g$ $d$ 
states of bulk Fe, has shown a significant improvement in the 
prediction of the equilibrium lattice parameter and of the bulk modulus of the
material with respect to LDA+U calculations with the Hubbard 
functional applied to all the $d$ states.
Similar results have also been obtained for a variety of metallic 
systems~\cite{leonov11,himmetoglu12_1}
with an analogous correction on localized $d$ states.

\begin{figure}[h!]
\includegraphics[width=0.47\textwidth]{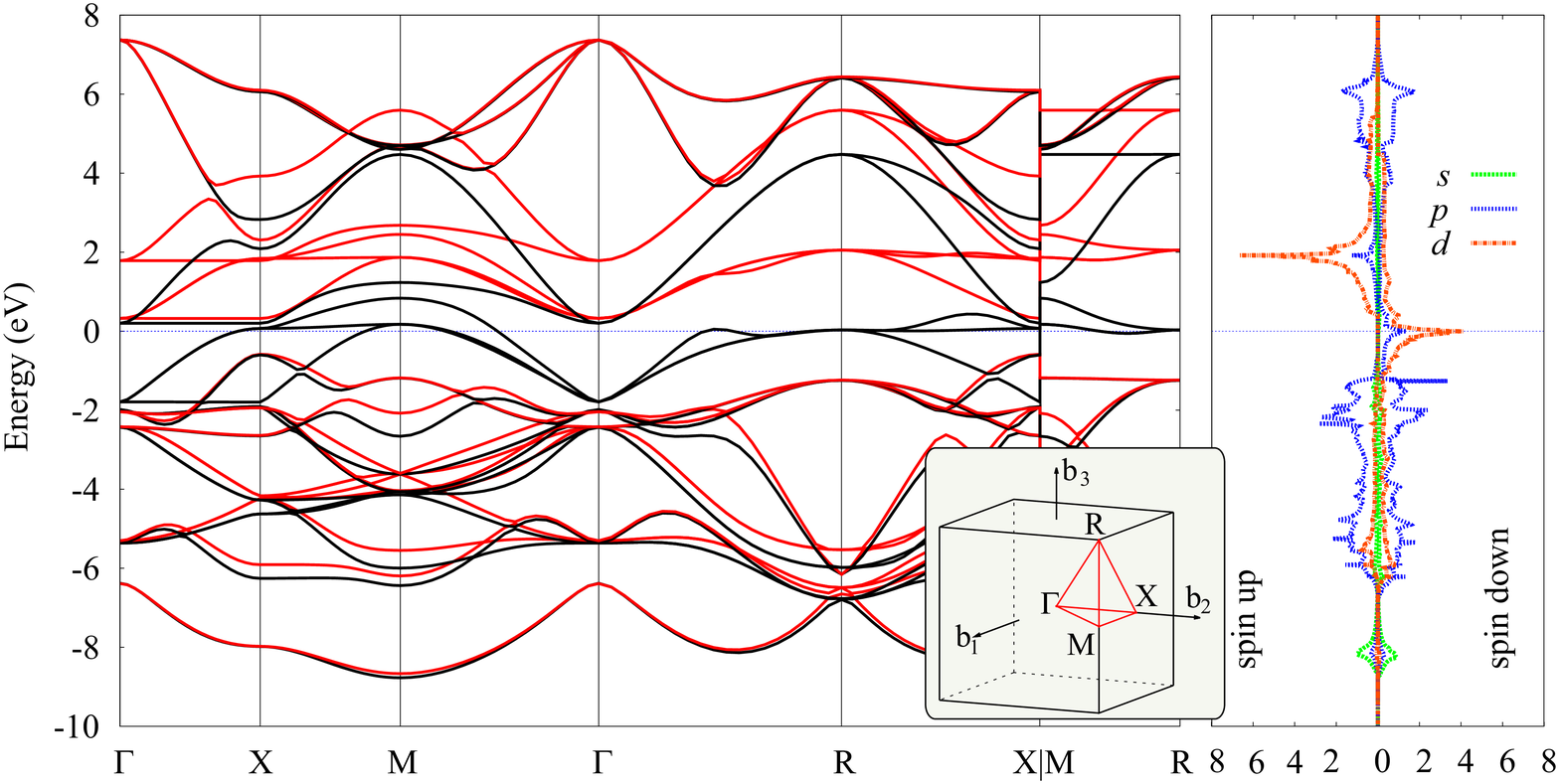}
\includegraphics[width=0.47\textwidth]{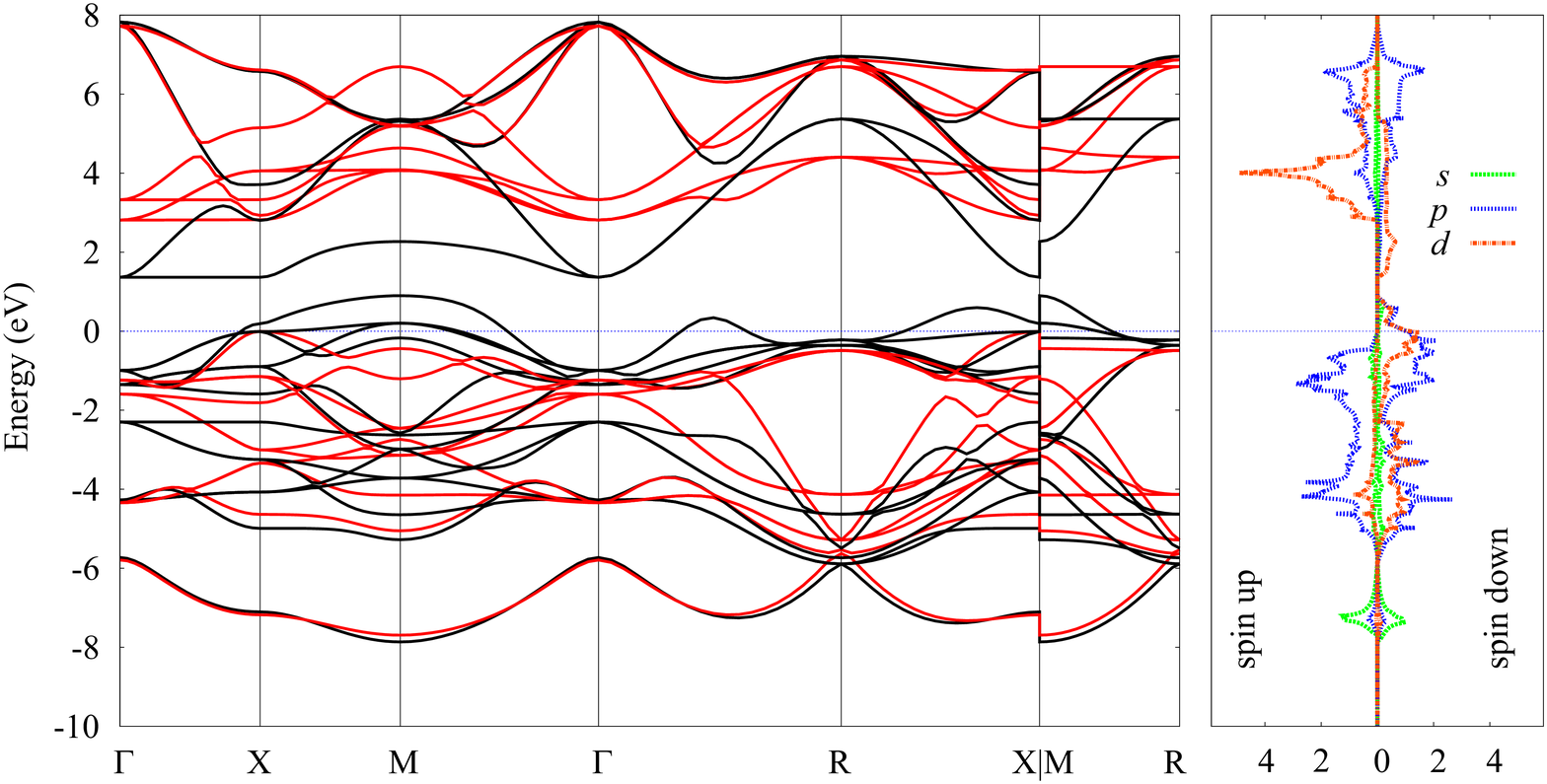}
\caption{(Obtained from \cite{kyang13}, color online) 
Electronic band structure (left) and density of states (right) of PbCrO$_3$. 
The high symmetry Brillouin zone path used for the electronic band structure 
calculation is shown in the inset. The upper panel shows PBE results, the 
lower ones, PBE+U. Red lines are for spin up states, black for spin down.
}
\label{pco1}
\end{figure}

As another example,
figure \ref{pco1} shows the band structure and the density of states
of PbCrO$_3$ and compares GGA (PBE) results (top panel) with GGA+U (bottom 
panel). The transitions this material undergoes between
antiferromagnetic (AFM), ferromagnetic (FM), and canted orders
have not yet been completely clarified
(see Refs. \cite{roth67,arevalo08,arevalo09} and references quoted therein). 
The calculations whose results are shown in Fig. \ref{pco1}
(performed within the AFLOW framework \cite{curta12,curta12_1,curta12_2})
assumed a ferromagnetic ground state. 
As evident from the figure, 
the FM order of spins yields a metallic band structure (half-metallic 
within GGA), even within LDA+U.
This is in contrast with the electrical resistance experiments 
on the material, suggesting (at least at teperatures above 200 K)
a semiconducting behavior \cite{arevalo09}. 
While opening a gap in the KS spectrum would probably require
splitting (through symmetry breaking) the strong peak of the
minority (down) spin $d$ states at the Fermi level in the GGA (PBE) DOS,
the Hubbard correction is still effective in shifting the 
energy of the $d$ states, increasing the overall separation between 
occupied and unoccupied levels of opposite spin. 
As a side effect, the Fermi level is pushed downward,
towards (and partly within) the occupied $p$ bands. 
The structural optimization of the crystal performed with LDA+U obtains
an equilibrium lattice parameter ($a=b=c=3.91$ \AA) in 
agreement with experimental measurements ($a=b=c=3.91$ \AA) 
and improving upon GGA results ($a=b=c=3.85$ \AA). 
While the accuracy of this result might be fortuitous, it corroborates the
idea that, due to the linearization of the energy with respect
to the occupation of localized states, the FLL 
LDA+U can actually improve the results of
approximate DFT functionals
even when the Hubbard correction does not significantly
modifies the energy of states in the immediate sorrounding
of the Fermi level. 
A first example of this use of LDA+U was provided in section \ref{feo},
discussing the rombohedral distortion of FeO under pressure.
%
A third case will be illustrated in the next section, focusing on the
intermetallic Heusler alloy Ni$_2$MnGa. 
This example will provide a more precise physical interpretation for the
shift in the single particle energies promoted by the Hubbard correction
and will illustrate its consequences on the strength of magnetic 
interactions and on the relative stability of different structural phases.

\subsection{Localization and magnetism in Ni$_2$MnGa}

Ni$_2$MnGa is one of the prototype representative of magnetic Heusler alloys.
Materials in this class are often characterized by martensitic transitions
occurring near room temperature associated with shape-memory effects. 
The particular appeal of systems in this family exhibiting a ferromagnetic
ground state consists in the possibility to couple structural
(martensitic) transitions with magnetic ones (e.g., magnetic ordering,
demagnetization, abrupt variations in magnetocrystalline anisotropy, etc)
that could lead to the development of applications of technological
interest (such as, e.g., actuators, sensors, energy conversion devices, etc)
\cite{srivastava11,magmech-1,magmech-2,magmech-3,magmech-4,sozinov02_1}. 
The design of alloys for which the martensitic and magnetic transitions 
are optimally coupled and occur at the same critical 
temperature has been pursued so far by varying the composition of these alloy
in a largely empirical way. The precise knowledge of the electronic mechanisms
controlling both types of transitions could thus greatly facilitate the 
search for a material with optimal coupling between these transitions
and a high degree of reversibility.

\begin{figure}[h!]
\includegraphics[width=0.4\textwidth]{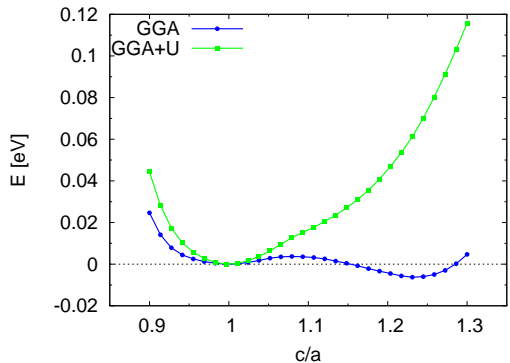}
\caption{(Adapted from Ref. \cite{himmetoglu12_1}; color online) 
The ground state energy as a function of $c/a$
for constant volume deformations in GGA and GGA+U.}
\label{figni2mnga}
\end{figure}

The stoichiometric Ni$_2$MnGa compound we focused on in a recent work 
\cite{himmetoglu12_1} has a cubic (FCC) austenite phase and is reported
to transform (at a temperature of about $200\, {\rm K}$) 
into a tetragonal martensite, characterized by a structural
modulation along the [110] direction
\cite{kokorin-modul,martynov-modulated,kaufmann-modulated,dai-eoa}. 
Since the Curie magnetic ordering temperature is $350\, {\rm K}$
\cite{brown-modulated}, both austenite 
and martensite phases are ferromagnetic across the structural transition.
Although some controversy still exists in literature, DFT calculations,
performed with GGA exchange correlation functionals, generally predict
the minimum of the energy in correspondance of a non-modulated
tetragonal structure that has only been observed for non-stoichiometric
alloys.
This result is illustrated by the blue line in 
Fig.~\ref{figni2mnga} 
where the energy profile as a function of the tetragonal distortion
$c/a$ is shown ($c/a=1$ corresponding to the cubic austenite).
As evident from the figure, the GGA functional achieves the minimum of 
the energy for an elongated tetragonal cell ($c/a\approx 1.1$) while experiments
\cite{kokorin-modul,martynov-modulated,dai-eoa,kaufmann-modulated} 
report a martensitic phase consisting of a modulated
tetragonal structure with $c/a \approx 0.97$. 
Since this specific value of $c/a$ is associated with
the modulation of the structure, 
the cubic austenite phase should be obtained as the most stable one,
when the modulation is neglected (as in the study presented here).
Instead, non modulated tetragonal phases, with $c/a \geq 1$, are 
reported in experiments for off-stoichiometric compounds 
characterized, e.g., by excess Mn \cite{kokorin-modul,sozinov02}. 

\begin{figure}[h!]
\includegraphics[width=0.4\textwidth]{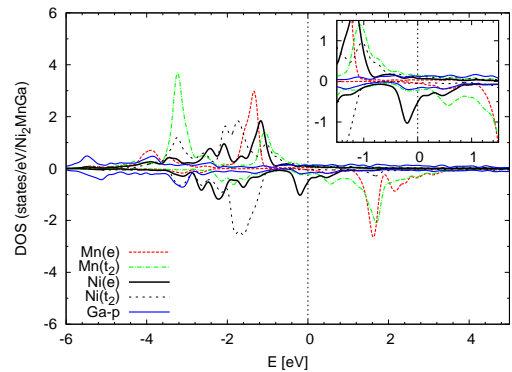}
\caption{(Adapted from Ref. \cite{himmetoglu12_1}, color online) 
Density of states of the 
austenite phase calculated using GGA. The relevant symmetry groups of d-orbitals
are explicitly indicated.}
\label{figni2mnga-dos}
\end{figure}

From the density of states of the austenite phase (the one for the martensite
would be similar), shown in Fig. \ref{figni2mnga-dos}, 
it is possible to observe that the states around the Fermi level (and thus 
responsible for the metallic character of the material) are mostly Ni
$d$ states, while Mn $d$ states are largely responsible for the
magnetization. In fact, the magnetic moments of Mn, Ni and Ga atoms are, 
respectively, $3.67\, \mu_B, \, 0.34\, \mu_B$ and $-0.13\, \mu_B$.
The discussion at the end of the previous section should clarify
the reason why in Ref. \cite{himmetoglu12_1} the computational study
of this material was performed with the FLL LDA+U corrective
functional, selectively applied to the $d$ states of Mn. 
This strategy results in a more accurate description of the material,
with larger magnetic moments on Mn atoms (see Table~\ref{tabmag}) and the austenite phase more
stable than the non-modulated martensite one.
%
\begin{table}[b]
\caption{\label{tabmag}
Calculated lattice parameter and magnetizations of each atom in the cubic (austenite) 
phase of Ni$_2$MnGa.}
\begin{ruledtabular}
\begin{tabular}{cccccc}
 & $a_0\,$ (\AA)  & $\mu_{\rm Mn}\, (\mu_B)$ & $\mu_{\rm Ni}\, (\mu_B)$ & $\mu_{\rm Ga}\, (\mu_B)$ &
$\mu_{\rm tot}\, (\mu_B/{\rm cell})$   \\
\hline
GGA & $5.83$ & $3.67$ & $0.34$ & $-0.13$ & $4.22$ \\
GGA+U & $5.83$\footnotemark[1] & $4.52$ & $0.16$ & $-0.13$ & $4.80$ \\
\end{tabular}
\end{ruledtabular}
\footnotetext[1]{Kept at the GGA calculated value.}
\end{table}
%
Fig. \ref{figni2mnga}  
compares the energy vs $c/a$ profiles obtained with GGA
and GGA+U and highlights how the Hubbard correction 
eliminates the spurious minimum at $c/a > 1$, predicting the cubic
(austenite) phase more stable than any non-modulated tetragonal structure.
In Ref. \cite{himmetoglu12_1} it is also shown that most
of the variation in the total
energy from the austenite to the martensite phase can be accounted for 
by a Heisenberg model of the total spin on Mn ions. 
In particular, computing (from constrained DFT calculations)
the interatomic magnetic couplings $J$ 
in dependence of the tetragonal elongation
and subtracting the corresponding 
Heisenberg term from the total energy, the
minimum corresponding to the (non-modulated) martensite phase is 
eliminated. 
This result, due to the significant renormalization of 
the magnetic couplings induced by the Hubbard correction, can be rationalized 
by describing the localized $d$ electrons on Mn atoms as Anderson
impurities interacting with a ``bath" of more delocalized Ni $d$ and 
Ga $p$ electrons. In fact, the Anderson 
model results in RKKY effective interactions between Mn magnetic
moments that have a clear dependence on the Hubbard $U$ (the parameter 
controlling
the energy splitting between full and empty localized Hubbard bands). 
The Hubbard $U$ (as imposed by the Hubbard correction) suppresses magnetic
interactions and reduces the energetic advantage associated with the 
deformation of the cubic austenite phase into a non modulated tetragonal
martensite. 
As shown in Ref. \cite{himmetoglu12_1}, the FLL LDA+U on Mn $d$
states (with the effective $U$ calculated from linear-response theory
as described in section \ref{linearr}) is also able to capture 
the stabilization of a non-modulated tetragonal phase in compounds with 
excess Mn with respect to the stoichiometric composition.

This example shows that the use of the FLL LDA+U on the localized
states of metals is not only possible, but actually very useful to capture, 
in certain cases, the effects of electronic localization, for example,
on the magnetic and the structural properties of a material. In these
instances, where 
a larger energy separation between Hubbard bands of localized states
is needed to improve the description of the material, the LDA+U 
probably represents a more convenient choice than more sophisticated
approaches (as DMFT) that are significantly more computationally 
demanding. 
On the other hand, the identification of the localized subset of states to be
corrected by the LDA+U functional might not be as trivial as in the case
described above, especially if the symmetry of the system is low.
In these cases a scheme to automatically select localized
states or able to treat localized and delocalized orbitals with equal
accuracy would be highly desirable.

\section{Extended functionals}
\label{extf}
\subsection{The LDA+U+V approach: when covalency is important}
\label{upv}
In this section we will
briefly discuss one of the latest extensions 
to the LDA+U functional: the LDA+U+V \cite{campo10}.
This modification is shaped on the ``extended" Hubbard model and 
includes both on-site and inter-site electronic interactions.
The extended formulation of the Hubbard Hamiltonian 
has been considered since the early days of this model 
\cite{hubbardIV,hubbardV} and can be expressed as follows:
\beq
H_{Hub} = t\sum_{\langle i,j \rangle,\sigma}(c^{\dagger}_{i,\sigma}
c_{j,\sigma} + h.c.) + U \sum_i n_{i,\uparrow}n_{i,\downarrow} 
+ V \sum_{\langle i,j \rangle}n_{i}n_{j}
\label{hubvm}
\eeq
where $V$ is the effective interaction between
electrons on neighbor atomic sites. 

The interest on the extended Hubbard model has been revamped in the 
last decades by the discovery of high T$_c$ superconductors
and the intense research activity focusing around them.
Whether the inter-site coupling $V$ has a determinant role
in inducing superconductivity is, however, still matter of debate
\cite{andersonsci87,andersonprl87,imada91,hirsch88,thakur07,jursa96,szabo96,demello99}.
Several studies have also demonstrated that the relative strength
of $U$ and $V$ controls many properties of the ground state of
correlated materials as, for example, the occurrence of possible phase separations~\cite{mancini09},
the magnetic order~\cite{morohoshi08,watanabe08},
the onset of charge-density and spin-density-wave regimes~\cite{vandogen94}.
In Refs.~\cite{anisimov97,anisimov96} the inter-site coupling
(between $d$ states) was recognized to be important to determine a
charge-ordered ground state in mixed-valence
systems, while in Ref.~\cite{verdozzi95} the extended Hubbard 
Hamiltonian was used 
to refine the Auger core-valence-valence line shapes of solids.
More recently, the extended Hubbard model has been used
to study the conduction and the structural properties of polymers and carbon
nano-structures and it was shown that 
the interplay between $U$ and $V$ 
controls the dimerization of graphene nanoribbons~\cite{zhu06}.

Our motivation to include inter-site interactions in the formulation of the
corrective Hubbard Hamiltonian was the attempt to define a more flexible and
general computational scheme able to precisely
account for (rather than just suppress)
the possible hybridization of atomic states on different atoms.
\index{hybridization}
In order to understand the implementation of the LDA+U+V \cite{campo10}
it is useful to start from a (mean-field-factorized)
second-quantization expression of the 
electronic interaction energy with a full set of 
site- and orbital- dependent interactions: 
\barr
\label{secq}
E_{int} &=& \frac{1}{2}\sum_{I,J,K,L}\sum_{i,j,k,l}\sum_{\sigma}
\:\langle \phi^{I}_{i}\phi^{J}_{j}|V_{ee}|\phi^{K}_{k}\phi^{L}_{l}\rangle\:
\times \nonumber \\
&&\left(n^{KI\sigma}_{ki}n^{LJ\sigma'}_{lj} - \delta_{\sigma\sigma'} n^{KJ\sigma}_{kj}n^{LI\sigma'}_{li}\right)
\earr
where the numbers $n^{KI\sigma}_{ki}$ represent the average values
of products of creation and annihilation fermion operators 
($\langle {c^{I\sigma}_i}^{\dagger}c^{K\sigma}_k\rangle$), to be
associated to generalized occupations, defined as:
\beq
\label{occupij}
n^{IJ\sigma}_{mm'}=\sum_{k,v}f_{kv}^{\sigma}\langle \psi_{kv}^{\sigma} |
\phi_{m'}^{J}
\rangle\langle \phi_{m}^{I} | \psi_{kv}^{\sigma}\rangle.
\eeq
In Eq.~(\ref{occupij})
the indices $m$ and $m'$ run over the angular momentum manifolds
that are subjected to the Hubbard
correction on atoms $I$ and $J$ respectively.
It is important to notice that the
occupation matrix defined in Eq.~(\ref{occupij})
contains information about all the atoms
in the same unit cell and the on-site occupations defined
in Eq.~(\ref{occup}) correspond to its diagonal blocks
(${\rm \bf n}^{I\sigma}={\rm \bf n}^{II\sigma}$).
Generalizing the approach described for the on-site case,
the $E_{Hub}$ term of the DFT+U+V can be obtained from Eq. (\ref{secq})
retaining only those terms that contain 
the interaction between orbitals belonging to 
couples of neighbor atomic sites:
$\langle \phi^{I}_{i}\phi^{J}_{j}|V_{ee}|
\phi^{K}_{k}\phi^{L}_{l}\rangle \to \delta_{IK}\delta_{JL}
\delta_{ik}\delta_{jl} V^{IJ} + \delta_{IL}\delta_{JK}\delta_{il}
\delta_{jk} K^{IJ}$.
Similarly to the on-site case,
the effective inter-site interactions
are assumed to be all equal to their atomic averages over the states of
the two atoms:
$\langle \phi^{I}_{i}\phi^{J}_{j}|V_{ee}|
\phi^{K}_{k}\phi^{L}_{l}\rangle \to \delta_{IK}\delta_{JL}
\delta_{ik}\delta_{jl}
V^{IJ} = \frac{\delta_{IK}\delta_{JL}
\delta_{ik}\delta_{jl}}{(2l_I+1)(2l_J+1)}\sum_{i',j'}
\langle \phi^{I}_{i'}\phi^{J}_{j'}|V_{ee}|
\phi^{I}_{i'}\phi^{J}_{j'}\rangle$. 
Within this hypothesis, and assuming that the inter-site exchange couplings
$K^{IJ}$ can be neglected, it is easy to derive the following expression 
($U^{I} = V^{II}$):
\barr
\label{hubv}
E_{Hub} &=& \sum_I \frac{U^I}{2} \left[\: (n^I)^2 -
\sum_\sigma Tr\left[({\rm \bf n}^{II\sigma})^2\right] \:\right] \nonumber \\
&+&\sum_{IJ}^{\hspace{5mm}\star} \frac{V^{IJ}}{2} \left[n^{I}n^{J} -
\sum_\sigma Tr({\rm \bf n}^{IJ\sigma} {\rm \bf n}^{JI\sigma})\right]
\earr
where the star in the second summation operator reminds that for each 
atom $I$, index $J$ covers all its neighbors up to a given shell. 
Generalizing the FLL expression of the on-site double-counting term
to include inter-site interactions, we arrive at the following expression:
\beq
\label{dcv}
E_{dc} = \sum_I \frac{U^I}{2}n^I(n^I-1) +
\sum_{I,J}^{\hspace{5mm}\star} \frac{V^{IJ}}{2} n^{I}n^{J}.
\eeq
Subtracting Eq. (\ref{dcv}) from Eq. (\ref{hubv}) it is easy to obtain:
\barr
\label{UV}
E_{UV}&=&E_{Hub} - E_{dc} = \sum_{I,\sigma} \frac{U^I}{2}Tr\left[{\rm \bf n}^{II\sigma}
\left({\bf 1}-{\rm \bf n}^{II\sigma}\right)\right] \nonumber \\
&-& \sum_{I,J,\sigma}^{\hspace{5mm}*}\frac{V^{IJ}}{2}
Tr\left[{\rm \bf n}^{IJ\sigma}{\rm \bf n}^{JI\sigma}\right].
\earr
To better understand the role of the inter-site part of the energy
functional it is convenient to derive the 
correction it contributes to the (generalized) KS potential: 
\barr
\label{VUV}
V_{UV}  
&=&\sum_{I,m,m'} \frac{U^I}{2}\left(\delta_{mm'} -
2n^{II\sigma}_{mm'}\right)\vert\phi^{I}_{m}\rangle\langle\phi^{I}_{m'}\vert \nonumber \\
&-& \sum_{I,J,m,m'}^{\hspace{5mm}*} V^{IJ}
n^{JI\sigma}_{mm'}\vert\phi^{I}_{m}\rangle\langle\phi^{J}_{m'}\vert.
\earr
From Eq.~(\ref{VUV}) it is evident that while the {\it on-site}
term of the potential is attractive for occupied states that are, at most,
linear combinations of atomic orbitals of the {\it same} atom
(resulting in on-site blocks of the occupation matrix,
${\rm \bf n}^{II\sigma}$, numerically dominant on others), 
the {\it inter-site} interaction stabilizes states that are linear
combinations of atomic orbitals belonging to {\it distinct} (neighbor) atoms
(e.g., molecular orbitals), that lead to large inter-site blocks
${\rm \bf n}^{JI\sigma}$ of the occupation matrix.
Thus, the two interactions give rise to competing tendencies, and
the character of the resulting ground state depends on 
the balance between them. Fortunately, the linear-response
calculation of the effective interactions, discussed in section
\ref{computeu} and in Ref. \cite{cococcioni05}, offers the  
possibility to compute both parameters simultaneously (and with no
additional cost with respect the on-site case).
In fact, the inter-site couplings $V^{IJ}$ correspond to the
off-diagonal terms of the interaction matrix defined in Eq. (\ref{ucalc}).

It is important to stress again that the trace operator in the on-site
functional guarantees the invariance of the energy only with respect
to rotations of atomic orbitals {\it on the same atomic site}.
In fact, the on-site corrective functional (Eq. (\ref{our1}))
is not invariant for unitary transformations of the atomic orbital
basis set that mix states from different atoms.
With the ``+U+V" corrective functional the invariance with respect to 
general rotations within the atomic basis set would be exactly recovered
in the limit situation with the same $U$ on all the atoms and equal to all
the $V$ between them ($U = V$). In fact, in this case, 
the sum of on-site and inter-site interactions in Eq. \ref{UV}
would be proportional to the trace of the square of the generalized 
occupation matrix in Eq. \ref{occupij}.
In the most general case, the differentiation between the 
interaction parameters would require full orbital dependence for the
corrective functional to be invariant.
Atomic center and angular momentum dependence of the corrective 
functional are implicitly
included in Wannier-function-based implementations of the LDA+U
\cite{oregan10,oregan12,fabris05,fabris05_1,mazurenko07,lechermann06,miyake08}.
In fact, even starting from an on-site only formulation, 
re-expressing Wannier functions on the basis of atomic wave functions
produces a variety of multi-center/multi-orbital interaction terms. 
The two approaches would thus
lead to equivalent results if all the relevant multiple-center
interactions parameters are included in the corrective functionals
and are computed consistently with the choice of the orbital basis.
While the use of Wannier-functions allows to minimize
the number of relevant electronic interactions to be
computed (especially
if maximally-localized orbitals are used \cite{marzari97}),
the atomic orbital representation provides a more intuitive
and transparent scheme to select relevant interactions terms (e.g.,
based on inter-atomic distances), and is more convenient to compute
derivatives of the energy as, for example, forces and stresses.

In the implementation of Eq.~(\ref{UV}) we have added
the possibility for the corrective functional to act
on two $l$ manifolds per atom as, for example,
the $3s$ and $3p$ orbitals of Si, or the $4s$ and $3d$ orbitals of Ni.
The motivation for this extension consists in the fact that
different manifolds of atomic states may require
to be treated on the same theoretical ground in cases where hybridization
is relevant (as, e.g., for bulk Si whose bonding structure
is based on the $sp^3$ mixing of $s$ and $p$ orbitals).

The new LDA+U+V functional
was first employed to study the electronic and structural
properties of NiO, Si and GaAs \cite{campo10}, prototypical representatives
of Mott or charge-transfer (NiO) and band insulators (Si and GaAs). 
The choice of these systems was made
to test the ability of the new functional to bridge the description of
the two kinds of insulators.
In fact, as discussed in previous sections (see Eq. (\ref{fgap1})),
the fundamental gap of a system is the sum
of the KS gap and the discontinuity in the xc potential (usually missing
in most approximate local or semi-local xc functionals) \cite{grosslibro}.
Since the ``+U" correction was designed
to reintroduce the discontinuity into the xc potential, 
LDA+U should be equally effective in improving the prediction
of the fundamental gap (from the KS spectrum) for both types of materials,
and it can be expected to improve the prediction of other properties too.

As other transition-metal oxides, NiO has a cubic rocksalt
structure with a rhombohedral symmetry brought about by its AF II ground
state. Because of the balance between crystal field and exchange
splittings of the $d$ states of Ni, nominally occupied by 8 electrons,
the material has a finite KS gap with oxygen $p$ states occupying
the top of the valence band. This gap, however, severely underestimates 
the one obtained from photoemission experiments (of about 4.3 eV 
\cite{sawatzky84}).
LDA+U has been used quite succesfully on this material (the spread
of results in literature is mostly due to the different values of $U$ used)
providing a band gap between 3.0 and 3.5 eV, and quite
accurate estimates for both magnetic moments and equilibrium
lattice parameter~\cite{dudarevnio00,bengone00,kressenio04}.
DFT+U has also been employed to compute the $k$-edge
XAS spectrum of NiO
using a parameter-free computational approach~\cite{maurinio09}
that has produced results consistent with experimental data.
The use of GW from the LDA+U ground state has provided a better estimate
of the energy gap compared to LDA+U, even though
other details of the density of states were almost unchanged
~\cite{kobayashi08}.

\begin{figure}[h!]
\vspace{-5mm}
\includegraphics[width=0.34\textwidth,angle=-90]{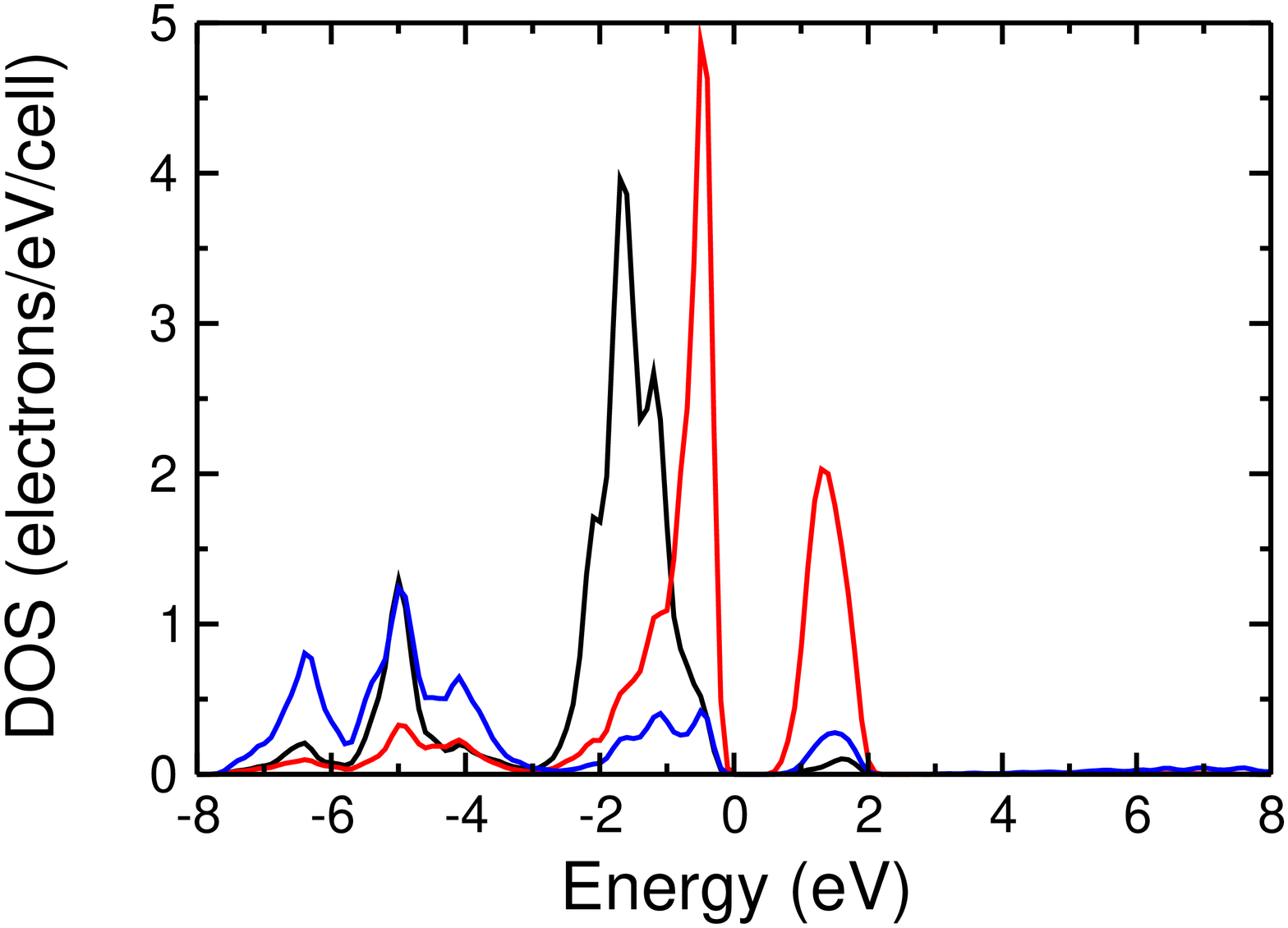}
\hspace{0mm}
\includegraphics[width=0.34\textwidth,angle=-90]{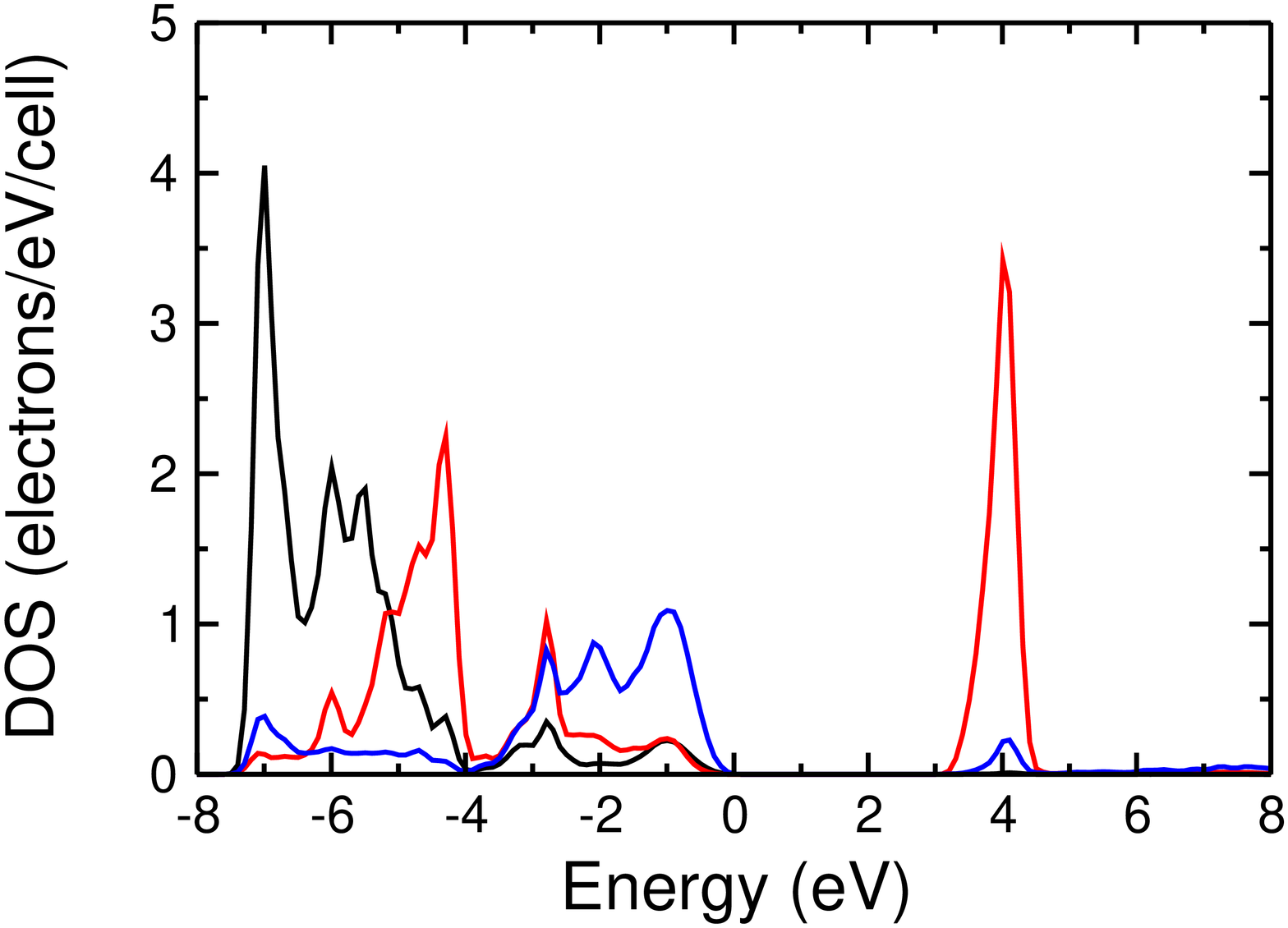}
\hspace{0mm}
\includegraphics[width=0.34\textwidth,angle=-90]{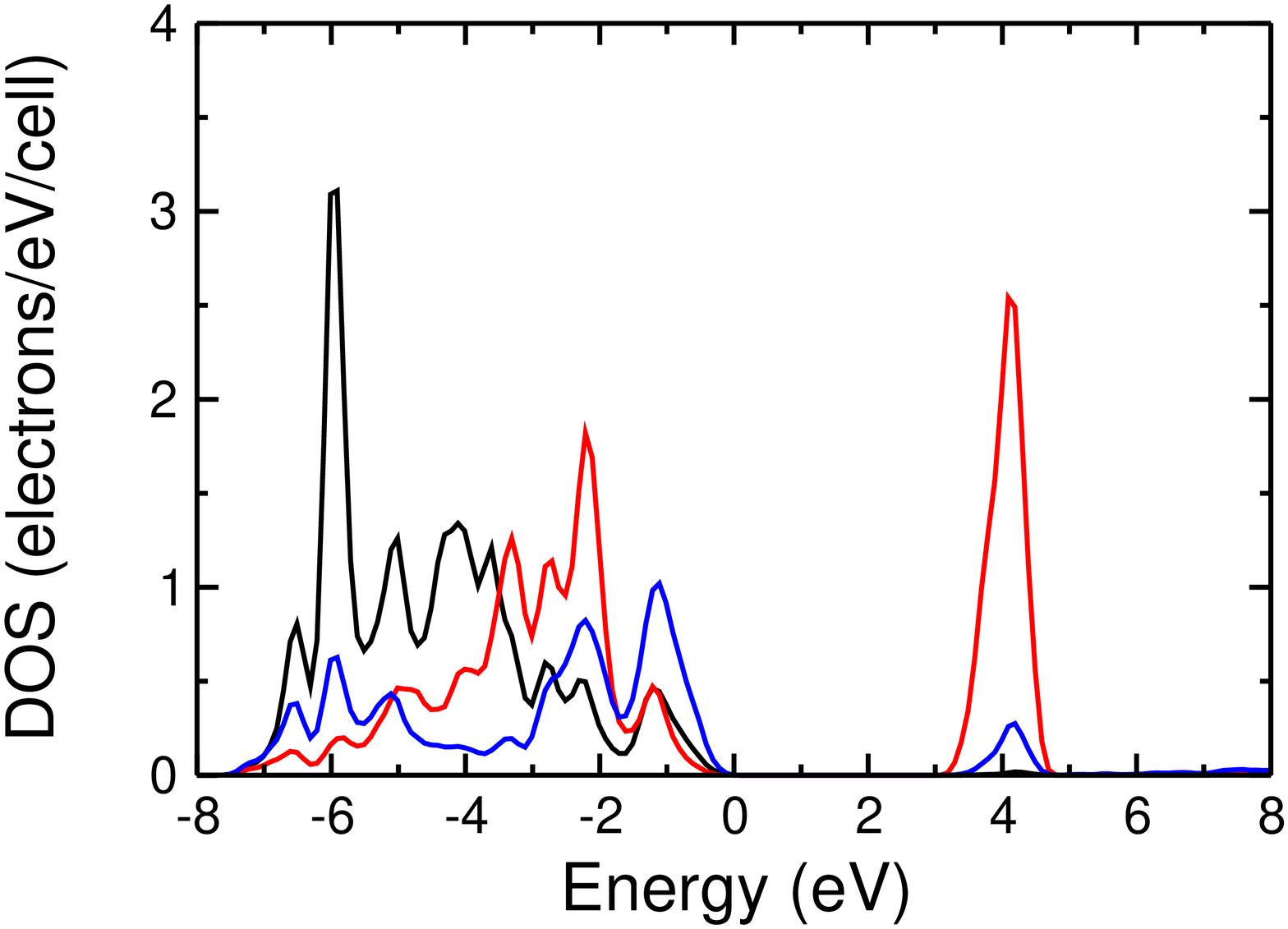}
\caption{\label{niodos} (Reprint of Fig. 2 of Ref. \cite{campo10}; color 
online) 
The density of states of NiO obtained
with different approximations: GGA (top); GGA+U (center);
GGA+U+V (bottom).
spin $d$ states of Ni, the blue line the $p$ states of O.
The energies were shifted for the top of the valence band to
correspond to the zero of the energy in all cases. The black line represents
majority spin $d$ states, the red line minority $d$ states, the blue line
oxygen $p$ states.}
\vspace{-3mm}
\end{figure}
Besides the on-site $U_{Ni}$, the LDA+U+V calculations 
also included the
interactions between nearest neighbor Ni and O ($V_{Ni-O}$) and between 
second nearest neighbor Ni atoms ($V_{Ni-Ni}$). The corrective functional
included interactions between the $d$ states of Ni, between Ni $d$ and O 
$p$ states and between $d$ and $s$ states of the Ni atoms. 
Other interactions were found to have a negligible effect
on the results and were neglected. The numerical values of the interaction 
parameters, all determined through the linear-response approach discussed above,
can be found in Ref. \cite{campo10}.
Fig. \ref{niodos} compares the density of states (DOS) of NiO as obtained from
GGA, GGA+U and GGA+U+V calculations. It is easy to observe that the 
GGA+U+V obtains a band gap of the same width as GGA+U, also maintaining 
the charge-transfer character of the material with O $p$ states at the
top of the valence band, as observed in photoemission experiments.
On the contrary, the GGA band gap is too small compared
with experiments and also has Ni $d$ states at the top of the
valence band.
As expected,
the inter-site interactions between Ni and O electrons also results
in a more pronounced overlap in energy between $d$ and $p$ states.
\begin{table}
\vspace{-4mm}
\caption{\label{niostruct}
The equilibrium lattice parameter, ($a$, in Bohr atomic radii),
the bulk modulus ($B$, in GPa),
and the band gap ($E_g$, in eV) of NiO obtained with different
computational approaches: GGA, ``traditional" GGA+U (with U only
on the $d$ states of Ni) and GGA+U+V 
with the interaction parameters computed ``self-consistently"
from the GGA+U+V ground state (see text). 
Comparison is made with experimental
results on all the computed quantities.}
\begin{center}
\begin{tabular}{|c|c|c|c|}
\hline
\hline
 & $a$ & $B$ & $E_g$  \\
\hline
GGA & 7.93 & 188 & 0.6  \\
\hline
GGA+U & 8.069 & 181 & 3.2   \\
\hline
GGA+U+V & 7.99 & 197 & 3.2  \\
\hline
Exp & 7.89 & 166-208 & 3.1-4.3  \\
\hline
\hline
\end{tabular}
\vspace{-8mm}
\end{center}
\end{table}
In table \ref{niostruct} a comparison is made between experiments and
calculations on the equilibrium lattice parameter, bulk modulus and
energy gap. It can be observed that while GGA provides the best estimate
of the experimental lattice parameter, GGA+U+V improves on the
result of GGA+U for the lattice parameter and the bulk modulus is also
corrected towards the experimental value.
Therefore, accounting for inter-site interactions not only is not
detrimental for the quality of 
the LDA+U description of the ground state of correlated materials 
but also has the potential to improve problematic aspects 
(e.g., structural properties)
counter-balancing the effects of eccessive electronic localization.

The application to Si and GaAs is, in some sense, the ``proof of fire" 
\index{correlation in band semiconductors}
for the LDA+U+V approach, as the insulating character of these materials
is due to the hybridization of $s$ and $p$ orbitals
from neighbor atoms which leads to a finite splitting between the
energy of fully occupied 
(bonding) and empty (anti-bonding) states. The eccessive stabilization
of atomic orbitals induced by the on-site $U$ suppresses the overlap
with neighbor atoms and tends to reduce the gap between valence and 
conduction states \cite{campo10}.
While providing a quite good description of the ground state properties
of these materials, the Kohn-Sham gap obtained from
LDA and GGA functionals is significantly smaller than the experimental band gap.
Although this is an expected result, corrective methods able to 
enforce the discontinuity to the xc potential and to improve the
size of the fundamental gap, are also beneficial for predicting 
other properties, and the same can be expected from using 
Hubbard corrections.
A more accurate estimate of the band gap of these materials
has been obtained
using SIC and hybrid functionals \cite{stadele99,nguyen09,heyd05} or with
the GW approach based on an LDA \cite{rohlfing93,aulbur99} or a
EXX \cite{aulbur00} ground state.

%
%
As mentioned above, for the LDA+U+V method to work on these systems and to 
capture the $sp^3$ hybridization their bonding structure is based on,
the Hubbard correction has to be applied to both $s$ and $p$ states and to
include a full spectrum of on-site ($U_{pp}$, $U_{ss}$, $U_{sp}$)
and inter-site interactions ($V_{pp}$, $V_{ss}$, $V_{sp}$).
The linear-response approach to calculate the effective Hubbard $U$,
described in section \ref{computeu}, allowed to reliably compute
all these intereaction parameters and to capture their dependence on
the volume of the crystal.
In Table \ref{table_comp}, the equilibrium lattice parameter, the
bulk modulus and the energy band gap obtained from GGA, GGA+U and GGA+U+V
calculations on Si and GaAs can be directly compared with the
results of experimental measurements (we refer to the data collected
in the web-database, Ref.~\cite{semicondbase}).
\begin{table}[h!]
\vspace{-2mm}
\caption{\label{table_comp}Comparative results for lattice
parameter ($a$, in ${\rm \AA}$), bulk modulus (B, in GPa)
and energy gap ($E_g$, in eV).}
\begin{center}
\begin{tabular}{ccc}
\hline
\hline
& Si & GaAs \\
\hline
& $a$\hspace{4.9mm} B\hspace{4.9mm} $E_g$ & $a$\hspace{4.9mm} B\hspace{4.9mm} $E
_g$ \\
\hline
GGA & 5.479,~83.0,~0.64 & 5.774,~58.4,~0.19 \\
GGA+U & 5.363,~93.9,~0.39 & 5.736,~52.6,~0.00  \\
GGA+U+V & 5.370,~102.5,~1.36& 5.654,~67.7,~0.90 \\
Exp. & 5.431,~98.0,~1.12 & 5.653,~75.3,~1.42  \\
\hline
\end{tabular}
\vspace{-5mm}
\end{center}
\end{table}
As it can be observed from this table,
the (on-site only) GGA+U predicts the equilibrium lattice
parameter in better agreement with the
experimental value than GGA for GaAs while it overcorrects GGA for Si;
however, the bulk modulus is improved by GGA+U with respect to
the GGA value only in the case of Si.
Due to the suppression of the interatomic hybridization, in both
cases, the energy band gap is lowered compared to GGA,
further worsening the agreement with experiments.
The use of the inter-site correction 
results in a systematic improvement for the evaluation of all these quantities.
In fact, encouraging the occupations of hybrid states, the inter-site
interactions not only enlarge
the splittings between full and empty levels (which increases
the size of the band gap), but also make bonds shorter (so that
hybridization is enhanced) and stronger, thus tuning both the equilibrium
lattice parameter and the bulk modulus of these materials to
values closer to the experimental data.
Calculations on GaAs explicitly included Ga $3d$ states in the valence
manifold as suggested by Ref. \cite{schultz08}. 
\begin{figure}[h!]
\vspace{-4mm}
\hspace{0mm}
\includegraphics[width=0.34\textwidth,angle=-90]{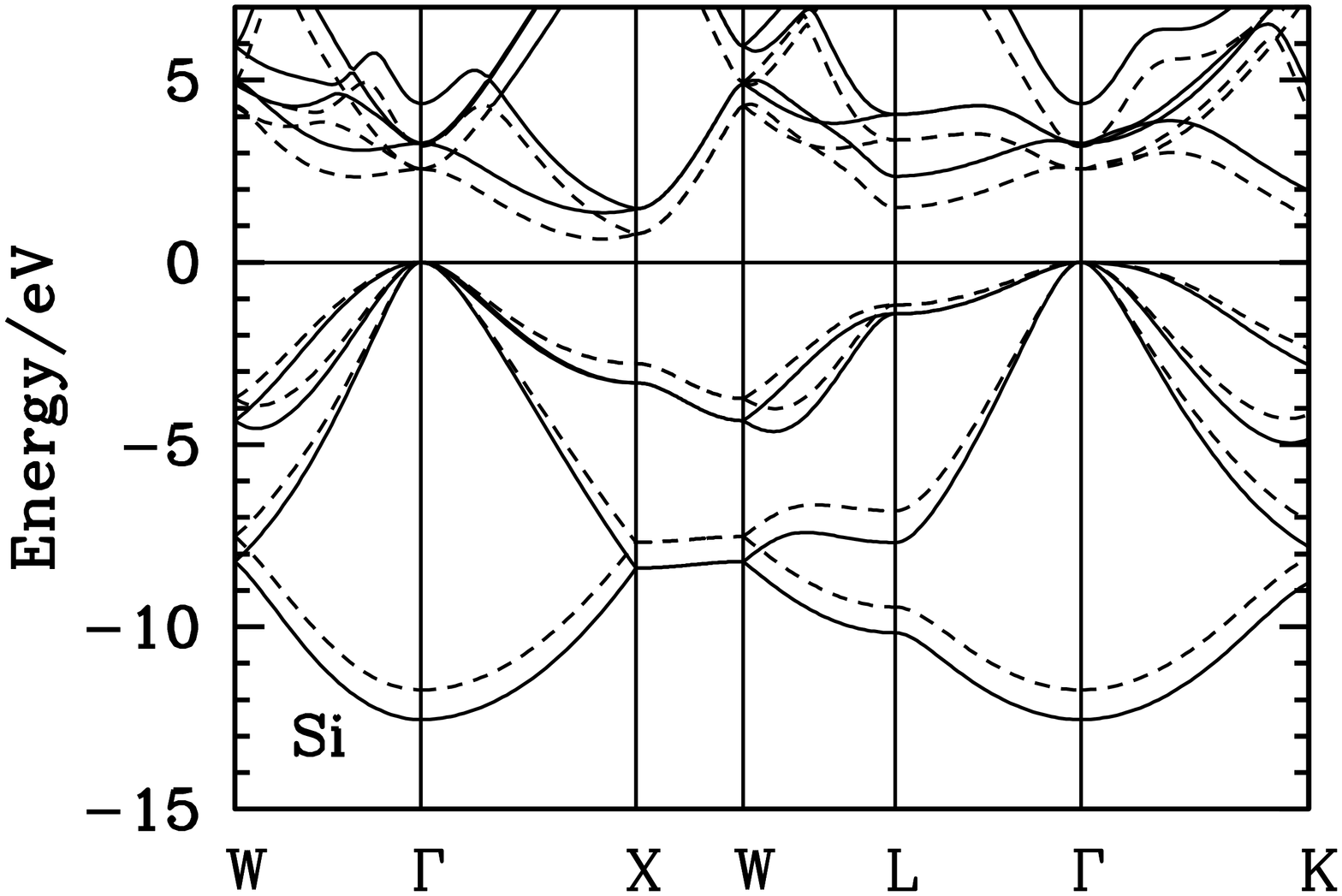}
\hspace{0mm}
\includegraphics[width=0.34\textwidth,angle=-90]{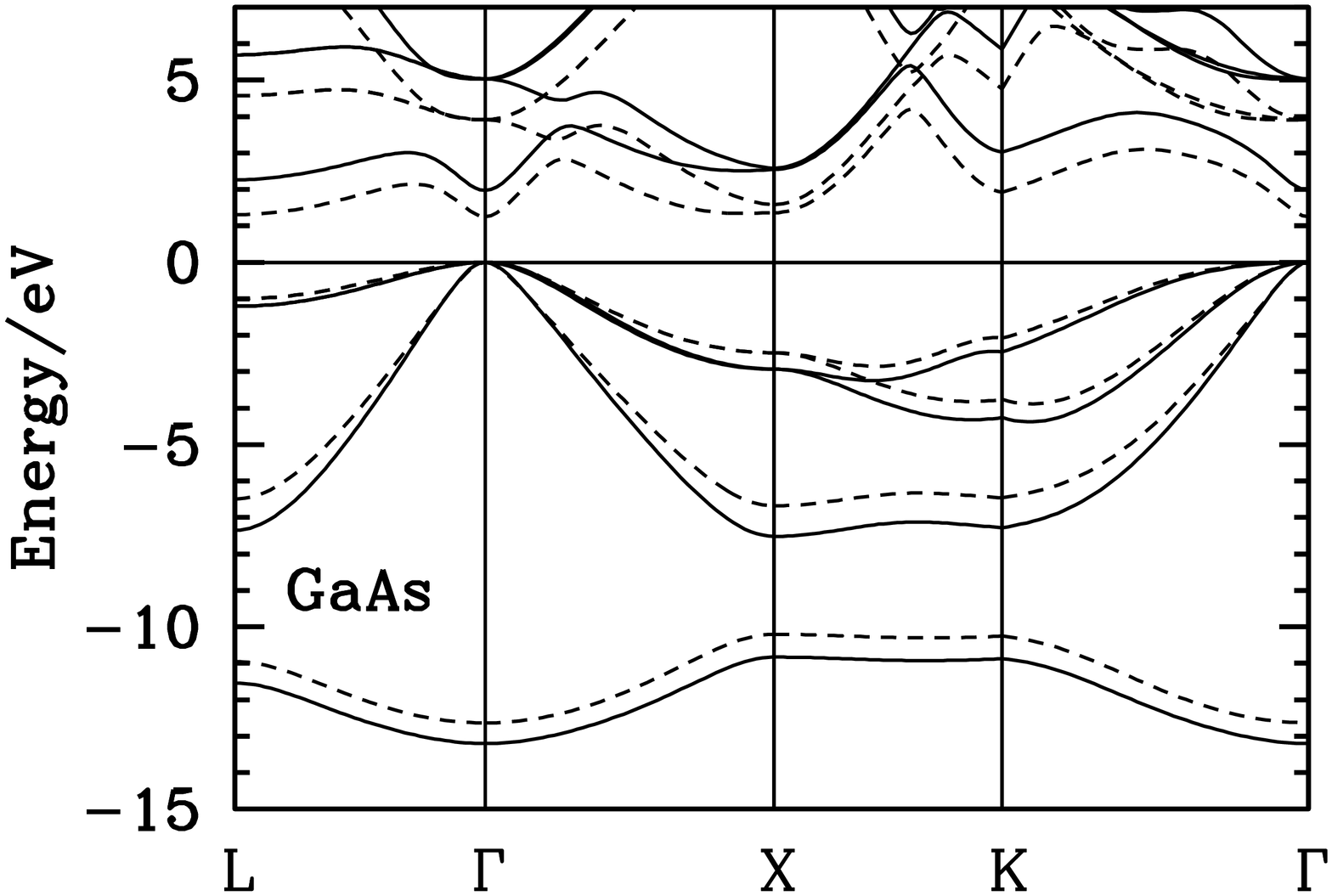}
\vspace{-3mm}
\caption{(From Fig. 4 of Ref. \cite{campo10}) The band structure of Si (top) and GaAs (bottom). Continuous lines represent GGA+U+V results and dashed lines represent standard GGA results. All energies were shifted so that the top of valence bands are at zero energy.}
\label{fig2}
\end{figure}

Fig. \ref{fig2} shows a comparison between the band structures of Si and GaAs
obtained with GGA and GGA+U+V. It is evident from the figure that 
the increase in the 
band gap obtained with the ``+U+V" correction
is the result of an almost uniform shift of electronic energies 
(downward for valence, upwards for conduction states) that maintains 
the overall dispersion pattern.

These results confirm that the extended Hubbard correction is able to 
significantly improve the description of band insulators and semiconductors
with respect to GGA, providing a more accurate estimate of structural
and electronic properties. In view of the fact that these systems are normally
treated with hybrid functionals or SIC approaches, the good results
obtained with LDA+U+V are the demonstration that this approach has similar
capabilities. The inaccuracy of the LDA+U (with on-site interactions only)
is not inherent to the reference model but rather to the 
approximations used in its final expression.
These results also clarify that, within the single particle KS representation
of the $N$-electron problem, band and Mott insulators can be treated with
similar corrective approaches.

The fact that LDA+U+V can be equally accurate in the description of band
and Mott insulators opens to the possibility to use it in a broad range
of intermediate situations where (Mott) electronic localization coexists with
or competes against the hybridization of atomic states from neighbor atoms, 
(as, e.g., in magnetic impurities in semiconductors or metals, 
high T$_c$ superconductors, etc), or in the description of 
processes (such as, e.g., electronic charge transfers excitation 
\cite{himmetoglu12_2}) involving a significant variation in the degree of
electronic localization.

In a recent work \cite{kulik11_1}
LDA+U+V was used to study transition-metal dioxide molecules
(e.g., MnO$_2$). The inclusion of the inter-site interaction was found to be 
crucial to predict the electronic configuration, the equilibrium structure
and its deformations in agreement with experiments.
The extended corrective functional has also been used as the starting point
of DFT+DMFT calculations and it has been demonstrated
that the inclusion of the inter-site interaction at a static mean-field
level (with the DMFT calculation performed on atomic impurities) produced
results of the same quality of more computationally
intensive cluster-DMFT calculations \cite{anisimov12}.


More recently, LDA+U+V has been used to calculate the lowest excited state
energies
of phosphorescent Ir dyes~\cite{himmetoglu12_2} 
using the $\Delta$-SCF method~\cite{ziegler77}.
These molecular complexes are widely used as sensitizers in
organic electronic devices~\cite{baldo98,holmes09}. 
In fact, the strong spin-orbit coupling that characterizes 
their metallic centers, is able to change the spin state of the photo-excited
electron-hole pair from singlet to triplet, thus extending its
life-time (the recombination process
is significantly inhibited by selection rules) and improving the efficiency
of the device.

\begin{figure}[!ht]\begin{minipage}[b]{0.49\linewidth}
\centering
\includegraphics[width=1.1\textwidth]{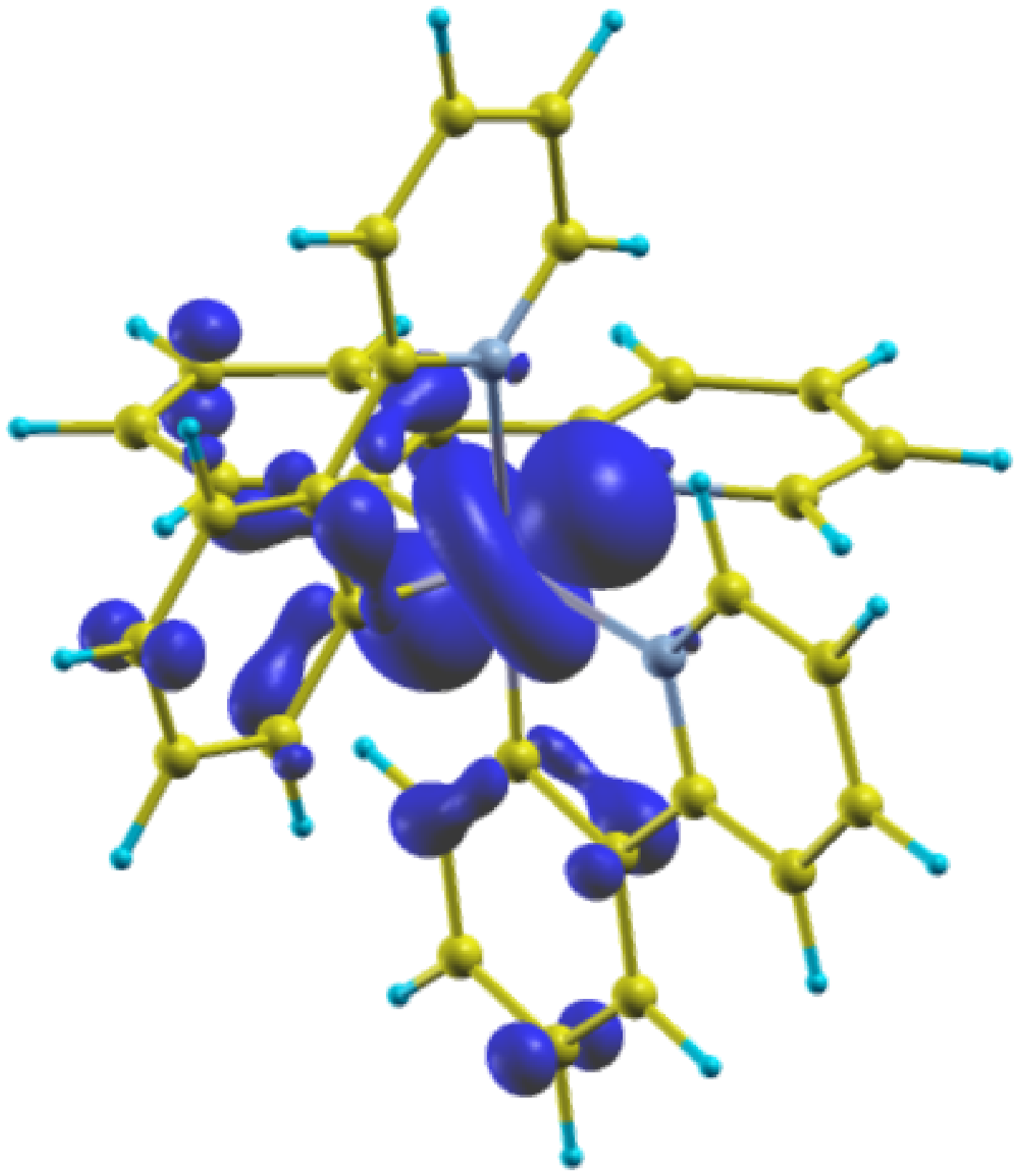}
\end{minipage}
\begin{minipage}[b]{0.49\linewidth}
\centering
\includegraphics[width=1.1\textwidth]{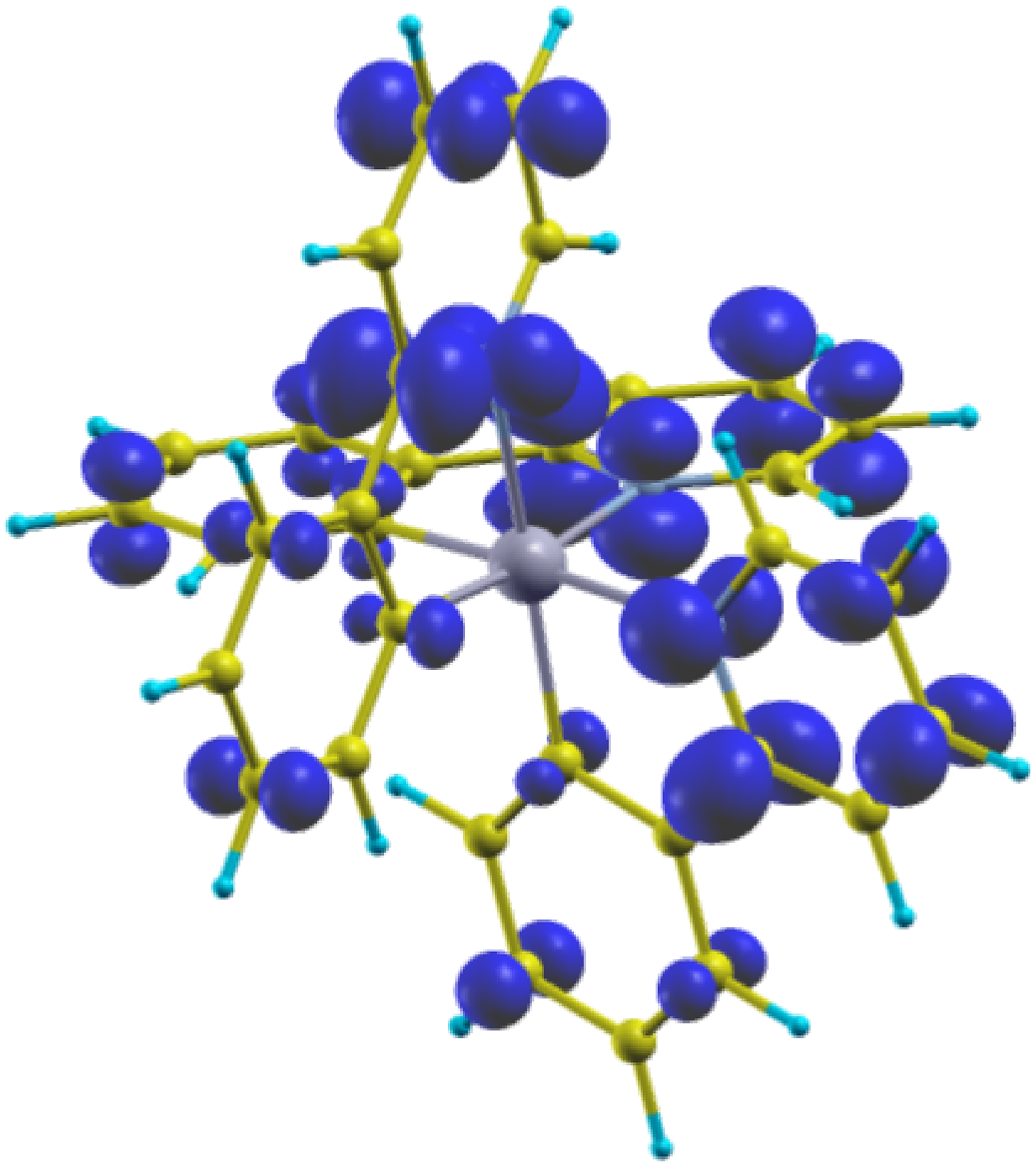}
\end{minipage}
\caption{\label{fig:irppy3}
(Adapted from Ref. \cite{himmetoglu12_2}; color online) Ir(ppy)$_3$ HOMO(left) and LUMO(right) calculated using the ground-state electronic density.}
\end{figure}

Approximate DFT functionals yield a
poor description of the electronic properties of these systems due to the 
localization
of electrons on Ir $d$ orbitals. The excited states of these molecules
are of metal-to-ligand charge transfer (MLCT) type, as illustrated in
Fig.~\ref{fig:irppy3} for one of the studied molecules, 
Ir(ppy)$_3$.
Consequently, an accurate calculation of excited state energies requires
that the functional used is able to capture 
the localization of electrons on the $d$
orbitals of Ir as well as their possible hybridization with the organic ligands.
In Ref \cite{himmetoglu12_2} it has been shown that
a straight use of the Hubbard $U$ correction on Ir $d$ orbitals
overlocalize electrons on the metal center,
suppresses their hybridization with organic
ligands, and results in a poor estimate of excited state energies. 
Instead, the Hubbard $V$ between the $d$ orbitals of 
Ir and those of neighbor atoms in the organic ligands corrects for the 
over-localization and gives 
results in good agreement with experiments~\cite{himmetoglu12_2}.
TDDFT can also be used to study the excitation energies of this system;
however, the MLCT character of these excitation imposes the use of hybrid
functionals that allow to capture the non-locality of the electronic
interactions.
%
\begin{table}[!ht]
\caption{\label{tabdscf} $\Delta$-SCF and TDDFT calculations of the 
lowest triplet and 
singlet states for three phosphorescent dyes: Ir(ppy)$_3$, FIrpic, and PQIr
(data from \cite{himmetoglu12_2}). 
All values are in $eV$ and measured from the ground state energy. The +U+V 
result is the one obtained after a preliminary structural optimization
of the molecule with this approach. TDDFT results were obtained with the 
Gaussian code \cite{g09}, using M06 \cite{m06} hybrid functionals.}
\begin{ruledtabular}
\begin{tabular}{cccccc}
Ir(ppy)$_3$ &  GGA  &  U  & TDDFT &  U+V  & Exp \\
\hline
T &  2.32  &  2.27  & 2.55  & 2.44 & 2.4\footnotemark[1] \\
S [NSP] &  2.50  &  2.90  & 2.79  & 2.73 & 2.6-2.7\footnotemark[2] \\
$\Delta\, E_{\rm TS}$ [NSP] & 0.18 & 0.62 & 0.24 & 0.29 & 0.2-0.3 \\
\hline
FIrpic &  GGA  &  U  & TDDFT & U+V & Exp \\
\hline
T &  2.46  &  2.79  & 2.66  & 2.52 & 2.6\footnotemark[3] \\
S [NSP] &  2.65  &  3.00  & 3.0  & 2.81 & 3.3-2.9\footnotemark[4] \\
$\Delta\, E_{\rm TS}$ [NSP] & 0.18 & 0.20 & 0.34  & 0.29 & 0.3-0.7 \\
\hline
PQIr &  GGA  &  U  &  TDDFT & U+V & Exp \\
\hline
T &  1.81  &  2.11  & 2.01   & 1.90 & 2.1\footnotemark[5]\\
S [NSP] &  2.01  &  2.34  & 2.37  & 2.18 & 2.3\footnotemark[6] \\
$\Delta\, E_{\rm TS}$ [NSP] & 0.19 & 0.23 & 0.36  & 0.28 & 0.2 \\
\end{tabular}
\end{ruledtabular}
\footnotetext[1]{From Refs. \cite{holmes09,baldo00,goushi04,eom09,adachi01,tsuboi03,
lamansky01,hofbeck10,colombo94,tsuboyama03,holzer05}.}
\footnotetext[2]{From Refs. \cite{holmes09,tsuboi03,lamansky01,holzer05}.}
\footnotetext[3]{From Refs. \cite{tsuboi08,ichimura87,adachi01-2,holmes03,tokito03,
you05,lee09,xiao11}.}
\footnotetext[4]{From Refs. \cite{tsuboi08,lee09}.}
\footnotetext[5]{From Refs. \cite{eom09,xiao11,su08}.}
\footnotetext[6]{From Refs. \cite{Dandrade04,Dandrade04-2,su08}.}
\end{table}
%
Table \ref{tabdscf} 
summarizes the excitation energies computed with LDA+U+V $\Delta$-SCF and TDDFT
techniques for three Ir phosphorescent complexes (Ir(ppy)$_3$, FIrpic, and 
PQIr) considered in Ref. \cite{himmetoglu12_2} and compares them 
with GGA and GGA+U $\Delta$-SCF calculations and available experimental data.
These results show that LDA+U+V represent a viable alternative to
more computationally intensive hybrid functionals 
to compute (either by $\Delta$-SCF or TDDFT) excitation energies that involve 
electron transfer processes. 
Ref. \cite{himmetoglu12_2} also rationalized the dependence of HOMO and LUMO 
energies in these molecules on the values of the Hubbard $U$ (on Ir $d$ states)
and $V$ (between Ir and its C and N nearest neighbors) showing a
predictable behavior of these excitation energies. These observations 
may provide valuable informations to tune the performance of these
molecules through the screening of substitutional impurities
in their ligand complexes.

\subsection{DFT+U+J: Magnetism and localization}
\label{dftpj}
While invariance
is unanimously recognized as a necessary feature of the corrective
functional, whether to use the full rotational invariant correction,
Eqs. (\ref{ub1}) and (\ref{ub2}), or its simpler version,
Eq. (\ref{our1}), has often appeared as a matter of taste and
has been dictated by the availability of either
implementation in current codes. 
In fact, the two corrective schemes give very
similar results for a large number of systems in which electronic
localization is not critically dependent on Hund's rule magnetism.
However, as mentioned in section \ref{simpler}, 
in some materials that have recently attracted considerable
interest, this equivalence does not hold and the explicit
inclusion of the exchange interaction ($J$) in the corrective functional 
appears to be necessary.
Examples of systems in this group include recently discovered Fe-pnictides 
superconductors \cite{nakamura09}, heavy-fermion \cite{bultmark09,pickett06},
non-collinear spin materials \cite{spaldin10}, or multiband metals,
for which the Hund's rule coupling, promotes, depending on the filling
of localized states,
metallic or insulating behaviors \cite{demedici11,georges11}.
In our recent work on CuO \cite{himmetoglu11} the necessity 
to explicitly include the Hund's coupling $J$
in the corrective functional was determined by a competition 
(likely to exist in other Cu compounds as well, such as 
high T$_c$ superconductors), 
between the tendency to complete the external 3$d$ shell and the 
onset of a magnetic ground state
(dictated by Hund's rule) with 9 electrons on the $d$ manifold.
The precise account of exchange interactions between
localized $d$ electrons beyond the simple approach of Eq. (\ref{our1})
(with $U_{eff} = U-J$) turned out to be crucial to predict the
electronic and structural 
properties of this material. 
In this work we used a simpler $J$-dependent corrective 
functional than the full rotationally invariant one to achieve this goal.
The expression of the functional can be obtained from the full
second-quantization formulation of the electronic interaction potential, 
given in Eq. (\ref{secq}),
by keeping only on-site terms that describe the interaction between up to
two orbitals. 
Approximating on-site effective interactions with the (orbital-independent)
atomic averages of Coulomb and exchange terms, 
$U^I = \frac{1}{(2 l + 1)^2}\, \sum_{i, j} \langle \phi_i^I \phi_j^I \vert V_{ee} \vert \phi_i^I \phi_j^I \rangle $
and
$J^I = \frac{1}{(2 l + 1)^2}\, \sum_{i, j} \langle \phi_i^I \phi_j^I \vert V_{ee} \vert \phi_j^I \phi_i^I \rangle $,
it is easy to derive the following expression:
\barr
E_{U+J}=E_{\rm Hub} - E_{\rm dc} &=& \sum_{I, \, \sigma}\, \frac{U^I - J^I}{2}\, {\rm Tr}[ {\bf n}^{I\, \sigma}\,
                                      ( {\bf 1} - {\bf n}^{I\, \sigma} ) ] 
\nonumber \\
            &+& \sum_{I, \, \sigma}\, \frac{J^I}{2}\,
                       {\rm Tr}[ {\bf n}^{I\, \sigma}\, {\bf n}^{I\, -\sigma} ]. 
                \label{dft_puj_1}
\earr
Comparing Eqs. (\ref{our1}) and (\ref{dft_puj_1}), one can see that the 
on-site Coulomb repulsion parameter ($U^I$) is effectively reduced
by $J^I$ for interactions between electrons of parallel spin and 
a positive $J$ term further discourages
anti-aligned spins on the same site, stabilizing magnetic ground states.
The second term on the right-hand side of equation (\ref{dft_puj_1}) can 
be explicated as $\sum_{I,\sigma}\,(J^I/2)\,\sum_{m,m'}  
n^{I\, \sigma}_{m\, m'}\, n^{I\, -\sigma}_{m'\, m}$ which shows how
it corresponds to an ``orbital exchange" between electrons of opposite spins 
(e.g., up spin electron going from $m'$ to $m$
and down spin electron from $m$ to $m'$). It is important to notice that this term is genuinely beyond Hartree-Fock. 
In fact, a single Slater determinant containing the four states $m \uparrow$ , $m \downarrow$, $m' \uparrow$ ,
$m' \downarrow$ would produce no interaction term like the one above. 
Thus, the expression of the $J$ term given in 
equation (\ref{dft_puj_1}), based on the product of
${\bf n}^{I\, \sigma}$ and ${\bf n}^{I\, -\sigma}$ is an 
approximation of a functional that would require
the calculation of the 2-body density matrix to be properly included.
However, 
the $J$ term in Eq. (\ref{dft_puj_1}) can be regarded as the
one needed to eliminate a term in the spurious curvature of the energy
deriving from the interaction between antiparallel spins. 
Therefore, its formulation and use in corrective functionals are legitimate.
Similar terms in the corrective functional have already been proposed
in literature \cite{demedici11,georges11,quan11,yoshitake11,pavarinibook11},
although within slightly different functionals.

Eq. (\ref{dft_puj_1}) represents a significant simplification with
respect to Eqs. (\ref{ub1}) and (\ref{ub2}) and proved crucial to predict
the insulating character of the cubic phase of CuO \cite{himmetoglu11}.

\begin{figure}[h!]
\includegraphics[width=0.3\textwidth]{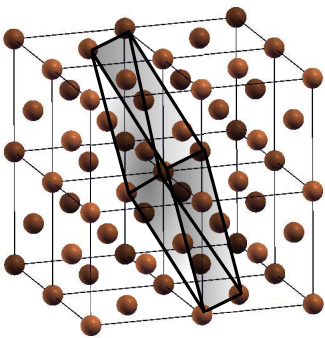}
\caption{ The triclinic supercell used for calculations. Only the Cu atoms are
shown for clarity. }
\label{figcuo_8}
\end{figure}

\begin{figure}[h!]
\includegraphics[width=0.4\textwidth]{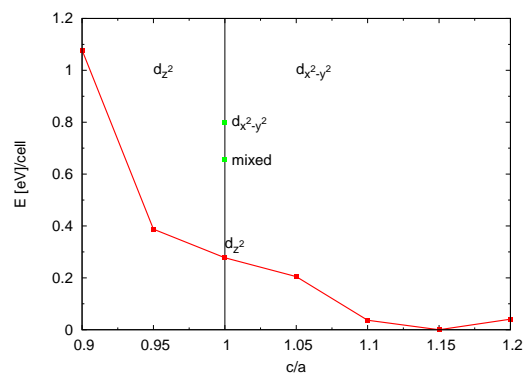}
\caption{(Adapted from Ref. \cite{himmetoglu11}; color online) 
Total state energy profile of CuO in the LDA+U+J insulating ground state
as a function of $c/a$.}
\label{figcuo-coa}
\end{figure}

Unlike other transition metal monoxides (all rhombohedral), 
CuO has a monoclinic unit cell. In addition, while exhibiting a
similar antiferromagnetic ground states (with ferromagnetic planes of Cu atoms 
alternating with opposite spins - the so-called AFII order),
its Ne\'el temperature is significantly lower than predicted from the
trend of other materials in this class, suggesting a weaker magnetic 
interaction between Cu ions. Although it is 
not the equilibrium structure of this 
system, the perfectly cubic crystal has been considered as a limiting case
of the tetragonal phase grown on selected substrates \cite{siemons2009tetragonal} 
or as a proxy system to study the electronic properties of cuprate
superconductors \cite{grant2008electronic}. 
While the analogy with other transition metal monoxides
would suggest an insulating behavior, the cubic phase
was invariably predicted to be metallic by approximate DFT, LDA+U
and hybrid functional calculations 
\cite{grant2008electronic,peralta2009jahn,chen2009hybrid}. 
This outcome is due to 
a ``double" (orbital and spin) degeneracy between the highest energy
$e_g$ states of each Cu atom where a hole should appear (Cu are nominally
in a 2+ oxidation state). As explained in section \ref{concept},
these degeneracies need to be lifted if a gap is to appear in the 
Kohn-Sham spectrum of the material. It is important to notice that the two 
degeneracies are mutually reinforcing: if 
spin states have the same energy the material is not 
magnetic and the symmetry of the 
crystal is perfectly cubic with an exact degeneracy between $e_g$ states.
The use of a triclinic super cell of the 
cubic structure, depicted in Fig. 
\ref{figcuo_8},
and an AFII magnetic order are sufficient to lift both these symmetries.
However, no insulating state is obtained if $J$ is set to 0.
In fact, the presence of the $p$ states of oxygen at the top of the
valence band together with 
Cu $d$ states makes the partial occupation of both manifolds
more energetically favorable than the localization of holes on
the $d$ states of Cu atoms (as necessary to obtain a finite magnetization).
In particular, a magnetic ground state with an inbalance of
population between Cu $d$ states of opposite spin and a (approximately)
complete O $p$ manifold has a slightly higher energy than a non magnetic ground
state with a larger number of electrons on Cu $d$ states (thus closer
to complete its $d$ shell) equally distributed between the two spin, and 
a hole spread between $d$ and $p$ levels that results in a metallic 
ground state.
A finite Hund's coupling $J$ favors the magnetic ground state, also resulting
in the complete localization of the hole on the Cu $d$ states and in the 
formation of a band gap.
The total energy of the cubic phase (insulating ground state)
as a function of the tetragonal distortion, shown in Fig. \ref{figcuo-coa}, 
presents a monotonic profile (although the material shows a different orbital
ordering with a hole occupying the $z^2$ state of Cu for $c/a < 1$ and 
the $x^2-y^2$ for $c/a > 1$).

The energy profile shown in Fig. \ref{figcuo-coa} was obtained recomputing
the interaction parameters for every value of $c/a$.
$U$, in particular, was calculated self-consistently, using the linear-response
approach illustrated in section \ref{linearr} on the LDA+U+J ground state
of the system (results are shown in Fig. 8 of Ref. \cite{himmetoglu11}).
The $J$, instead, was computed (through a generalization of
the linear-response technique based on the perturbative potential
of Eq. (\ref{dvm})),
only from the non magnetic GGA ground state of the cubic
phase and the same value was used for all the considered $c/a$
and lattice parameters.
In fact, as explained in section \ref{linearr}, the linear
response procedure applied to $m^I$ is less reliable and technically 
more difficult to apply when atoms assume a finite magnetization.
Furthermore, the smooth variation of $U_{sc}$ 
(about 1 eV from $c/a = 0.9$ to $c/a = 1.2$),
shown in Fig. 8 of Ref. \cite{himmetoglu11} 
suggests that the variation of $J$ with the lattice parameter
and $c/a$ can be safely neglected.

The monotonic decrease of the energy of the system with the tetragonal
distortion of its unit cell, is in contrast with previous studies 
\cite{peralta2009jahn,chen2009hybrid}
that obtained a double-well energy profile
with two minima corresponding to 
tetrahedral phases with $c/a < 1$ and $c/a > 1$, respectively,
but in agreement with experiments that reported a stable structure
only for $c/a > 1$.
The LDA+U+J functional was also able to predict the existence
of several orbital-ordered states (with the hole hosted on different
$d$ orbitals on different Cu atoms) whose energy is not much higher 
than the ground state one, as indicated in Fig. \ref{figcuo-coa}.
We argue that these states can play an important role at finite temperature
in this and similar Cu-O-based materials.

The study of CuO illustrates the effectiveness of the LDA+U+J in capturing
the ground state of systems where (intra-atomic, Hund's type)
magnetic interactions play an important role in determining the localization
of electrons on strongly correlated orbitals.
The simplicity of the formulation of the ``+U+J" corrective functional
of Eq. (\ref{dft_puj_1}) greatly facilitates its use and
the implementation of other algorithms (such as, for example, the
calculation of forces, stresses \cite{cococcioni10}
and phonons \cite{floris11}), discussed in section \ref{derivae},
that will be crucial to study, for example, the lattice vibration
of systems characterized by a strong Hund's coupling on strongly
localized electrons.

More complex aspects of magnetism in strongly correlated materials 
require the LDA+U to be extended to a non-collinear spin formalism,
as done, for example, in Ref. \cite{spaldin10}. This implementation is, in fact,
crucial to study correlated systems characterized by
canted magnetic moments, strong spin-orbit
interactions and magnetic anisotropy (quite common in rare earth 
compounds) \cite{pickett06}, spin-wave excitations (magnons)
\cite{essenberger11}. 
Due to the quantity of both formal and physical aspects that would
be necessary to discuss for its thorough tractation, this
extension of the LDA+U method will not be presented in this work, and
the interested reader is encouraged to refer to the above
mentioned literature for details.

\section{Energy derivatives}
\label{derivae}
One of the most important advantages brought about by the simple
formulation of the LDA+U corrective functional is 
the possibility to easily and efficiently 
compute total energy derivatives, as forces, 
stresses, dynamical matrices, etc. These are crucial quantities 
to identify and characterize
the equilibrium structure of materials under different 
physical conditions, to study the vibrational properties, to perform 
molecular dynamics calculations and
to account for finite ionic temperature effects (typically dominant 
in insulators).
In this section we will review the formalism to calculate forces, stresses, 
and second derivatives from a LDA+U ground state
(Refs. \cite{cococcioni10,floris11} for details). 
For most of its part we will assume that the variation of $U$ 
with the ionic positions and/or the lattice parameters
can be neglected; the importance of varying $U$
will be discussed in Sec. \ref{derivu}.
In the derivation of the Hubbard contribution to 
total energy derivatives, 
we will also assume a corrective functional based on atomic
orbitals as localized basis sets. This derivation can be easily generalized
to other basis sets, provided that the derivative of their localized
orbitals can be computed analytically.

\subsection{The Hubbard forces}
\label{force_hub} \index{LDA+U: calculation of forces}
The Hubbard forces are defined as the negative 
derivatives of the Hubbard energy with
respect to the atomic displacements. 
Taking the derivative of $E_U$ in Eq. (\ref{our1}) it is easy to arrive at 
the expression:
\begin{eqnarray}
\label{fh}
F^{U}_{\alpha, i} &=& - \frac{\partial E_{U}}{\partial
\tau_{\alpha i}} =
- \sum_{I,m,m',\sigma} \frac{\partial E_{U}}{\partial
n^{I\sigma}_{m,m'}}
\frac{\partial n^{I\sigma}_{m,m'}}{\partial \tau_{\alpha i}}\nonumber \\ 
&=&
- \frac{U}{2} \sum_{I,m,m',\sigma}(\delta_{mm'}-
2 n^{I\sigma}_{m'm})
\frac{\partial n^{I\sigma}_{m,m'}}{\partial \tau_{\alpha i}}
\end{eqnarray}
where $\delta \tau_{\alpha i}$ is the atomic displacement and
$n^{I\sigma}_{m,m'}$ are 
the occupation matrices, defined in
Eq. (\ref{occup}).
Since the Hellmann-Feynman theorem applies for energy first order derivatives,
no response of the electronic
wave function has to be taken into account and the symbol $\partial$
indicates only the explicit (``bare")
derivative with respect to atomic positions.
Based on the definition in Eq. (\ref{occup}) it is easy to work out the 
formula for the derivative of the atomic occupations:
\begin{eqnarray}
\label{dndt}
\frac{\partial n^{I\sigma}_{m,m'}}{\partial \tau_{\alpha i}} &=&
\sum_{k,v}f_{kv} [\frac{\partial}{\partial \tau_{\alpha i}}
(\bra \psi^{\sigma}_{kv}|\phi^{I}_{m'k}\ket) \bra \phi^{I}_{mk}|\psi^{\sigma}_{kv}\ket \nonumber \\
&+&
\bra \psi^{\sigma}_{kv}|\phi^{I}_{m'k}\ket \frac{\partial}
{\partial \tau_{\alpha i}}
\left( \bra \phi^{I}_{mk}|\psi^{\sigma}_{kv}\ket \right)
]
 \end{eqnarray}
where $k$ and $v$ are  k-point and band indexes, respectively.
The problem is then reduced to calculate
the quantities
\begin{equation}
\label{dpdt}
\frac{\partial}{\partial \tau_{\alpha j}}
\bra \phi^{I}_{m}|\psi^{\sigma}_{kv}\ket
\end{equation}
By virtue of Hellmann-Feynman theorem
the quantities in Eq. (\ref{dpdt})
are calculated considering only
the derivative of the atomic wavefunctions, which are explicitely dependent
on the ionic positions $\tau_{\alpha i}$:
\begin{eqnarray}
\label{dpdt1}
\frac{\partial}{\partial \tau_{\alpha j}}
\bra \phi^{I}_{m}|\psi^{\sigma}_{kv}\ket =
\bra \frac{d \phi^{I}_{m}}{d \tau_{\alpha j}}
|\psi^{\sigma}_{kv}\ket.
\end{eqnarray}
As the KS state $\psi^{\sigma}_{kv}$
is a Bloch function,
the only non-zero Fourier components are at ${\rm \bf k} + {\rm \bf G}$, 
where ${\rm \bf k}$ is a vector of the Brillouin zone (BZ) and
${\rm \bf G}$ is a reciprocal lattice vector.
Therefore, the reciprocal space representation of 
the product between atomic orbitals and KS wave functions reads:
\beq
\label{pfou}
\bra \phi^{I}_{m}|\psi^{\sigma}_{kv}\ket = 
\sum_{{\rm \bf G}} \left[c^I_m({\rm \bf k} + {\rm \bf G})\right]^*
a^{\sigma}_v({\rm \bf k} + {\rm \bf G})
\eeq
where $c^I_m$ and $a^{\sigma}_v$ are, respectively, 
 the atomic orbital $\phi^{I}_{m}$ and $\psi^{\sigma}_{kv}$ Fourier components.
The explicit dependence on the atomic positions is simply
obtained via a change of variable in the integral that defines the
Fourier transform, and it can be easily demonstrated that
\beq
 c^I_m({\rm \bf k} + {\rm \bf G}) = e^{-i({\rm \bf k} + {\rm \bf G}) \cdot
\tau_I}c^0_m({\rm \bf k} + {\rm \bf G})
\eeq
where $c^0_m$ is the Fourier component of the atomic wave function
of the atom $I$ when centered at the 
origin of the Cartesian system of reference. 
The structure factor  $e^{-i({\rm \bf k} + {\rm \bf G}) \cdot
\tau_I}$  is what determines the dependence of the product in Eq. (\ref{dpdt})
on the atomic positions. The following expression is easily obtained:
\barr
\label{dpfou}
&&\frac{\partial}{\partial \tau_{\alpha j}}
\bra \phi^{I}_{m}|\psi^{\sigma}_{kv}\ket = \nonumber \\
&&i~\delta_{I\alpha}\sum_{{\rm \bf G}} \left(c^I_m({\rm \bf k} + {\rm \bf G})\right)^*
a^{\sigma}_v({\rm \bf k} + {\rm \bf G})~({\rm \bf k} + {\rm \bf G})_j
\earr
where $({\rm \bf k}+{\rm \bf G})_{j}$ is the component of the vector
$({\rm \bf k}+{\rm \bf G})$
along the direction $j$ and the ``$i$" in front of the sum is the imaginary unit.
Due to  the Kronecker $\delta_{I\alpha}$,
this derivative 
is non zero only if the involved atomic
function is centered on the atom which is displaced. 
As a result, the Hubbard corrective functional only
contributes to forces on ``Hubbard" atoms. 
This conclusion does not hold for calculations using
ultrasoft pseudopotentials \cite{vanderbilt90}, that produce finite Hubbard
contributions also to forces on non-Hubbard atoms.
This special case is not explicitly treated in this review.

\subsection{The Hubbard stresses}
\label{force_hub} \index{LDA+U: calculation of stress}
\label{stres_hub}
From  the expression of the Hubbard energy functional in
Eq. (\ref{our1}) the Hubbard 
contribution to the stress tensor can be computed as: 
\begin{equation}
\label{dede}
\sigma_{\alpha \beta}^{U} =
-\frac{1}{\Omega}\frac{\partial E_{U}}{\partial \varepsilon_
{\alpha \beta}}
\end{equation}
where $\Omega$ is the volume of the unit cell (the energy $E_{U}$ is
referred to a single unit cell) and $\varepsilon_{\alpha \beta}$ is 
the strain tensor. This quantity can be defined from the deformation
of the crystal as follows:
\begin{equation}
\label{def}
{\bf r}_{\alpha}\rightarrow { \bf r'}_{\alpha} =
\sum_{\beta}(\delta_{\alpha \beta}+
\varepsilon_{\alpha \beta}){\bf r}_{\beta}
\end{equation}
where ${\bf r}$ is the space coordinate internal to the unit cell.
The calculation of the stress proceeds along the same
steps as  for the Hubbard forces
[Eqs. (\ref{fh}), (\ref{dndt})]. 
The problem is thus reduced to evaluating the derivative
\begin{equation}
\label{dpde1}
\frac{\partial}{\partial \varepsilon_{\alpha \beta}}
\bra \phi^{I}_{mk}|\psi^{\sigma}_{kv}\ket.
\end{equation}
This calculation will follow the procedure  presented 
in Ref. \cite{nielsen85}, where a theory for stress and 
force in quantum mechanical systems was introduced.
The functional dependence of atomic and KS wave functions
on the strain can be determined by deforming the lattice according
to Eq. (\ref{def}) and studying how these wave functions are
modified by the applied distortion (while preserving their normalization), 
in the assumptions this is
small enough to justify first order expansions around ${\mathbf \epsilon} = 0$.
The mathematical details of this calculation
can be found in Ref. \cite{cococcioni10} and won't be
repeated here.
The final expression of the derivative in Eq. (\ref{dpde1})
is:
\begin{eqnarray}
\label{der1}
\frac{\partial}{\partial \varepsilon_{\alpha \beta}}
\bra \phi^{I}_{m}|\psi^{\sigma}_{kv}\ket
|_{\varepsilon = 0} &=&
-\frac{1}{2}\delta_{\alpha \beta}
\bra \phi^{I}_{m}|\psi^{\sigma}_{kv}\ket \nonumber \\
&-& \sum_{G}e^{i({\bf k}+{\bf G})\cdot \tau_{I}}
a^{\sigma}_{v}({\bf k}+{\bf G})\times \nonumber \\
&\times& \partial_{\alpha}
[c^{I}_{m}({\bf k}+{\bf G})]^{*}
({\bf k}+{\bf G})_{\beta}.
\end{eqnarray}
The derivative of the Fourier components of the atomic wavefunctions 
($\partial_{\alpha} c^{I}_{m}({\bf q}) \equiv \partial c^{I}_{m}({\bf q})/\partial
q_{\alpha}$) depends on the
particular definition of the atomic orbitals. As its expression can
  vary according to different implementations, it will will not be detailed here. 

\subsection{Phonons and second energy derivatives}
\label{force_hub} \index{LDA+U: calculation of phonons}
\label{phon_hub}
Second (and higher) order derivatives of the total energy 
are crucial to characterize the vibrational properties of materials 
and a large number of connected quantities  
like Raman spectra, electron-phonon interactions, 
thermal conductivity, etc. Effective Born charges,  
dielectric and piezo-electric tensors are also evaluated considering total 
energy derivatives. 
It is therefore important to 
have the capability to compute second and higher order energy 
derivatives from first principles. 
This task has represented a considerable challenge when studying correlated
systems, for which corrective schemes beyond standard DFT approximations 
have to be usually employed.
In most of these schemes,  
due to the complexity of the corrective functional and the
consequent difficulty in computing derivatives analytically, 
frozen phonon techniques are normally used.
However, these supercell-based techniques are efficient only 
at high-symmetry points of the reciprocal space, but prohibitively
expensive elsewhere, rendering somewhat problematic 
the convergence of quantities that depend on
sums over the entire BZ. 
For these types of calculations affordable linear-response approaches 
are highly needed \cite{giannozzi91,gonze95,
gonze95_1,baroni01}. 
A linear response calculation of phonons
based on Green's functions (from a DFT+DMFT calculation)
has also been proposed in Ref. \cite{kotliar03} but the computational cost
of this method is prohibitively high for most systems.

In this section we review a recent extension \cite{floris11}
of the density functional perturbation theory (DFPT) \cite{baroni01} 
to the LDA+U energy functional. 
Thanks to the low cost of the LDA+U method and the efficiency of 
DFPT calculations at arbitrary phonon {\bf q} vectors, 
this implementation offers an excellent
compromise between accuracy and computational efficiency 
to calculate vibrational spectra of materials and properties 
related  to the  higher order 
derivatives of the Hubbard-corrected total energy. 

According to the $2n+1$ theorem, the $n^{th}$ order derivative of the 
many-body wavefunction give access to total energy derivatives 
up to the  $2n+1$ order. Therefore, in a DFT framework, 
second (and higher) order derivatives of the energy
require the calculation of the first
order derivative of the ground state density.
This is the main quantity computed by DFPT that obtains it 
from the self-consistent solution 
of linear-response equations applied to
the DFT ground state.
We refer to Ref. \cite{baroni01} for an extensive tractation
and for the definition of the notation used here.
In the following, we specifically treat 
total energy  second derivatives with respect to atomic
positions for the calculation of phonons, but the results can be  
extended to derivatives with respect to any 
couple of parameters the Hamiltonian depends on.

The displacement $\lambda\equiv\{L\alpha\}$ of 
an atom $L$ along the direction $\alpha$ from
its equilibrium position induces 
a response $\Delta^{\lambda} n({\rm \bf r})$ of the electronic charge density
that leads to a variation 
$\Delta^{\lambda}V_{SCF}$ of the self-consistent 
KS potential $V_{SCF}$.
The Hubbard potential $V_{Hub} = \sum_{I\sigma mm'}U^{I}\left[\frac{\delta_{mm'}}{2}-n^{I\sigma}_{mm'}\right]|\phi^I_{m'}\rangle\langle \phi^I_{m}|$,
also responds to the atomic displacements and its variation, to 
be added to $\Delta^{\lambda}V_{SCF}$, reads:
\begin{eqnarray}
\label{dvu}
\Delta^{\lambda} V_{Hub} &=&
\sum_{I\sigma mm'} U^{I}
\left[\frac{\delta_{mm'}}{2}-n^{I\sigma}_{mm'}\right] \times \nonumber \\
&& \times \left[ |\Delta^{\lambda} \phi^I_{m'}\rangle \langle\phi^I_{m}| + |\phi^I_{m'}\rangle \langle \Delta^{\lambda}\phi^I_{m}| \right] \nonumber \\
&-& \sum_{I\sigma mm'}U^{I} \Delta^{\lambda} n^{I\sigma}_{mm'}|\phi^I_{m'}\rangle
\langle\phi^I_{m}|
\end{eqnarray}
where 
\begin{eqnarray}
&&\Delta^{\lambda}  n^{I\sigma}_{mm'} = \nonumber \\
&&\sum_i^{occ} \{\langle 
\psi_i^{\sigma} |\Delta^{\lambda} \phi^I_{m}\rangle \langle \phi^I_{m'}|\psi_i^{\sigma}\rangle
+ \langle \psi_i^{\sigma}|\phi^I_{m}\rangle  \langle \Delta^{\lambda} \phi^I_{m'}|
\psi_i^{\sigma}\rangle \}  + \nonumber \\
&&\sum_i^{occ} \{ \langle  \Delta^{\lambda} \psi_i^{\sigma} |\phi^I_{m}\rangle  \langle\phi^I_{m'}|\psi_i^{\sigma}\rangle
+  \langle \psi_i^{\sigma}|\phi^I_{m}\rangle \langle\phi^I_{m'}|\Delta^{\lambda}\psi_i^{\sigma}\rangle \}.
\label{dni}
\end{eqnarray}
In Eq. (\ref{dni})
$|\Delta^{\lambda} \psi_i^{\sigma}\rangle$ is the linear response of the KS state
$|\psis \rangle$ to the atomic displacement and is computed 
self-consistently with  $\Delta^{\lambda}V_{SCF}$ (contaning the 
Hubbard contribution), during the solution of the DFPT 
equations \cite{baroni01}.

%
Once  $\Delta^{\lambda} n({\rm \bf r})$ is obtained,
the dynamical matrix of the system can be computed to calculate the phonon
spectrum and the vibrational modes of the crystal.
The Hubbard energy contributes to the dynamical matrix with the term
\barr
\Delta^{\mu}(\partial^{\lambda} E_{\rm Hub}) &=& \sum_{I\sigma m m'} U^I\left[
\frac{\delta_mm'}{2} - n^{I\sigma}_{mm'}\right] 
\Delta^{\mu}\left(\partial^{\lambda} n^{I\sigma}_{mm'}\right) \nonumber \\
&-&\sum_{I\sigma m m'} U^I \Delta^{\mu}n^{I\sigma}_{mm'}\partial^{\lambda}n^{I\sigma}_{mm'}
\label{d2e}
\earr 
which is the total derivative of the Hellmann-Feynman Hubbard 
forces [Eq. (\ref{fh})]. Again, in Eq. \ref{d2e},
the symbol $\partial^{\lambda}$ indicates ``bare"
derivatives, while $\Delta^{\mu}$ also includes linear-response contributions
(i.e., variations of the KS wave functions).

In ionic insulators and semiconductors a 
non-analytical term $C_{I\alpha,J\beta}$ must be added to the dynamical
matrix to account for the coupling of longitudinal
vibrations with the macroscopic electric field generated 
by the ion displacement \cite{bornlibro54,cochran62}. 
This term, responsible for the LO-TO splitting at 
${\rm \bf q} = \Gamma$, depends on the Born effective charge
tensor ${\bf Z}^*$ and the high-frequency dielectric tensor
$\varepsilon^{\infty}$: $C_{I\alpha,J\beta} = \frac{4\pi e^2}{\Omega}
\frac{({\rm \bf q}\cdot{\rm \bf Z}^*_I)_{\alpha}
({\rm \bf q}\cdot{\rm \bf Z}^*_J)_{\beta}}{{\rm \bf q} \cdot
\stackrel{\leftrightarrow}{\varepsilon^{\infty}} \cdot {\rm \bf q}}$.
These quantities can be computed from the  transition amplitude 
$\langle \psi_{c,{\rm \bf k}}|[H_{SCF},{\rm \bf r}]|
\psi_{v,{\rm \bf k}}\rangle$ \cite{giannozzi91},
where $c$ and $v$ indicate 
conduction and valence bands, respectively.
Due to its non-locality, the Hubbard potential 
contributes to this quantity with the following term:
\barr
\label{comhub}
&&\langle \psi_{c,{\rm \bf k}}|[V_{\rm Hub}^{\sigma},{\bf r}]|
\psi_{v,{\rm \bf k}}\rangle = \nonumber \\
&&\sum_{Imm'} U^I\left[ \frac{\delta_{mm'}}{2}
-n^{I\sigma}_{mm'}\right] \times \nonumber \\
&&\times\left[-i\langle\psi_{c,{\rm \bf k}}|\frac{d}{d{\bf k}}\left(|\phi^I_{m,{\rm \bf k}}
\rangle \langle \phi^I_{m',{\rm \bf k}}|\right)|\psi_{v,{\rm \bf k}}\rangle\right]
\earr
where $\phi^I_{m,{\rm \bf k}}$ are Bloch sums of atomic wave functions. 

Eqs. (\ref{dvu}), (\ref{d2e}) and (\ref{comhub}) completely define
the extension of DFPT to LDA+U, introduced in Ref. \cite{floris11}
and implemented in the 
PHONON code of the {\sc Quantum ESPRESSO} package
\cite{giannozzi09}. 

\begin{figure}[h]
 \vspace{-5mm}
 \centering
 \includegraphics[width=0.37\textwidth,angle=270]{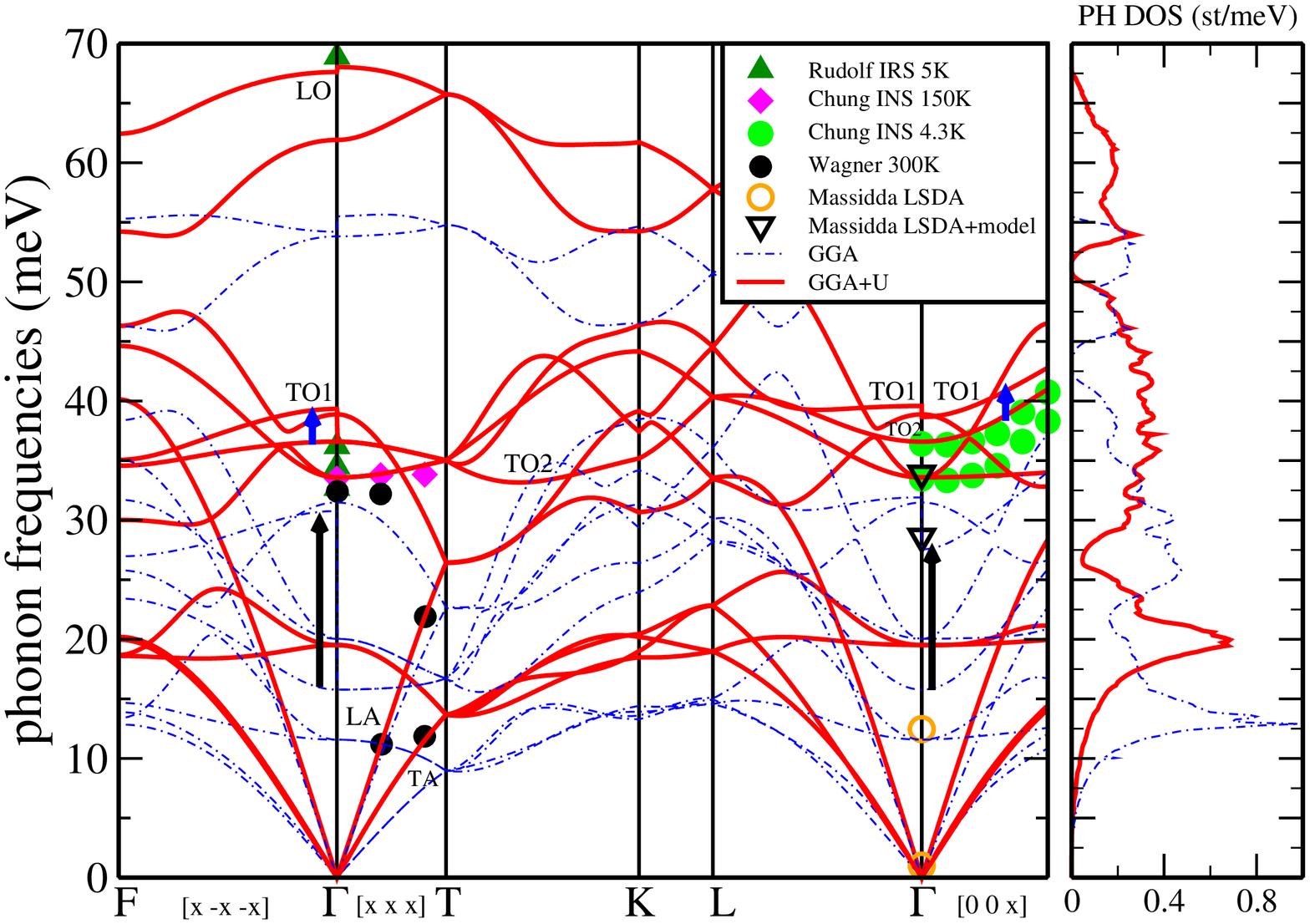}
 \hspace{0mm}
 \includegraphics[width=0.37\textwidth,angle=270]{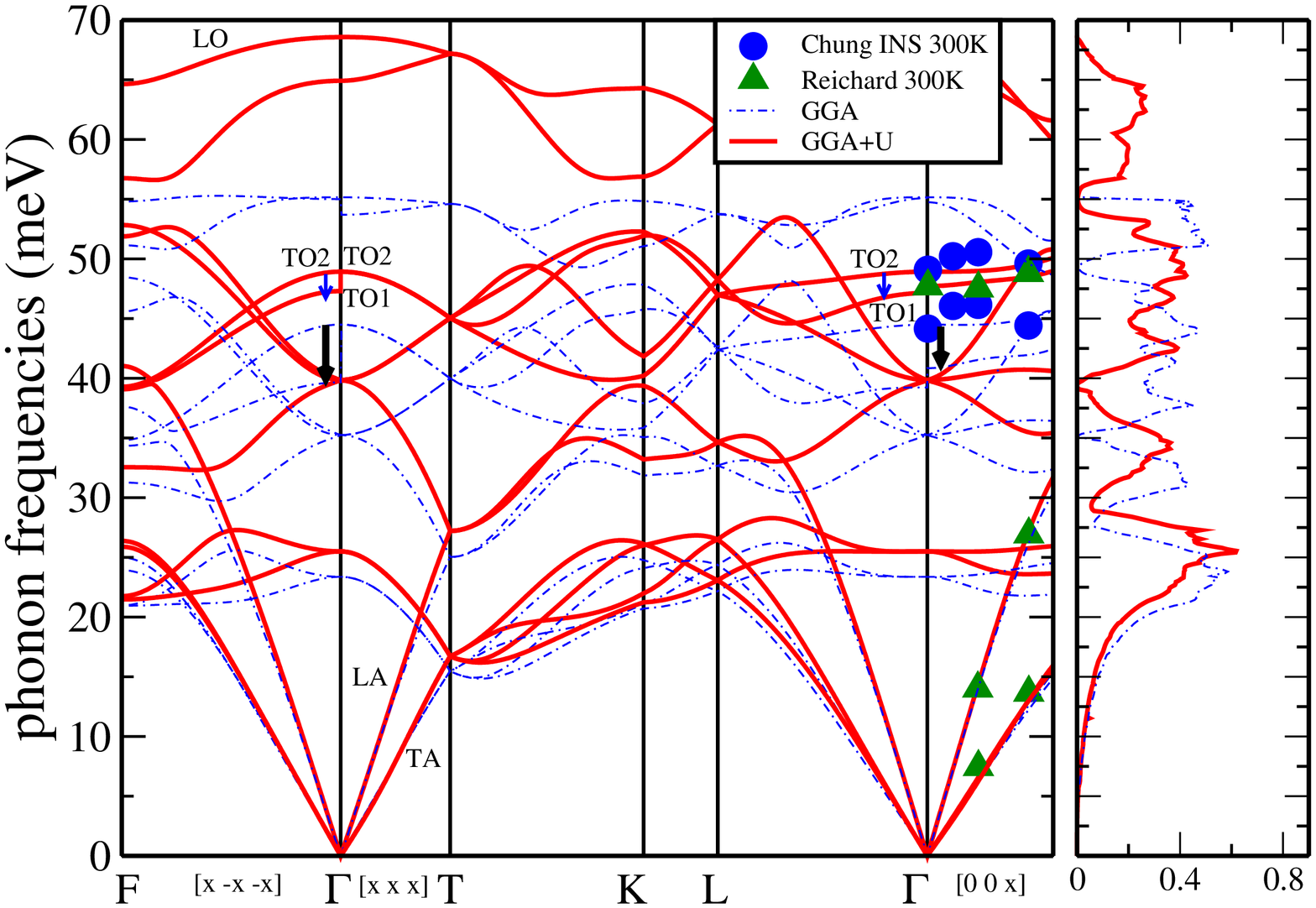}
 \vspace{-2mm}
 \caption{(Adapted from \cite{floris11}; color online) 
Phonon dispersion and
vibrational DOS of MnO (top) and NiO (bottom), calculated with 
GGA (dashed blue lines) and GGA+U (thick solid red lines). Blue 
(black) arrows mark the GGA+U (GGA) magnetic splittings and their sign.
Upper plot: filled symbols represent experimental data 
\cite{chung03,rudolf08,haywood71}, and
open symbols the results from other calculations (at zone center) 
\cite{massidda99}.
Lower plot: symbols represent experimental data \cite{chung03,reichardt75}.}
\label{phu}
 \vspace{-3mm}
\end{figure}

As a case study we present below the 
calculations of the vibrational spectrum of MnO and
NiO (Fig. \ref{phu}) \cite{floris11}. The Hubbard $U$ for both systems
was computed using the linear-response method discussed in 
section \ref{linearr} and resulted 5.25 eV for Mn and 5.77 eV for Ni.

MnO and NiO crystallize in the 
cubic rock-salt structure but acquire a rhombohedral symmetry due to their
antiferromagnetic order consisting of ferromagnetic planes of cations
alternating with opposite spin along the [111] direction. 
Because of the lower symmetry, the cubic diagonals loose their equivalence which leads to 
the splitting of the transverse optical modes (with oxygen and metal
sublattices vibrating against each-other) around the zone center 
\cite{massidda99}.
\index{transition-metal oxides: vibrational properties}
Fig. \ref{phu} compares the vibrational spectrum of MnO and NiO 
obtained with GGA and GGA+U.
As evident from the figure, the Hubbard correction produces an overall
increase in the 
phonon frequencies of both materials, significantly 
improving the agreement with available experimental results 
\cite{chung03,rudolf08,haywood71,reichardt75}. 
In particular, it recovers the agreement for the 
TO and LO modes, strongly underestimated ($\approx$ 15 meV) by GGA.
The phonon frequencies computed from the GGA+U ground state
lead to a reduced splitting between TO modes compared
to GGA, which is also in better agreement with experimental data. 

These results demonstrate that
 electronic correlations have significant effects on the 
structural and  vibrational properties and that the calculation of
quantities involving phonons and their interaction with other excitations
(e.g., Raman spectra or electron-phonon couplings) or requiring
the integration of the vibrational frequencies across the Brillouin
zone (e.g., to account for finite ionic temperature effects) must 
employ  corrective functionals accounting for strong 
correlation.

\subsection{Derivatives of U}
\label{derivu}
In all the previuos sections, while discussing the contribution of the 
Hubbard functional to energy derivatives,
the Hubbard $U$ was kept fixed. In fact, its dependence
on the atomic positions and/or the cell parameters is usually assumed to
be small and neglected. This is, of course, an approximation, 
whose validity 
should be tested carefully, case by case. 
In fact, some recent works have shown that accounting for the variation 
of $U$ with the ionic positions and 
with lattice parameters can be important to obtain 
quantitatively predictive results.
In Ref. \cite{hsu09}, focused on the properties of the low-spin
ground state of LaCoO$_3$ under pressure, the Hubbard $U$ was 
recalculated for every volume explored. This structurally-consistent $U$
proved to be crucial to predict the value of pressure-dependent 
structural parameters (such as, lattice spacing, rhombohedral angle, 
Co-O distance and bond angles) in good agreement with experimental data. 
In Ref. \cite{tsuchiya06} the linear-response calculation of the $U$
as a function of the unit cell volume (or the applied pressure) 
allowed for a precise evaluation
of the pressure-induced high-spin to low-spin transition 
in (Mg$_{1-x}$Fe$_{x}$)O Magnesiow\"ustite for different Fe concentrations.

The complexity of the analytic expression of the Hubbard $U$ makes it
difficult to account for its variation with the atomic position and
lattice parameters. However, a recent article \cite{kulik11} has 
introduced a method to efficiently compute the
derivative $dU^I/dR^J$ that allows to capture (at least at first order)
the variation of $U$ with the ionic position.
This extension is based on the linear-response approach 
to compute $U$ \cite{cococcioni05} discussed in section
\ref{computeu}. Starting from Eq. \ref{ucalc}, it is easy to work out
the formal expression of
the derivative of the Hubbard $U$ with respect to the atomic 
positions (atomic and direction indexes are dropped for simplicity): 
\beq
\label{dudr}
\frac{dU}{d{\bf \rm \tau}} = \frac{d}{d{\bf \rm \tau}}\left(\chi_0^{-1} - 
\chi^{-1}\right) = \chi^{-2}\frac{d\chi}{d{\bf \rm \tau}} - 
\chi_0^{-2}\frac{d\chi_0}{d{\bf \rm \tau}}.
\eeq
The derivative of the respose functions can be also evaluated starting
from their definition:
\beq
\label{dchidr}
\frac{d\chi}{d{\bf \rm \tau}} = \frac{d}{d{\bf \rm \tau}}\frac{dn}{d\alpha} =
\frac{d}{d\alpha}\frac{dn}{d{\bf \rm \tau}}\approx
\frac{d}{d\alpha}\frac{\partial n}{\partial{\bf \rm \tau}}.
\eeq
The approximation made in the last equality of this expression 
allows to use the ``bare" derivative of the atomic occupations to compute the
derivative of the response matrices. This is the quantity used in 
the calculation of the Hubbard forces and expressed in Eq. \ref{dndt}.
In practice, the derivative of the response matrices $\chi$ and $\chi_0$
can be obtained from the same linear response calculation used to
compute their values, evaluating the response
of the derivative of the atomic occupations (Eq. \ref{dndt})
to the perturbation added to the
potential, Eq. \ref{dv} (refer to Ref. \cite{kulik11} for details). 
In Ref. \cite{kulik11} this approach is used to account for the
variation of $U$ with atomic positions during chemical reactions involving
bi-atomic molecules. It is demonstrated that a configuration-dependent
effective interaction parameter significantly improves the quantitative 
description of the potential energy surfaces the system explores
with respect to calculations (quite common in literature) using 
the same (average) value of $U$ for all the configurations. 

The promising results obtained in Ref. \cite{kulik11} give hope that analogous
implementations could actually be used to estimate the dependence of $U$
on the strain of a crystal, 
$dU^I/d\epsilon_{\alpha \beta}$,
(analogous formulas apply) 
to be used in the calculation of the stress,
and also to compute the contribution to second derivatives. 
If the derivatives of the Hubbard $U$ could be evaluated automatically
(possibly from the expression of the effective interaction in terms
of atomic orbitals) its inclusion in calculations
would greatly improve the accuracy
of molecular dynamics simulations based on LDA+U \cite{sit06,sit07}. 

\section{Summary and outlook}
\label{summary}

Based on a simple corrective functional modeled on the Hubbard Hamiltonian,
the LDA+U method is one of the most widely used computational 
approaches to correct the inaccuracies of approximate DFT exchange
correlation functionals in the description of systems characterized
by strongly correlated (and typically localized) electrons.
Much of the popularity this method has gained in the scientific
community employing DFT as main computational tool
is certainly related to the fact that it is straightforward 
to implement in existing DFT codes, is very simple to use, 
and carries a very marginal additional computational cost with
respect to ``standard" functionals. 
These features, together with the possibility to tune the strength of
the Hubbard correction through the numerical value of a single, easy-to-control
interaction parameter, have contributed to establish LDA+U as
a semiemprical method, or as a tool for a
rough (and mostly qualitative) assessment of the 
effects of electronic correlation on the physical properties of a
given system.
As a consequence, LDA+U has often been regarded as a semiquantitative
approach (as is the Hubbard model it is based on) or, at most,
as a first order correction to approximate DFT functionals
to provide a starting point for more
sophisticated and supposedly more accurate
approaches. 

In this article we have discussed the LDA+U method 
with the aim to assess its potential and to clarify the conditions
under which it can be expected to provide a quantitative description
of correlated materials.
The analysis was based on the review of the theoretical foundation
of this method, 
a description of its most common 
formulations and implementations, and a discussion of its framing in 
the context of DFT, highlighting the typical problems it aims to address,
and the quality of the correction it provides.
The discussion about open issues of LDA+U (e.g., the use 
of specific localized basis sets, the invariance of the corrective functional, 
the choice of the double counting term) and the illustration of the results
obtained from the study of specific materials
gave us the possibility to explore the descriptive potentials 
of this corrective approach and to remark and understand its limits.
Through a review of the existing literature it was shown, for example,
how LDA+U improves the evaluation of the band gap of insulators,
the description of the structural and magnetic properties of correlated
systems, the energetics of electron transfer processes and chemical
reactions. We also pointed out many 
difficulties this method encounters in describing
the properties of metals or, in general, systems with more delocalized
electrons.
Particular emphasis was put on the necessity to have a method to 
compute the effective electronic interactions from first 
principles. A linear response approach to this problem
was described in detail and compared with other methods
present in literature.
Recent extensions to the formulation of the Hubbard corrective functional
were then presented, along with the beneficial effects they have on the accuracy
of LDA+U and on the range of systems and phenomena
that can be reliably modeled by this method.
In particular, it was shown that the inclusion of inter-site Coulomb 
interactions in the Hubbard corrective functional
enables the description of systems where electronic localization
occurs on hybridized orbitals or for which excitations from the
ground state may be accompanied by electron-transfer processes.
At the same time, the extension of the Hubbard corrective Hamiltonian
to explicitly include magnetic (Hund's rule) couplings was discussed 
highlighting the very significant improvements it brings to 
the treatment of materials where the onset
of a magnetic ground state favors or competes with electronic localization or
where spin and charge degrees of freedom are intimately coupled in determining
the physical behavior of electrons on strongly localized orbitals.
Finally, it was shown that the possibility to easily define and
implement energy derivatives of the corrective functional renders
LDA+U a quite unique tool to capture the effects of electronic
correlation and Coulomb-driven localization on the structural properties
of materials (such as, e.g., equilibrium lattice parameters, elastic
constants, etc), to perform molecular dynamics simulations, to compute
the vibrational spectrum and related quantities (Raman spectra, electron-phonon
couplings, etc).

Notwithstanding the inherent limits of this 
approach (such as, e.g., its static
character and the consequent inability to capture dynamical, frequency 
dependent effects), we argue that LDA+U represents a very useful
computational tool to model correlated systems that, while significantly
improving the accuracy of approximate DFT energy functionals, presents
marginal computational costs, thus enabling the possibility of 
calculations that would be impossible or overly expensive 
if based on a more accurate description of the quantum many-body features
of the electronic ground state.

LDA+U can provide quantitatively
precise predictions about the physical beavior of various systems, provided
the appropriate formulation of the corrective functional is used and the 
effective electronic interactions are precisely computed for all the
atomic sites the Hubbard correction acts on.
Further extensions and refinements to the Hubbard corrective functional 
are however highly desirable to improve the accuracy and the numerical
efficiency of the LDA+U method.
The automatic calculation of the effective electronic interactions (e.g.,
using the expression in Eq. (\ref{umat0})) would certainly represent a useful
extension to current implementations as it would avoid the need
of separate (e.g., linear-response) calculations to achieve
this result, and would allow to account for the variation of the 
electronic couplings with the atomic positions, the magnetic configuration,
and the physical conditions the material is subject to.
A more flexile expression of the corrective functional (including,
e.g., selected multi-site and multi-orbital interaction terms of the Hubbard
Hamiltonian) would be also very important to improve its descriptive capability
and its numerical accuracy, as demonstrated already by the definition of
the LDA+U+V and the LDA+U+J approaches. Extensions of this kind
require, however, a very substantial theoretical work that should identify: 
$i)$ the precise conditions under which the mean-field-like Hubbard functional
of LDA+U can be expected to contribute the necessary 
corrections for a precise representation of the many-body features of
electronic interactions; $ii)$ the extensions to the expression
of the corrective functional necessary to make it work effectively in more
general cases (e.g., metals, paramagnetic insulators, etc).
We hope that this work will be completed in the near future and 
will result in the definition of a generalized corrective functional able
to describe 
degenerate ground states of correlated systems without 
breaking/lowering the symmetry of the electronic charge density,
to automatically distinguish genuine
metallic ground states from the overlap 
of degenerate insulating solutions,
and to describe the main effects of possible interplays between charge 
and spin degrees of freedom.

\section{Acknowledgements}
Authors would like to thank Dr. S. Fabris and Dr. K. Yang for providing figures
2 and 6 respectively.
M. C. acknowledges support from the NSF CAREER award, DMR 1151738. 

\appendix
\section{Linear response calculation of {\bf U}}
\label{app1}
In this appendix linear-response theory will be used to derive the explicit
expression of the Hubbard $U$, as computed in Eq. (\ref{ucalc}).
To this purpose it is useful to start from a generalization of the
perturbing potential in Eq. (\ref{dv}) that is based on a non-diagonal
projector operator:
\beq
\Delta V_{ext}^{IJ} = \alpha^{IJ}_{ij} |\phi^I_i\rangle\langle\phi^J_j|
\label{deltavext}
\eeq
where the upper and lower case letters label atomic sites and electronic
states, respectively.
In linear response-theory the variation of a (Kohn-Sham) wave function
can be computed as follows (comprehensive indexes $m$ and $n$ are 
understood to run over k-points and bands while spin is omitted for simplicity):
\beq
\label{dpsi}
\Delta\psi_n = \sum_{m \neq n} \psi_m \frac{\langle\psi_m |
\Delta V_{scf} |\psi_n\rangle}{\epsilon_n - \epsilon_m }.
\eeq
In this equation $\Delta V_{scf}$ is the total variation of the
self-consistent potential containing, besides $\Delta V_{ext}$, a term
originating from the response of the electronic system that reads:
\beq
\Delta {\hat V}_{\rm resp} ({\rm \bf r}) = 
\left( \Delta V_H ({\rm \bf r}) + \Delta V_{\rm xc} ({\rm \bf r}) \right)\  {\hat K}({\rm \bf r}) \,\,\, , \,\,\, 
\label{new-dvresp}
\eeq
where  ${\hat K}({\rm \bf r}) = \vert {\rm \bf r} \rangle \langle {\rm \bf r} \vert$ and 
\barr
\Delta V_H ({\rm \bf r}) = \int \frac{\Delta \rho({\rm \bf r}')}{\vert {\rm \bf r}-{\rm \bf r}' \vert} d{\rm \bf r}' \,\,\, , \,\,\, \nonumber \\
 \Delta V_{\rm xc} ({\rm \bf r}) = \int \frac{\delta v_{\rm xc}({\rm \bf r})}{\delta \rho({\rm \bf r}')}\, \Delta \rho({\rm \bf r}') d{\rm \bf r}' .
\earr
To proceed we will assume that the atomic basis set
$\{\phi^I_i\}$ is orthogonal and complete. Therefore the identity
can be resolved as:
\beq
\mathbf{1} = \sum_{Ii} |\phi^I_i\rangle \langle\phi^I_i|.
\label{unit}
\eeq
Also, a generalized occupation matrix, analogous to the one defined in
Eq. (\ref{occupij}), is needed: 
\beq
n^{IJ}_{ij}=\sum_{n}f_{n}
\langle \phi_{i}^{I} | \psi_{n}\rangle
\langle \psi_{n}|\phi_{j}^{J}\rangle.
\label{occup1}
\eeq
For the sake of simplicity, 
the spin index is dropped for KS
states (or can be considered included in their comprehensive index $n$).
The $f_n$ coefficients
represent the occupations of the KS states (determined from the
Fermi-Dirac distribution of the corresponding eigenvalues around the Fermi
energy). 
in this part only the case of semiconductors/insulators
will be discussed (metals would require to explicitly
account for the response of the $f_n$, as shown,
e.g., in Ref. \cite{baroni01}).
Using Eq. (\ref{dpsi}) the response of the occupation matrix to
the external perturbation can be expressed as follows:
\barr
\Delta n^{IJ}_{ij} &=&
\sum_n f_n \left[ \langle \psi_n |\phi_{j}^{J}\rangle\langle \phi_{i}^{I} | \Delta \psi_{n}\rangle +
\langle \Delta \psi_n |\phi_{j}^{J}\rangle\langle \phi_{i}^{I} | \psi_{n}\rangle \right] \nonumber\\
&=& 2 ~Re \sum_{n \in {\rm v}} \sum_{m \in {\rm c}} 
\frac{
\langle \phi_{i}^{I} |\psi_m\rangle
\langle \psi_m|\Delta V_{scf}|\psi_n\rangle 
\langle \psi_n |\phi_{j}^{J}\rangle 
}
{\epsilon_n - \epsilon_m}.
\label{dn}
\earr
The last equality is a consequence of the fact that only the case of
semiconductors/insulators is discussed in this part 
(metals would require to explicitly
account for the response of the $f_n$, as shown,
e.g., in Ref. \cite{baroni01}) which allows to classify KS states as
valence ($v$) or conduction ($c$) states according to
their occupation ($f_n = 1$ and $f_n = 0$, respectively). 
If the system were non interacting $\Delta V_{resp} = 0$, $\Delta V_{scf} =
\Delta V_{ext}$. In this case the interacting and non-interacting 
response matrices coincide: ${\mathbf \chi} = {\mathbf \chi}_0$.
These conditions are verified 
if the Hartree and xc potential are the same as
in the non perturbed ground state of the system
(when the self-consistent solution of the KS equations
has been reached) or in the perturbed calculation (which starts
from the unperturbed potential and wave functions) at the
very first iteration of the diagonalization process when 
these terms of the electron-electron interaction potential
have not yet responded to the perturbation in the external
potential.
In fact, in these circumstances, the electronic system responds as 
an independent electron gas of the
same density as the real one. Substituting $\Delta V_{scf}$ with
$\Delta V_{ext}$ in Eq. (\ref{dn}), $\Delta_0 n^{IJ}_{ij}$ 
is obtained. The following equalities define the response
matrices (tensors) $\chi$ and $\chi_0$:
\barr
\label{dn1}\Delta n^{IJ}_{ij} &=& \sum_{KLkl}(\chi_0)^{IJLK}_{ijlk}\langle \phi_k^K | \Delta 
V_{scf}
| \phi_l^L \rangle \nonumber \\
&=& \sum_{KLkl}\chi^{IJLK}_{ijlk}\langle \phi_k^K | \Delta V_{ext}
| \phi_l^L \rangle \nonumber \\
&=& \sum_{KLkl}\chi^{IJLK}_{ijlk} 
\alpha^{KL}_{kl}
\earr
Based on the discussion above, the last equality can be used to 
evaluate $\chi_0$
at the first iteration of the perturbed run. 
From Eqs. (\ref{deltavext}), (\ref{unit}), (\ref{dn}) and (\ref{dn1}) 
the following expression can be easily derived:
\beq
(\chi_0)^{IJLK}_{ijlk} = 2 ~Re \sum_{n \in v} \sum_{m \in c} \frac{\langle \phi_{i}^{I} |\psi_m\rangle\langle \psi_n|\phi^J_j\rangle 
\langle \phi^L_l |\psi_n\rangle
\langle \psi_m |\phi_{k}^{K}\rangle}
{\epsilon_n - \epsilon_m}.
\label{chi0}
\eeq
In order to obtain the explicit expression of the Hubbard $U$ it is
convenient to rewrite Eq. (\ref{ucalc}) in a Dyson-like form:
\beq
\label{dyson}
\chi = \chi_0 + \chi_0~{\rm \mathbf U}~\chi
\eeq
where ${\rm \mathbf U}$ represent the interaction matrix (tensor) of
elements $U^{IJKL}_{ijkl}$. In order to compute the total
response matrix $\chi$ we can use the last equality of Eq. (\ref{dn1}):
\barr
\label{dn2}
\Delta n^{IJ}_{ij} &=& \sum_{KLkl}\chi^{IJLK}_{ijlk}\langle \phi_k^K | \Delta V_{ext} | \phi_l^L \rangle \nonumber \\
&=& \sum_{KLkl}\chi^{IJLK}_{ijlk} \alpha^{KL}_{kl}.
\earr
Therefore, by definition, 
$\chi^{IJLK}_{ijlk} = \Delta n^{IJ}_{ij}/\alpha^{KL}_{kl}$
(to be understood as a finite-difference approximation of a derivative).
Note the inversion in the order of indexes from $\chi$ to $\alpha$: as
will be evident from the formulas below, this is
required for the covariance of the theory with respect to 
rotations of the basis set.
Using the completeness of the localized orbital basis set (Eq. (\ref{unit})),
it is convenient to rewrite Eq. (\ref{dn}) as follows:
\begin{widetext}
\beq
\label{dn3}
\Delta n^{IJ}_{ij} = 2\,
~Re \sum_{n \in v} \sum_{m \in c} \sum_{OP}\sum_{op}
\frac{\langle \psi_n |\phi_{j}^{J}\rangle\langle \phi
_{i}^{I} |\psi_m\rangle\langle \psi_m|\phi^O_o\rangle\langle
\phi^O_o|\Delta V_{scf}|\phi_p^P\rangle \langle \phi_p^P|\psi_n\rangle}
{ \epsilon_n - \epsilon_m}.
\eeq
Using Eqs. (\ref{deltavext}) and (\ref{new-dvresp}) the quantity
$\langle\phi^O_o|\Delta V_{scf}|\phi_p^P\rangle$ can be easily
expressed as follows:
\beq
\label{dvscf}
\langle\phi^O_o|\Delta V_{scf}|\phi_p^P\rangle = \alpha^{OP}_{op} +
\langle\phi^O_o|\Delta V_{resp}|\phi_p^P\rangle = \alpha^{OP}_{op} +
\int \int (\phi^O_o({\rm \bf r}))^*\phi_p^P({\rm \bf r})
f_{Hxc}({\rm \bf r},{\rm \bf r}')
\Delta \rho({\rm \bf r}')~d{\rm \bf r}~d{\rm \bf r}'
\eeq
where $f_{Hxc}$ is the Hartree and exchange-correlation interaction kernel:
$f_{Hxc}({\rm \bf r},{\rm \bf r}') = \frac{1}{|{\rm \bf r}-
{\rm \bf r}'|}+\frac{\delta v_{xc}({\rm \bf r})}{\delta \rho({\rm \bf r}')}$.
At this point we need the explicit expression of $\Delta \rho({\rm \bf r})$.
Using again the completeness condition, Eq. (\ref{unit}), Eq. (\ref{dpsi}) 
can be rewritten as:
\beq
\label{dpsi1}
\Delta\psi_n({\rm \bf r}) = \sum_{m \neq n} \sum_{I i} \phi^I_i({\rm \bf r})
\frac{\langle \phi^I_i |\psi_m\rangle\langle\psi_m |
\Delta V_{scf} |\psi_n\rangle}{\epsilon_n - \epsilon_m}.
\eeq
Using Eq. (\ref{dn3}) we easily obtain: 
\begin{eqnarray}
\label{drho}
\Delta \rho({\rm \bf r}) &=& \sum_n f_n \left[  \psi_n^*({\rm \bf r})\Delta\psi_n({\rm \bf r}) +
\psi_n({\rm \bf r})\Delta\psi_n^*({\rm \bf r})\right] =
2 ~Re \sum_{RrSs}\sum_{n \in v}\sum_{m \in c} \phi^S_s({\rm \bf r})^*
\phi^R_r({\rm \bf r})\frac{\langle\phi^R_r|\psi_m\rangle\langle \psi_n |\phi_s^S\rangle\langle\psi_m |
\Delta V_{scf} |\psi_n\rangle}{\epsilon_n - \epsilon_m}\nonumber \\
&=&\sum_{RrSs} \phi^S_s({\rm \bf r})^*
\phi^R_r({\rm \bf r}) \Delta n^{RS}_{rs} =
\sum_{Rr}\sum_{Ss}\sum_{KLkl} \phi^S_s({\rm \bf r})^* \phi^R_r({\rm \bf r})
\chi^{RSLK}_{rslk}\alpha^{KL}_{kl}.
\end{eqnarray}
Inserting this expression in Eq. (\ref{dvscf}) we obtain:
\beq
\label{dvscf1}
\langle\phi^O_o|\Delta V_{scf}|\phi_p^P\rangle = \alpha^{OP}_{op} +
\sum_{Rr}\sum_{Ss}\sum_{KLkl} \left[ \int \int (\phi^O_o({\rm \bf r}))^*\phi_p^P({\rm \bf r})
f_{Hxc}({\rm \bf r},{\rm \bf r}')
\phi^S_s({\rm \bf r}')^* \phi^R_r({\rm \bf r}')~d{\rm \bf r}~d{\rm \bf r}'\right]
\chi^{RSLK}_{rslk}\alpha^{KL}_{kl}.
\eeq
Once this expression is used in Eq. (\ref{dn3}) we finally arrive at:
\begin{eqnarray}
\label{dn4}
\Delta n^{IJ}_{ij} &=& 2\,
~Re \sum_{n \in v} \sum_{m \in c} \sum_{OPop}
\frac{\langle \psi_n |\phi_{j}^{J}\rangle\langle \phi
_{i}^{I} |\psi_m\rangle\langle \psi_m|\phi^O_o\rangle
\langle \phi_p^P|\psi_n\rangle}
{\epsilon_n - \epsilon_m}\alpha^{OP}_{op} \nonumber \\
&+& 
2\,~Re \sum_{n \in v} \sum_{m \in c} \sum_{OPop}\sum_{RSrs}\sum_{KLkl}
\frac{\langle \psi_n |\phi_{j}^{J}\rangle\langle \phi
_{i}^{I} |\psi_m\rangle\langle \psi_m|\phi^O_o\rangle
\langle \phi_p^P|\psi_n\rangle}
{\epsilon_n - \epsilon_m}
\times \nonumber \\
&\times&
\left[ \int \int (\phi^O_o({\rm \bf r}))^*\phi_p^P({\rm \bf r})
f_{Hxc}({\rm \bf r},{\rm \bf r}')
\phi^S_s({\rm \bf r}')^* \phi^R_r({\rm \bf r}')~d{\rm \bf r}~d{\rm \bf r}'\right
]
\chi^{RSLK}_{rslk}\alpha^{KL}_{kl}.
\end{eqnarray}
Taking the derivative of both members with respect to $\alpha^{KL}_{kl}$
and using Eq. (\ref{chi0}), the following Dyson-like
expression is finally arrived at:
\beq
\label{umat}
\chi^{IJLK}_{ijlk} = (\chi_0)^{IJLK}_{ijlk} + \sum_{OPop}\sum_{RSrs}
(\chi_0)^{IJPO}_{ijpo} \left[ \int \int (\phi^O_o({\rm \bf r}))^*\phi_p^P({\rm \bf r})
f_{Hxc}({\rm \bf r},{\rm \bf r}')
\phi^S_s({\rm \bf r}')^* \phi^R_r({\rm \bf r}')~d{\rm \bf r}~d{\rm \bf r}'\right
] \chi^{RSLK}_{rslk}.
\eeq
Direct comparison with Eq. (\ref{dyson}) yields:
\beq
\label{umat1}
U^{OPSR}_{opsr} = \int \int (\phi^O_o({\rm \bf r}))^*\phi_p^P({\rm \bf r})
\left[\frac{1}{|{\rm \bf r}-
{\rm \bf r}'|}+\frac{\delta v_{xc}({\rm \bf r})}{\delta \rho({\rm \bf r}')}\right]
\phi^S_s({\rm \bf r}')^* \phi^R_r({\rm \bf r}')~d{\rm \bf r}~d{\rm \bf r}'
\eeq
\end{widetext}
which is the central result of this appendix.
Eq. (\ref{umat1}) highlights that the effective interaction computed through
the linear response approach illustrated above is nothing else than
the expectation value of the $bare$ Hartree and xc kernels on
quadruplets of wave functions belonging to the chosen basis set (e.g.,
orthogonalized atomic orbitals).

In the linear-response calculation of $U$ introduced in Ref. \cite{cococcioni05}
the perturbation is applied uniformly to all the localized orbitals of
the same atom:
\beq
\label{cpert}
\Delta V_{ext}^I = \alpha^I \sum_i |\phi_i^I\rangle\langle\phi^I_i|
\eeq
and the response of the system is studied through the variations of the
trace of the occupation matrix on each site:
\beq
\label{cdn}
\Delta n^I = \sum_i \Delta n_{ii}^{II} =
\sum_i\sum_{Rr}\chi^{IIRR}_{iirr}\alpha_{rr}^{RR} = \sum_R \alpha^R
\sum_{ir}\chi^{IIRR}_{iirr}.
\eeq
In Eq. (\ref{cdn}) the last equality is justified by the fact that in the
procedure illustrated in Ref. \cite{cococcioni05} the strength of the perturbation
is uniform over all the states of the same atoms.
The response matrix computed by the procedure in Ref. \cite{cococcioni05} can
thus be expressed as:
\beq
\label{cchi}
\tilde \chi^{IR} = \sum_{ir} \chi^{IIRR}_{iirr}
\eeq
and, from Eq. (\ref{umat}), the following equation easily follows:
\barr
\label{cu}
\tilde\chi^{IR} &=& \tilde\chi^{IR}_0 + \sum_{ir} \sum_{KQTZ}\sum_{kqtz}
\left(\chi_0\right)^{IIQK}_{iiqk} U^{KQTZ}_{kqtz}\chi^{ZTRR}_{ztrr}
\nonumber \\
&\equiv&
\tilde\chi^{IR}_0 + \tilde\chi^{IQ}_0\tilde U^{QZ} \tilde\chi^{ZR}
\equiv \tilde\chi^{IR}_0 + A^{IR}
\earr
(repeated indexes are summed over).
Unfortunately, Eq. (\ref{cu}) is not closed in $\tilde\chi$ and $\tilde\chi_0$
and the last two equalities (second line) define the quantities 
$\tilde {\rm \bf U}$ and ${\rm \bf A}$.
The computed Hubbard $U$ thus results:
\begin{widetext}
\beq
\label{cu1}
\tilde U^{QZ} = (\tilde\chi_0^{-1})^{QR} A^{RS} (\tilde\chi^{-1})^{SZ} =
\sum_{RS}
(\tilde\chi_0^{-1})^{QR} \left[\sum_{rs} \sum_{KSTM}\sum_{kstm}\left(\chi_0\right)^{RRSK}_{rrsk} U^{KSTM}_{kstm}\chi^{MTSS}_{mtss}\right] (\tilde\chi^{-1})^{SZ}.
\eeq
\end{widetext}
In matrix notation this equation can be expressed as:
\beq
\label{umat1I}
\tilde {\rm \bf U} = ({\rm \bf \tilde \chi_0})^{-1}{\rm \bf A}~{\rm \bf \tilde \chi}^{-1}.
\eeq
From the definition given in Eq. (\ref{cu}) we also have:
\beq
\label{umat2}
\tilde {\rm \bf U} = ({\rm \bf \tilde \chi_0})^{-1} - {\rm \bf \tilde \chi}^{-1}
\eeq
Combining this equation with Eq. (\ref{umat1I}) to eliminate $\tilde \chi^{-1}$
we obtain:
\barr
\label{umat3}
\tilde{\rm \bf U} &=& (\tilde \chi_0)^{-1}{\rm \bf A}((\tilde \chi_0)^{-1} - \tilde{\rm \bf U}) \nonumber \\
&=& (\tilde\chi_0)^{-1}{\rm \bf A}(\tilde\chi_0)^{-1} - (\tilde\chi_0)^{-1}{\rm \bf A}\tilde{\rm \bf U}
\earr
from which, solving for $\tilde {\rm \bf U}$, gives: 
\beq
\label{uexp}
\tilde {\rm \bf U} = \frac{(\tilde \chi_0)^{-1}
{\rm \bf A}(\tilde \chi_0)^{-1}}{1+(\tilde \chi_0)^{-1}{\rm \bf A}}
\eeq
that shows a formal resemblance to the effective interaction, Eq. (\ref{crpa}), 
computed from cRPA. A more detailed discussion on the comparison
between linear-response and cRPA results is offered in the article,
after Eq. (\ref{ars}).
Notice that, with respect to the notation
of section \ref{computeu}, the meaning of quantities with and without 
``$~\tilde{}~$"
is reversed; for example, there $\tilde U$ indicated the fully orbital
dependent interaction parameter, here it represents the atomically averaged
(and orbital independent) one obtained from actual linear response calculations.

In order to understand how the unscreened interaction $U^{KSTM}_{kstm}$, 
whose value is typically in the 15-30 eV range, is screened
to an effective $\tilde U^{QZ}$ in the 2-6 eV range,
it is appropriate to analyze the expression in Eq. (\ref{cu1}).
To this purpose it is useful to separate in the response matrices
a site- and state- ``bi-diagonal" term from off-diagonal ones as follows:
\beq
\label{chi42}
\chi^{ZTSS}_{ztss} = \delta_{tz}\delta_{TZ} \chi^{TTSS}_{ttss} +
\bar\chi^{ZTSS}_{ztss}
\eeq
(the same decomposition is assumed for $\chi_0$). Assuming the dominance
of diagonal terms over off-diagonal ones in the occupation matrices, it is fair
to deduce that $\bar\chi$ is small with respect to the diagonal terms.
Also, due to the fact that the increase in the diagonal parts of the 
occupation matrices usually decreases off-diagonal ones, the
two terms of the right hand side of Eq. (\ref{chi42})
generally have opposite signs. Inserting Eq. (\ref{chi42}) in Eq. (\ref{cu1})
one easily obtains:
\begin{widetext}
\begin{eqnarray}
\label{cu2}
\tilde U^{QZ} &=& 
\sum_{RS}
(\tilde\chi_0^{-1})^{QR} \left[\sum_{rs} \sum_{KT}\sum_{kt}\left(\chi_0\right)^{RRKK}_{rrkk} U^{KKTT}_{kktt}\chi^{TTSS}_{ttss}\right] (\tilde\chi^{-1})^{SZ}
\nonumber \\
&+& \sum_{RS} 
(\tilde\chi_0^{-1})^{QR} \left[\sum_{rs} \sum_{KSTM}\sum_{kstm}\left(\bar\chi_0\right)^{RRSK}_{rrsk} U^{KSTM}_{kstm}\chi^{MTSS}_{mtss}\right] (\tilde\chi^{-1})^{SZ}
\nonumber \\
&+& \sum_{RS}
(\tilde\chi_0^{-1})^{QR} \left[\sum_{rs} \sum_{KSTM}\sum_{kstm}\left(\chi_0\right)^{RRSK}_{rrsk} U^{KSTM}_{kstm}\bar\chi^{MTSS}_{mtss}\right] (\tilde\chi^{-1})^{SZ}
+ \mathcal{O}(\bar\chi_0\bar\chi).
\end{eqnarray}
If the unscreened interaction can be safely approximated by its atomic average,
$U^{KKTT}_{kktt}\sim \bar U^{KKTT} \def \bar U^{KT}$, Eq. (\ref{cu2}) becomes: 
\begin{eqnarray}
\label{cu3}
\tilde U^{QZ} &=& \bar U^{QZ} 
+ \sum_{RS} 
(\tilde\chi_0^{-1})^{QR} \left[\sum_{rs} \sum_{KSTM}\sum_{kstm}\left(\bar\chi_0\right)^{RRSK}_{rrsk} U^{KSTM}_{kstm}\chi^{MTSS}_{mtss}\right] (\tilde\chi^{-1})^{SZ}
\nonumber \\
&+& \sum_{RS}
(\tilde\chi_0^{-1})^{QR} \left[\sum_{rs} \sum_{KSTM}\sum_{kstm}\left(\chi_0\right)^{RRSK}_{rrsk} U^{KSTM}_{kstm}\bar\chi^{MTSS}_{mtss}\right] (\tilde\chi^{-1})^{SZ}
+ \mathcal{O}(\bar\chi_0\bar\chi).
\end{eqnarray}
As explained above, the second and third terms on the r.h.s. 
of Eq. (\ref{cu3}) are negative (and dominant on the last) and are 
responsible for the significant difference in value between
$\tilde U^{QZ}$ and $\bar U^{QZ}$.
\end{widetext}


\newpage
\bibliographystyle{unsrturl}
\bibliography{refstot}

\begin{thebibliography}{100}
\bibitem{kyang13}
{Figure \ref{pco1} was kindly contributed by Dr. K. Yang, one of the developers
  of the AFLOW framework \cite{curta12,curta12_1,curta12_2}, that was employed
  in the structural optimization of the material. PbCrO$_3$ corresponds to the
  ICSD structure n. 160196.}
\end{thebibliography}

\end{document}